\documentclass[journal]{IEEEtran}

\usepackage{amsmath,amssymb}
\usepackage{graphicx}
\usepackage{cite}
\usepackage{url}
\usepackage{multirow}
\usepackage{booktabs}
\usepackage{subcaption}
\usepackage{algorithm}
\usepackage{algorithmicx}
\usepackage{xcolor}
\usepackage{algpseudocode}
\usepackage{array}
\usepackage{svg}

\title{
On Network-Aware Semantic Communication and
Edge-Cloud Collaborative Intelligence Systems
}

\author{
Murdadha Nasif, and Ahmed~Refaey~Hussein,~\IEEEmembership{Senior Member,~IEEE}
\thanks{M. Nasif, and A. R. Hussein are with the Department of Electrical and Computer Engineering, University of Guelph, Canada.}
}

\begin{document}
\maketitle

\begin{abstract}
Semantic communication and edge-cloud collaborative intelligence are increasingly recognized as foundational enablers for next-generation intelligent services operating under stringent bandwidth, latency, and resource constraints. By shifting the communication objective from bit-perfect delivery toward the transmission of task-relevant semantic representations, semantic communication enables adaptive tradeoffs among communication overhead, inference accuracy, computational load, and end-to-end latency. This survey provides a comprehensive and system-level synthesis of recent advances in semantic communication at the edge-cloud interface, encompassing architectural models for collaborative intelligence, representation learning and semantic abstraction techniques, network-aware and resource-adaptive semantic encoding strategies, and learning-driven optimization and orchestration mechanisms. Beyond efficiency considerations, the survey situates semantic communication within practical operational contexts, including security, trust, resilience, and scalability, drawing connections to zero-trust networking, physical-layer security, and emerging edge-cloud control paradigms. Finally, open challenges and research directions are identified, highlighting the role of semantic communication as a key building block for AI-native networking and 6G-ready intelligent systems.
\end{abstract}


\section{Problem Context and Semantic Communication Paradigm}
\label{sec:intro_survey}
\noindent
This section establishes the problem context and motivation for semantic communication and collaborative intelligence at the edge-cloud interface, emphasizing network variability, latency constraints, and the need for task-driven transmission and computation.


\label{ch01:intro}
    As cellular networks evolve to support an increasingly diverse set of data-intensive and latency-sensitive applications, next-generation (xG) networks are expected to fundamentally transform how information is produced, transported, and consumed. 
    Unlike previous generations, which primarily focused on maximizing channel capacity for reliable bit transmission, xG networks face a sustainability challenge driven by the exponential growth of data traffic. 
    Global mobile data volumes are projected to increase by nearly 200\% between 2024 and 2030, reaching zettabyte scales largely due to video streaming, immersive media, and Extended Reality (XR) services~\cite{Ericsson2024}. Under these conditions, the traditional ``bit-pipe'' communication paradigm is approaching its physical, energetic, and economic limits.
    
    Semantic Communication (SemCom) has emerged as a promising alternative to address this challenge by shifting the objective of communication from the faithful reproduction of symbols to the effective delivery of meaning. 
    By leveraging advances in Artificial Intelligence (AI) and Deep Learning (DL), semantic communication systems can identify, extract, and transmit only the information that is relevant to the receiver’s interpretation or task execution. 
    This paradigm enables substantial reductions in bandwidth consumption while maintaining or even enhancing Quality of Service (QoS), thereby supporting emerging xG applications such as immersive XR, collaborative robotics, and intelligent cyber-physical systems.

\subsection{The Semantic Paradigm Shift}
\label{ch01:sem_shift}
    The transition toward semantic-aware networking necessitates revisiting the foundations of classical information theory. 
    Since Shannon’s seminal work in 1949, communication system design has focused almost exclusively on optimizing symbol transmission under noise and bandwidth constraints. 
    While this framework has enabled remarkable progress, it fundamentally treats all transmitted bits as equally important, regardless of their semantic relevance. 
    As xG networks increasingly serve intelligent and task-driven applications, this assumption becomes a limiting factor. 
    Consequently, the communications community has begun to adopt a hierarchical interpretation of communication that explicitly accounts for meaning and task effectiveness, marking a shift toward what is often referred to as the "post-Shannon" era.

    \subsubsection{The Shannon-Weaver Model}
        The conceptual basis for semantic communication originates from Warren Weaver’s classification of communication problems, introduced alongside Claude Shannon’s mathematical theory. Weaver identified three distinct levels of communication~\cite{xin_semantic_2024}:
    
        \begin{itemize}
            \item \textbf{Level A (Technical Problem):} How accurately can symbols be transmitted over a channel? This level emphasizes bit reliability, error control, and Signal-to-Noise Ratio (SNR) optimization and has dominated the design of communication systems from 1G through 5G.
            \item \textbf{Level B (Semantic Problem):} How precisely do transmitted symbols convey the intended meaning? This level prioritizes preserving semantic information (e.g., the concept of an object or action) rather than the exact bit sequence.
            \item \textbf{Level C (Effectiveness Problem):} How effectively does the received meaning influence the desired action or outcome? Often referred to as goal-oriented communication, this level evaluates success based on task completion or utility.
        \end{itemize}

    Semantic communication directly addresses Level B and provides a foundation for Level C, making it particularly relevant for intelligent, task-driven xG applications.

\subsection{Deep Learning Enabled Semantic Architectures (DeepSC)}
\label{ch01:dl_semcom}
    Deep Learning has become the primary enabler for realizing semantic communication in practice. 
    Deep Learning Enabled Semantic Communication (DeepSC) systems employ neural networks to learn compact, task-relevant semantic representations directly from data, allowing the network to ``understand'' information before transmission rather than treating it as an opaque bit stream.
    
    \subsubsection{DeepSC for Text and Speech}
        The foundational DeepSC architecture leverages Transformer-based encoder-decoder models to process textual data. 
        Unlike conventional codecs, such as Huffman or arithmetic coding, which rely on statistical symbol frequencies, DeepSC learns semantic feature vectors that capture contextual meaning. 
        At the transmitter, a semantic encoder maps an input sentence $s$ into a dense latent representation, which is transmitted over the physical channel. 
        The receiver employs a semantic decoder to reconstruct the estimated sentence $\hat{s}$ from the received representation.
        
        Training is typically performed using a composite loss function that combines cross-entropy loss, ensuring syntactic correctness, with mutual information-based objectives that preserve semantic content even under low SNR conditions~\cite{9398576}. 
        This framework has been extended to speech through DeepSC-ST, where attention mechanisms and Recurrent Neural Networks (RNNs) extract linguistic and paralinguistic features such as tone and emotion. 
        By transmitting semantic tokens instead of raw audio samples, DeepSC-ST achieves intelligible reconstruction at significantly lower bitrates than conventional vocoders~\cite{weng2023deeplearningenabledsemantic}.
    
    \subsubsection{Generative and Knowledge-Base Enabled Semantic Communication}
        Recent advances further enhance semantic efficiency by integrating shared Knowledge Bases (KBs) and generative AI models. 
        In this paradigm, both the transmitter and receiver maintain synchronized repositories of domain knowledge, including object classes, textures, or environmental context. 
        Rather than transmitting complete visual data, the transmitter sends compact semantic prompts (e.g., "a red car turning left"), while the receiver employs local generative models, such as diffusion-based architectures, to reconstruct high-fidelity content.
        
        Only residual information not captured by the shared knowledge base is transmitted, enabling compression ratios that exceed classical entropy limits~\cite{Zhang_2025}. 
        This approach effectively trades communication bandwidth for computation, a principle that is particularly relevant in edge-cloud systems.

\subsection{Motivation}
\label{ch01:mot}
    Despite the promise of semantic communication, practical deployment in xG networks introduces new challenges. 
    Future networks are expected to support massive numbers of intelligent agents, real-time inference pipelines, and immersive applications under strict latency and reliability constraints. 
    While data volumes continue to grow toward zettabyte scales, spectral efficiency improvements at the physical layer have begun to plateau, exposing a fundamental bottleneck traditionally associated with the Shannon capacity limit.
    
    Conventional syntactic communication frameworks, based on separate source and channel coding, are ill-suited for this environment. 
    These systems treat all bits with equal importance, regardless of their contribution to task utility. 
    For example, in an autonomous driving video stream, background pixels are protected with the same rigor as safety-critical objects such as pedestrians. 
    This indiscriminate allocation of resources leads to inefficient bandwidth utilization, increased end-to-end latency, and poor scalability.
    
    Furthermore, deploying semantic models at the network edge introduces additional complexity. 
    Edge devices exhibit heterogeneous compute capabilities, memory constraints, and network conditions. 
    Ignoring these factors often results in violations of Service Level Agreements (SLAs), particularly for latency-sensitive applications. 
    Consequently, semantic communication systems must be designed not only for semantic efficiency but also for deployment realism and SLA compliance.

\subsection{Research Gap}
\label{ch01:research_gap}
    Although the theoretical foundations of semantic communication have matured rapidly, a substantial gap persists between existing semantic communication models and their practical deployment in edge-cloud systems. 
    Most prior work has focused on conceptual architectures, centralized inference, or wireless-only scenarios, often relying on idealized assumptions that do not hold in operational xG networks. 
    As a result, several critical challenges remain insufficiently addressed, limiting the applicability of semantic communication beyond controlled experimental settings.

    \begin{itemize}
        \item \textbf{Lack of multi-tier deployment models:} 
        The majority of existing studies assume a two-tier architecture (device-to-cloud), overlooking the increasingly dominant three-tier paradigm (device-edge-cloud) adopted in modern xG infrastructures. 
        This omission neglects the role of edge servers as intermediate semantic processing and orchestration points, where latency, compute capacity, and placement decisions critically influence end-to-end performance.
    
        \item \textbf{Unvalidated feasibility on edge hardware:} 
        State-of-the-art semantic communication models, including vision transformers and foundation-model-based encoders, are typically evaluated on GPU-equipped platforms. 
        Their computational feasibility on CPU-limited or resource-constrained edge devices is rarely analyzed, leaving a gap in understanding whether such models can meet real-time or interactive latency requirements under realistic deployment constraints.
    
        \item \textbf{Absence of SLA-driven performance modeling:} 
        Existing semantic communication frameworks seldom incorporate explicit end-to-end latency decomposition or Service Level Agreement (SLA) constraints into their design or evaluation. 
        Without SLA-aware modeling, it is difficult to assess whether semantic gains translate into actionable performance guarantees for latency-sensitive applications.
    
        \item \textbf{Limited network-aware adaptivity:} 
        Most adaptive semantic and joint source-channel coding schemes rely on physical-layer feedback (e.g., SNR) or offline retraining to handle changing conditions. 
        Such approaches are ill-suited to edge-cloud links, where performance variability is dominated by transport-layer bandwidth and round-trip latency, and where real-time, deterministic adaptation is required without retraining or feedback from the radio layer.
    \end{itemize}
    
    Addressing these gaps is essential to transition semantic communication from a primarily theoretical paradigm into a deployable, scalable, and SLA-compliant service for edge-cloud systems in beyond-5G and 6G networks.

\subsection{survey Contributions}
\label{ch01:contrib}
    This survey addresses the fundamental challenges of deploying semantic communication in edge-cloud systems under realistic compute and network constraints. 
    In contrast to prior work that primarily focuses on centralized or wireless-only scenarios, this work emphasizes deployability, SLA awareness, and transport-layer variability at the edge-cloud interface. 
    The main contributions of this survey are summarized as follows:

    \subsubsection{Distributed Swin Transformer Architecture for Edge-Cloud Semantic Communication}
        This survey proposes a distributed semantic communication architecture based on the Swin Transformer, explicitly partitioned across the edge and cloud tiers. 
        In the proposed design, semantic feature extraction is performed at the edge, while semantic decoding and task-driven inference are offloaded to the cloud. 
        This partitioning is motivated by the heterogeneous compute capabilities of edge devices and cloud servers and is aligned with emerging multi-tier edge-cloud deployments.
        
        Unlike monolithic semantic encoders, the proposed architecture enables high-capacity semantic representations while respecting edge resource constraints. 
        Through extensive latency and compute profiling, this work demonstrates that transformer-based semantic encoders introduce a non-negligible computational overhead at the edge, which fundamentally limits SLA compliance when deployed naively. 
        By explicitly modeling and evaluating this trade-off, the survey provides deployment-level insight into when semantic encoding should be executed at the edge versus offloaded, thereby bridging the gap between semantic communication theory and practical edge-cloud systems.
    
    \subsubsection{Network-Aware Semantic Transcoding for Latency-Constrained Edge-Cloud Links}
        This survey introduces a deterministic, network-aware semantic transcoding mechanism that dynamically adapts the transmitted semantic payload to backbone network conditions. 
        The proposed approach selectively transmits the most informative latent semantic channels based on real-time network telemetry, including available bandwidth and round-trip latency, without relying on physical-layer feedback or retraining.
        
        The transcoding mechanism enables a controllable latency-fidelity trade-off by reducing semantic payload size under constrained links and exploiting additional bandwidth to improve reconstruction quality when capacity is available. 
        Unlike fixed-rate semantic codecs or SNR-driven JSCC schemes, the proposed method directly targets the dominant impairments of edge-cloud communication, namely transport latency and throughput variability. 
        Experimental results demonstrate that network-aware semantic masking significantly reduces transmission latency and bounds bandwidth utilization while maintaining high perceptual reconstruction quality across a wide range of network conditions.
        
        These contributions establish semantic communication as an adaptive, telemetry-driven service suitable for SLA-governed edge-cloud deployments and provide a foundation for future extensions toward task-aware, multimodal, and tail-latency-sensitive semantic networking.

\subsection{survey Organization}
\label{ch01:thesis_org}
    The remainder of this survey is organized as follows. 
    section~\ref{ch02} reviews related work and presents the system model underlying the proposed framework. 
    section~\ref{ch03} introduces an SLA-driven performance prediction model for edge semantic communication deployments. 
    section~\ref{ch04} presents the semantic transcoding framework and its integration with fine-grained latency modeling across the edge-cloud continuum. 
    section~\ref{ch05} introduces a network-aware deterministic modulation strategy to optimize transmission latency further while maintaining near-full image fidelity. 
    section~\ref{ch06} concludes the survey and outlines directions for future research.

\subsection{section Summary}
\label{ch01:summary}
    This section positioned the survey within the broader evolution toward meaning-centric, latency-aware communication systems for xG networks. By highlighting the limitations of traditional bit-centric architectures and identifying key deployment gaps in existing semantic communication approaches, it established the motivation for the proposed distributed, SLA-aware semantic communication framework developed in the subsequent chapters.

\section{Learning Models for Semantic Representation and Transcoding}
\label{sec:models}
\noindent
This section surveys representation learning and generative modeling techniques that enable semantic encoding, semantic masking, and adaptive transcoding. Particular emphasis is placed on scalable vision backbones and transformer-based architectures for controlling the latency-fidelity operating point.


\label{ch02}

\subsection{Introduction and Scope of the section}
\label{ch02:intro}
    The objective of this section is to establish the theoretical, architectural, and network-level foundations necessary for the design and evaluation of semantic communication systems in edge-cloud environments. 
    While semantic communication has gained significant attention as a means to reduce bandwidth consumption and improve robustness, its deployment in real-world xG networks introduces new challenges that extend beyond physical-layer considerations. 
    In particular, edge-cloud systems are constrained by heterogeneous compute resources, transport-layer latency variability, and Service Level Agreement (SLA) requirements that are not adequately captured in existing semantic communication models.
    
    This section therefore, serves three complementary purposes. 
    First, it reviews the evolution of semantic communication, from early language-centric adaptive computation models to modern deep learning architectures for visual data. 
    Second, it examines the network substrate required to operationalize semantic communication, focusing on xG infrastructure, O-RAN disaggregation, and telemetry-enabled control. 
    Third, it formalizes end-to-end (E2E) latency modeling and SLA compliance as first-class design objectives, motivating the system architecture introduced at the end of the section.
    
    The material presented here directly informs the SLA-driven performance analysis in section~\ref{ch03}, the semantic transcoding framework in section~\ref{ch04}, and the network-aware adaptive mechanisms introduced in section~\ref{ch05}.

\subsection{Semantic Communication: From Theory to Practice}
\label{ch02:semcom}
    Semantic communication (SemCom) represents a fundamental shift away from classical communication paradigms by prioritizing the transmission of \emph{meaning} rather than strict syntactic correctness~\cite{qin2022semantic}. 
    Traditional communication systems treat all bits as equally important and are designed to maximize symbol fidelity, regardless of the semantic relevance of the information being conveyed. 
    In contrast, semantic communication explicitly accounts for the utility of transmitted information with respect to a downstream task, aiming to preserve only what is necessary for correct interpretation or decision-making.
    
    Enabled by advances in Artificial Intelligence (AI), Natural Language Processing (NLP), and representation learning, semantic transmitters can learn compact latent representations that capture intent, context, and task-relevant semantics~\cite{LIU2023,yang2022semantic}. 
    By operating in this learned semantic space, communication systems can significantly reduce payload size, computational overhead, and end-to-end latency—properties that are particularly critical for edge-cloud applications, where transport delay and compute heterogeneity dominate system performance.
    
    Crucially, semantic communication reduces latency not only through improved compression efficiency, but by avoiding unnecessary communication altogether. 
    Information that does not contribute meaningfully to the downstream task can be filtered or suppressed early in the pipeline, preventing redundant data from traversing bandwidth-constrained or latency-sensitive network segments. 
    This early elimination of semantically irrelevant information reduces both transport delay and downstream processing cost, providing a systems-level advantage that extends beyond conventional source coding.
    
    \subsubsection{Adaptive Computation as a Foundation for Semantic Communication}
        The conceptual roots of semantic communication can be traced to early work in text and speech processing, where adaptive computation naturally aligns with semantic complexity. 
        In these modalities, not all inputs require the same level of processing to extract meaning, motivating architectures that dynamically allocate computational resources.
        
        The Universal Transformer (UT), proposed by Dehghani \emph{et al.}, exemplifies this approach by combining self-attention with recurrent computation~\cite{dehghani2019}. 
        Unlike fixed-depth Transformer architectures, UT refines its representations over multiple recurrent steps, allowing more complex or ambiguous inputs to receive additional computation while simpler inputs converge earlier. 
        This dynamic refinement mechanism establishes a direct link between semantic complexity and computational effort.
        
        Building on this idea, Zhou \emph{et al.} introduced the Adaptive Universal Transformer (AUT) for communication systems by incorporating Adaptive Computation Time (ACT)~\cite{graves2017adaptivecomputationtimerecurrent,zhou2021semanticcommunicationadaptiveuniversal}. 
        ACT enables the model to halt computation once a sufficient level of semantic confidence is reached, thereby explicitly tying computation to semantic richness and channel conditions. 
        In communication settings, this mechanism demonstrates that both bandwidth usage and computational effort can be adaptively controlled based on semantic importance rather than fixed-rate constraints.
        
        Collectively, these works establish a foundational principle that underpins modern semantic communication systems: communication and computation resources should be allocated according to \emph{semantic utility} rather than raw data volume.
        
        This principle extends naturally beyond language and speech to visual data, where semantic relevance is inherently non-uniform across the input~\cite{10158528,9685667}. 
        In images, a small subset of regions—such as objects, faces, and salient structures—typically carries the majority of task-relevant information, while large portions of the background contribute marginally to semantic understanding. 
        Consequently, treating all visual content with equal transmission and processing priority leads to inefficient use of both network and computational resources.
        
        Recognizing and exploiting this semantic imbalance motivates the design of visual semantic communication architectures that selectively encode and transmit informative features while suppressing semantically redundant content. 
        This insight forms the conceptual foundation for the edge-cloud semantic communication models developed in this survey and motivates the use of adaptive, attention-based mechanisms in subsequent sections.

\subsection{Visual Semantic Communication Architectures}
\label{ch02:visual_semcom}
    Extending semantic communication from language and speech to visual data introduces additional challenges stemming from high dimensionality, spatial redundancy, and strict latency constraints. 
    Unlike textual modalities, images and video streams contain orders of magnitude more data, much of which is semantically redundant or weakly correlated with downstream task objectives. 
    Consequently, effective visual semantic communication architectures must simultaneously achieve three goals: extract task-relevant semantics, scale efficiently with image resolution, and remain computationally feasible for deployment at the network edge. 
    These requirements have driven a progression from convolutional joint source-channel coding models toward attention-based Transformer architectures, which offer greater flexibility in manipulating and prioritizing semantic features. 
    This section reviews this architectural evolution and motivates the selection of Transformer-based models—particularly the Swin Transformer—as the backbone for edge-cloud semantic communication systems.
    
    \subsubsection{From CNN-Based JSCC to Transformer-Based Models}
        The application of semantic communication to images gained momentum with deep learning–based Joint Source-Channel Coding (JSCC). 
        Bourtsoulatze et al. introduced DeepJSCC, a CNN-based framework that directly maps images to channel symbols, bypassing separate compression and channel coding stages~\cite{Bourtsoulatze_2019}. 
        While DeepJSCC demonstrates robustness to channel noise, it is typically trained offline and lacks explicit mechanisms for adapting to heterogeneous network and compute conditions.
        
\begin{figure}[h]
            \centering
            \includegraphics[width=0.5\linewidth]{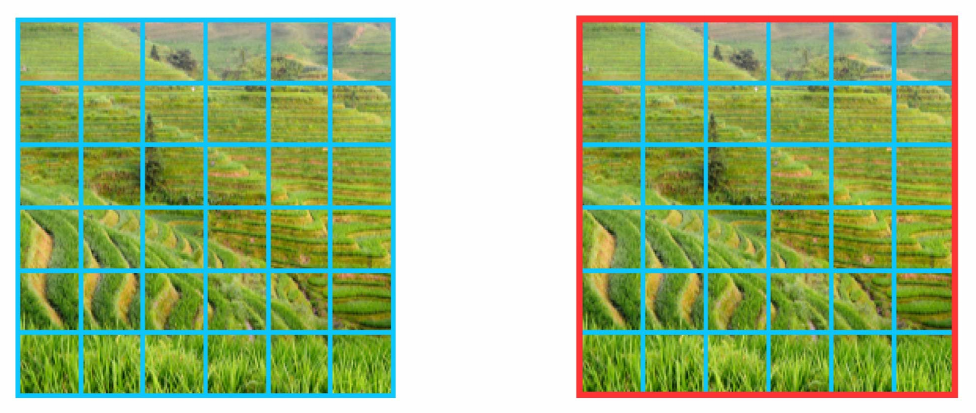}
            \caption{Window-based Multi-head Self-Attention (W-MSA) across image patches in ViT}
            \label{fig:ch02:w_msa}
        \end{figure}

        Transformer-based architectures address several limitations of CNN-based approaches by enabling global context modeling and flexible feature manipulation.
        Hence, the Vision Transformer (ViT) is introduced as a visual transformer that adapts self-attention, and it works by splitting an input image to fixed-size patches~\cite{dosovitskiy2021}, also known as windows, but it suffers from a significant computational bottleneck. 
        ViT attention mechanism works by dividing the image to $16 \times 16$ patches and compute the relationship of each patch with the rest of the patches.
        Fig~\ref{fig:ch02:w_msa} demonstrates a simplified depiction of the process.
        This approach improves the semantic correlation between the different regions in a given image; however, this global attention incurs a quadratic complexity $\mathcal{O}(N^2)$, making ViT incapable of scaling to high-resolution images in hardware-constrained systems.

    \subsubsection{Swin Transformer for Scalable Semantic Representation}
\begin{figure}[h]
            \centering
            \begin{subfigure}{0.48\textwidth}
                \includegraphics[width=\linewidth]{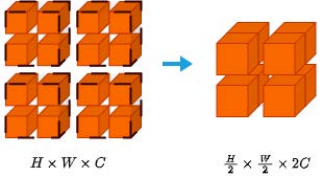}
                \caption{Spatial downsampling}
            \end{subfigure}
            \hfill
            \begin{subfigure}{0.48\textwidth}
                \includegraphics[width=\linewidth]{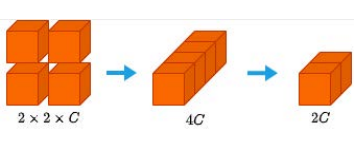}
                \caption{Channel projection}
            \end{subfigure}
            \caption{The Patch Merging mechanism in the Swin Transformer. (a) Spatial Reduction: Neighboring $2 \times 2$ patches are merged to downsample the resolution by a factor of 2. (b) Dimension Reduction: The concatenated features ($4C$) are linearly projected to a lower dimension ($2C$)}
            \label{fig:ch02:patch_merging}
        \end{figure}

        To address scalability, Liu et al. proposed the Shifted Window (Swin) Transformer, a hierarchical vision architecture that serves as a replacement to the traditional ViT.
        Unlike ViT single window attention, the Swin Transformer constructs a hierarchical feature map by merging patches in deeper layers, as illustrated in Fig.~\ref{fig:ch02:patch_merging}.
        This process mimics how CNNs reduces input resolution while increasing receptive fields, allowing the model to maintain linear computational complexity $\mathcal{O}(N)$ with respect to the image size~\cite{liu2021swin}.
        
\begin{figure}[h]
            \centering
            \includegraphics[width=0.85\linewidth]{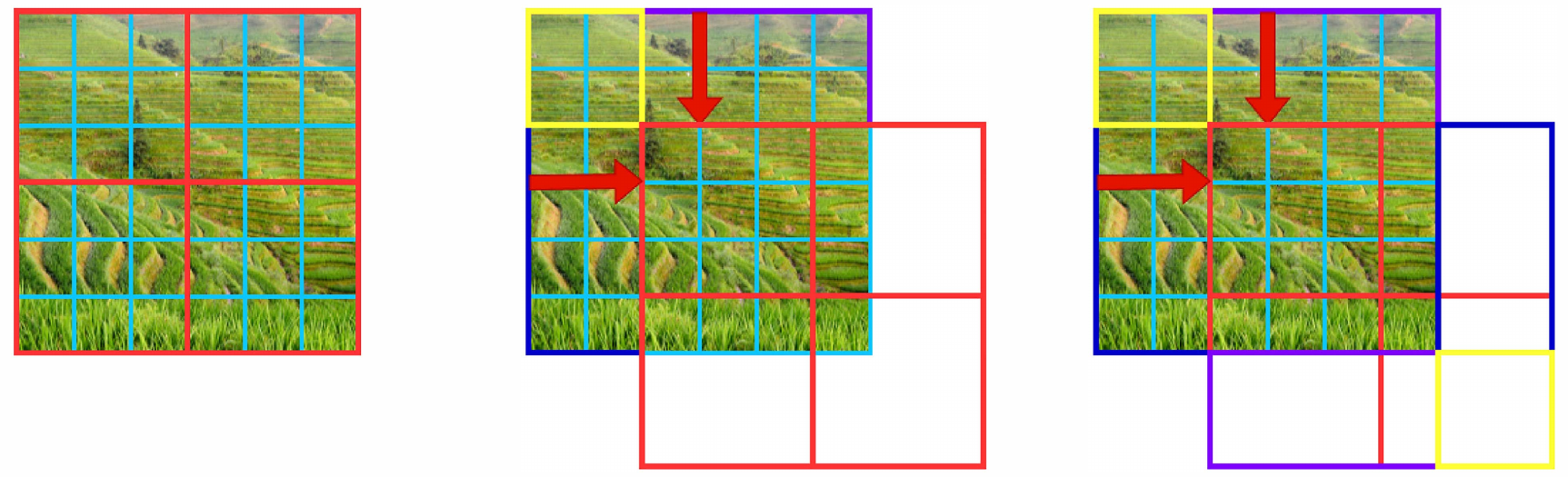}
            \caption{Illustration of the window shifting mechanism moving diagonally given some stride defined in the Swin Architecture}
            \label{fig:ch02:sw_msa}
        \end{figure}

        The core difference between ViT and Swin is in Swin's shifted window partitioning scheme, where the model computes self-attention only within local non-overlapping windows ($7 \times7$ based on Liu et al.'s work) to reduce cost.
        To prevent information isolation, the window partition is shifted by half the window size in consecutive layers, creating a cross-window connections that allow semantic information to propagate globally without the heavy overhead of global attention. Fig.~\ref{fig:ch02:sw_msa} demonstrates this shifting.
        
        These mechanisms are encapsulated within the Swin Transformer Block, which replaces the standard single Multi-Head Self-Attention (MSA) with window-based modules.
        Crucially, these blocks compute in consecutive pairs to facilitate the shifting mechanism described earlier.
        The first module in a single Swin Block utilizes a Window-based Multi-Head Self-Attention (W-MSA) module, while the second employs the Shifted Window configuration (SW-MSA).

\begin{figure*}
            \centering
            \begin{subfigure}{0.7\textwidth}
                \centering
                \includegraphics[width=\linewidth]{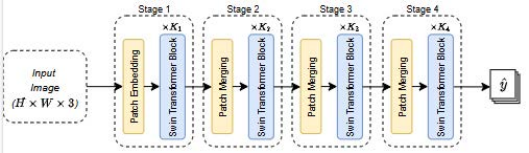}
                \caption{Swin Architecture}
                \label{fig:ch02:spatial_reduction}
            \end{subfigure}
            \hfill
            \begin{subfigure}[b]{0.29\textwidth}
                \centering
                \includegraphics[width=\linewidth]{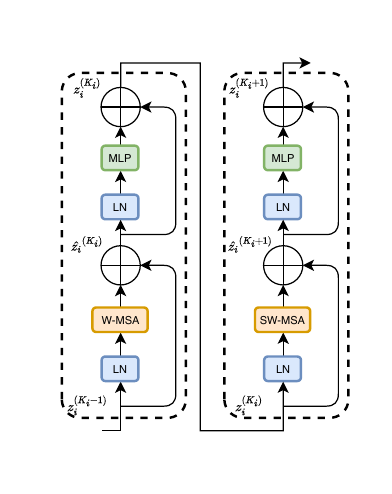}
                \caption{Swin Block}
                \label{fig:ch02:dimension_reduction}
            \end{subfigure}
            \caption{Swin Architecture diagram in (a) and the composition of the Swin Transformer Block in (b)}
            \label{fig:ch02:swin_architecture}
        \end{figure*}

        As depicted in Fig.~\ref{fig:ch02:swin_architecture}, in every Swin Block, each attention module is followed by a two-layer Multi-Layer Perceptron (MLP) with GELU non-linearity~\cite{liu2021swin}.
        LayerNorm (LN) is applied before each MSA and MLP module, and a residual connection is applied after each module, ensuring stable gradients during deep training.
        The overall architecture is organized into hierarchical stages, where patch merging layers reduce resolution between stages to produce a pyramid of feature maps, enabling the model to detect visual entities at various scales effectively.

        These properties make Swin particularly attractive for semantic communication in edge-cloud systems, where high-resolution inputs, predictable scaling behavior, and controllable compute overhead are essential.
    
    \subsubsection{Semantic Communication Systems Based on Swin}
        Several recent semantic communication frameworks adopted Swin as a backbone to achieve robustness in dynamic environments, such as wireless image transmission systems.
        As such the Wireless Image Transmission Transformer (WITT) introduced by Yang et al., build upon the Swin Transformer and incorporates a spatial modulation module called \texttt{Channel ModNet}.
        This module acts as a plug-in controller that scales the latent representations based on the Channel State Information (CSI)~\cite{yang2022_witt}.
        It consists of several fully connected (FC) layers that map the Signal-to-Noise Ratio (SNR) to a modulation vector, which is then multiplied by the semantic features.
        This allows a single model to adapt to varying channel noise without retraining.

        Building on WITT, SwinJSCC introduces a dual adaptation mechanism by adding a transmission rate modulator called\texttt{Rate ModNet} alongside the Channel ModNet~\cite{yang2023_swinjscc}.
        While the Channel ModNet adapts to noise, the Rate ModNet adapts to some bandwidth constraints.
        It takes a target transmission rate $R$ as input and generates a channel-wise mask that prunes the features of the latent representation by importance.
        This enables the system to adapt to tight wireless channels and adjust its compression rate on the fly.
        
        While effective, these systems primarily target wireless physical-layer adaptation. 
        They do not explicitly address transport-layer variability, edge compute constraints, or SLA-driven deployment considerations that dominate edge-cloud systems.

\subsection{Semantic Communication in Edge-Cloud Contexts}
\label{ch02:edge_cloud_context}
    Semantic communication in edge-cloud systems operates under fundamentally different constraints than those encountered in wireless access links. 
    While wireless channels are dominated by noise, fading, and interference, edge-cloud communication typically traverses high-capacity wired backhaul and core networks where bit error rates are negligible. 
    In these environments, system performance is instead governed by transport-layer effects, including throughput variability, queuing delay, round-trip latency, and the heterogeneity of compute resources across edge and cloud tiers~\cite{yang2023_swinjscc}. 
    As a result, classical physical-layer robustness objectives play a secondary role, and end-to-end latency emerges as the primary performance metric.
    
    This shift fundamentally alters the role of semantic communication. 
    Rather than compensating for channel impairments, semantic compression in edge-cloud systems serves as a mechanism for controlling payload size and, by extension, transport latency and congestion. 
    However, the benefits of aggressive semantic compression must be carefully balanced against the computational overhead introduced by deep neural encoders at the edge. 
    Unlike cloud servers, edge nodes operate under constrained CPU, memory, and energy budgets, making it infeasible to deploy arbitrarily complex semantic models without violating latency or resource constraints.
    
    Despite these differences, many semantic communication systems continue to adopt distortion minimization as a core optimization objective:
    \begin{equation}
        (\phi, \theta) = \arg \min_{\phi, \theta}
        \mathbb{E}_{x}\left[d(x, \hat{x})\right],
    \end{equation}
    where $\phi$ and $\theta$ denote the encoder and decoder parameters, respectively, and $d(x,\hat{x})$ measures semantic distortion after compression and reconstruction. 
    In the edge-cloud context, this objective implicitly captures a three-way trade-off between semantic fidelity, computational cost, and transport latency. 
    Reducing distortion generally requires deeper encoding and richer latent representations, which increase edge-side compute time, while aggressive compression lowers payload size at the cost of reconstruction quality.
    
    Recent research has begun to explore mechanisms for navigating this trade-off. 
    Region-of-interest–aware semantic coding prioritizes semantically salient regions, reducing transmission cost while preserving task-relevant information~\cite{10901610}. 
    Computation-aware adaptation techniques selectively activate model components to trade encoding depth for runtime efficiency~\cite{10832517}. 
    These approaches underscore the necessity of jointly optimizing computation and communication rather than treating them as independent subsystems.
    
    Nevertheless, existing solutions often assume static or offline adaptation and do not explicitly incorporate real-time network telemetry or SLA constraints. 
    This limitation motivates the development of semantic communication frameworks that are explicitly designed for edge-cloud deployment, where transport latency, compute feasibility, and deterministic adaptation must be jointly considered to ensure predictable end-to-end performance.

\subsection{Modern xG Infrastructure and O-RAN Enablement}
\label{ch02:oran}
    The practical deployment of semantic communication systems is inseparable from the evolution of modern xG network infrastructure. 
    To simultaneously support heterogeneous service classes such as enhanced Mobile Broadband (eMBB), ultra-Reliable Low-Latency Communications (uRLLC), and massive machine-type communications (mMTC), contemporary Radio Access Networks (RAN) are undergoing a fundamental architectural shift. 
    Traditional monolithic base stations, tightly integrated and vendor-specific, are increasingly replaced by disaggregated and virtualized architectures that expose programmable interfaces and elastic compute resources~\cite{10024837}. 
    This transition is a prerequisite for embedding intelligence directly into the network fabric, including semantic encoding and adaptive communication pipelines.

    Historically the Radio Access Network (RAN) relied on proprietary equipment where the Baseband Unit (BBU) and Remote Radio Head (RRH) were supplied by a single vendor using proprietary interfaces (e.g., CPRI)~\cite{gavrilovska2020}.
    5G introduced the concept of functional splits, allowing the BBU to be decomposed into logical units.
    This shift is essential for semantic communications, as it transforms the base station from a fixed signal processor into a flexible cluster of compute nodes capable of hosting complex neural networks like the Swin Transformer.
    
\begin{figure}[htbp]
        \centering
        \includegraphics[width=\linewidth]{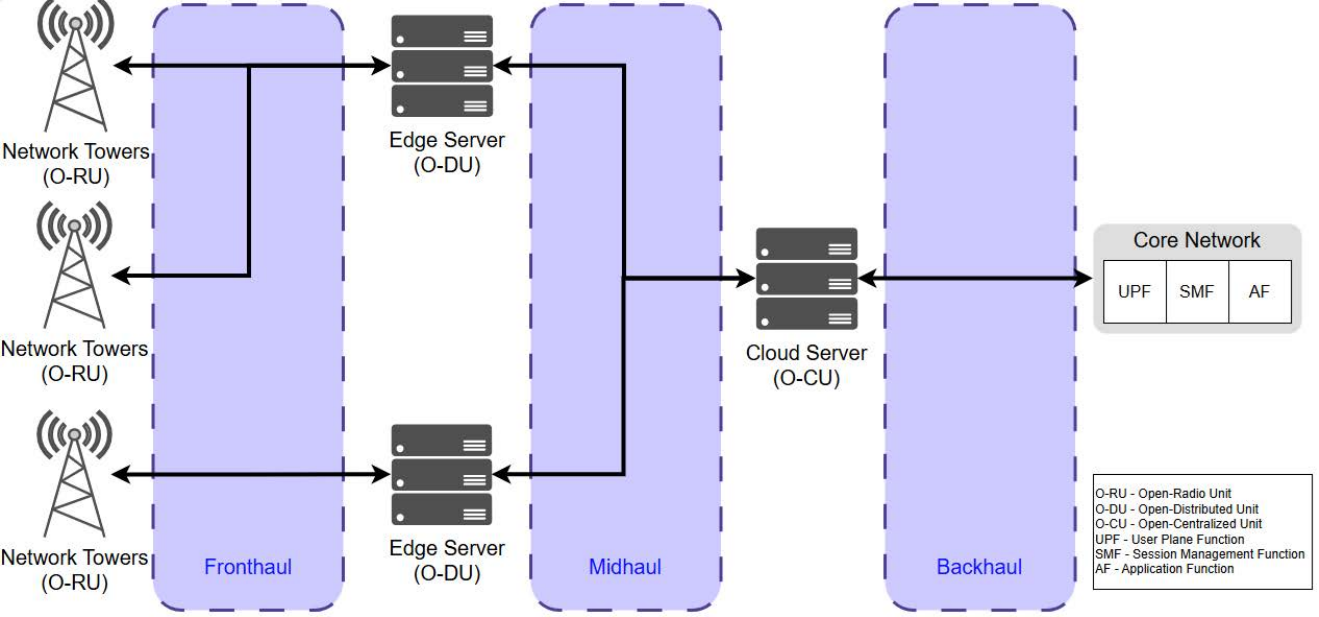}
        \caption{O-RAN architecture with functional disaggregation into O-RU, O-DU, and O-CU components}
        \label{fig:ch02:o_ran}
    \end{figure}

    Open Radio Access Network (O-RAN) represents a key realization of this transformation. 
    O-RAN decomposes the base station into three logic segments, as demonstrated in Fig.~\ref{fig:ch02:o_ran}.
    At the physical edge resides the Radio Unit (O-RU), which handles Radio Frequency (RF) signals, amplification, and analog-to-digital conversion.
    It connects via the Open Fronthaul interface to the Distributed Unit (O-DU), a node responsible for real-time Layer~2 functions such as Radio Link Control (RLC) and Medium Access Control (MAC).
    Due to its proximity to the user and access to hardware accelerators (e.g., GPU or FPGAs), the O-DU is the ideal location for deploying DL-based models, for example a semantic encoder to compress data before it enters a capacity-constrained transport network~\cite{8479363}.
    Further upstream is the Centralized Unit (O-CU), which manages non-real-time Layer~2 and 3 functions like Packer Data Convergence Protocol (PDCP) and Radio Resource Control (RRC).
    The O-CU is typically virtualized in a regional cloud, making it suitable for hosting the semantic task-driven platform for applications like image segmentation, classification, or image reconstruction.
    
    From the perspective of semantic communication, O-RAN fundamentally redefines the role of the edge. 
    By colocating general-purpose compute accelerators (CPU, GPU, or FPGA) with the O-DU or MEC platform, O-RAN enables the execution of deep learning models—such as Transformer-based semantic encoders—directly within the access network. 
    This allows semantic compression to be applied \emph{before} traffic enters the transport and core networks, where latency and congestion effects dominate end-to-end performance. 
    As a result, O-RAN provides a natural deployment substrate for edge-based semantic encoding that reduces payload size, transport delay, and core-network load.
    
    A central enabler of this capability is the RAN Intelligent Controller (RIC), which introduces standardized control loops for AI/ML-driven optimization. 
    The non-real-time RIC (operating on time scales above one second) supports policy learning and long-term optimization through the Service Management and Orchestration (SMO) layer, while the near-real-time RIC (10~ms–1~s) hosts xApps that perform fast adaptation based on real-time telemetry. 
    These telemetry streams include channel state indicators, throughput measurements, queue occupancy, and latency statistics, which are directly relevant for semantic communication control.
    
    Crucially, this telemetry-driven control plane transforms the RAN and edge infrastructure into a programmable environment where semantic encoding policies can be dynamically adjusted in response to observed network conditions. 
    Instead of static or offline compression strategies, semantic encoders can adapt their effective rate, feature selection, or computational depth based on real-time bandwidth and latency signals. 
    This capability aligns naturally with the requirements of SLA-driven edge-cloud systems, where predictable end-to-end latency and resource utilization must be maintained under fluctuating network conditions.
    
    In general, O-RAN enables the physical deployment of semantic communication models at the edge and also provides the telemetry, control interfaces, and orchestration mechanisms required for deterministic, network-aware adaptation. 
    These properties make O-RAN a foundational enabler for the edge-cloud semantic communication frameworks developed in the subsequent chapters.

\subsection{End-to-End Latency Modeling and SLA Constraints}
\label{ch02:latency_sla}
    Meaningful evaluation and control of semantic edge-cloud systems require an explicit decomposition of end-to-end (E2E) latency rather than treating the network as a monolithic black box. 
    In practical xG deployments, latency arises from a combination of computation delays at distributed processing nodes and communication delays across heterogeneous access, transport, and core networks. 
    Failing to isolate these components obscures the true performance bottlenecks and limits the effectiveness of adaptive control strategies.
    This proposal adopts the methodology proposed by Coll-Perales et al., which decomposes latency into granular components rather than treating the network as a black box~\cite{CollPerales2023_E2E-V2X}.
    
    In the context of semantic communications, latency is categorized into computational and transmission domains.
    The computational delay represents the data processing at both the edge ($l_{\mathrm{edge}}$) and the cloud ($l_{\mathrm{cloud}}$).
    Once the data is encoded at the edge, the data incurs transmission delay across the wireless link ($l_{\mathrm{raio}}$) and/or the wired backhaul ($l_{\mathrm{tx}}$)
    The magnitude of $l_{\mathrm{tx}}$ is sensitive to the semantic compression ratio at the edge.
    Aggressive masking reduces the payload size, directly alleviating transport latency.
    Finally, the data traverses the Core Network ($l_{\mathrm{CN}}$) and the link between the User Plane Function (UPF) and the Application Server ($l_{\mathrm{UPF-AS}}$) before reaching the cloud for further processing, such as task-driven operations.
    Table~\ref{tab:ch02:lat_def} summarizes the principal latency components considered in this survey.
    
    \begin{table}[htbp]
        \centering
        \caption{End-to-end latency components in semantic edge-cloud systems}
        \label{tab:ch02:lat_def}
        \begin{tabular}{ll}
            \toprule
            $l_{\mathrm{edge}}$ & Edge-side semantic encoding and processing latency \\
            $l_{\mathrm{radio}}$ & Wireless access latency between device and RAN \\
            $l_{\mathrm{tx}}$ & Transport and backhaul transmission latency \\
            $l_{\mathrm{CN}}$ & Core network traversal latency \\
            $l_{\mathrm{UPF-AS}}$ & Latency between UPF and application server \\
            $l_{\mathrm{cloud}}$ & Cloud-side decoding and task execution latency \\
            $l_{\mathrm{pp}}$ & Peering-point latency in multi-domain routing \\
            \bottomrule
        \end{tabular}
    \end{table}
    
    For a single-operator, single-domain deployment, the total E2E latency ($l_{\mathrm{E2E}}$) can be expressed as:
    \begin{equation}
        l_{\mathrm{E2E}} = l_{\mathrm{edge}} + l_{\mathrm{radio}} + l_{\mathrm{tx}} + l_{\mathrm{CN}} + l_{\mathrm{UPF-AS}} + l_{\mathrm{cloud}}.
        \label{eq:ch02:e2e_latency}
    \end{equation}
    
    In cross-domain scenarios involving inter-operator routing, an additional peering delay term $l_{\mathrm{pp}}$ must be included, further increasing sensitivity to transport conditions.

    With the latency model defined, the critical challenge becomes ensuring that $l_{\mathrm{E2E}}$ remains within acceptable limits.
    The deployment of DL-based semantic encoders at the network edge introduces a complex dependency on computational resources, making strict adherence to Service Level Agreements (SLAs) a significant challenge.
    SLA is a formal contract between the service provider and a consumer that guarantees specific Quality of Service (QoS) metrics, such as maximum latency, minimum throughput, and/or uptime assurance~\cite{10.1145/2428955.2429005}.
    For mission-critical xG applications, such as autonomous vehicles or remote surgical procedures, an SLA violation (e.g., $l_{\mathrm{E2E}} > l_{\mathrm{threshold}}$) is not merely a performance dip but a potential system failure.

    Resource allocation in semantic communications is essentially a coupled optimization problem involving the "Compute-Communication Continuum"~\cite{8016573,10896925}.
    Unlike traditional communications, where latency is dominated primarily by transmission, semantic systems introduce a tunable trade-off between the processing delay and the transmission delay.
    A high semantic compression ratio reduces the payload size, thereby minimizing the transmission latency and easing congestion.
    However, achieving this compression requires deeper neural network execution, which increases the computational burden on the edge server and drives up the latency at the edge.
    Conversely, shallower neural networks offering a more shallow encoding may reduce the encoding time, but at the cost of a larger payload size or, worse, a less accurate representation of the data, risking not only higher transmission delays but also worse service.
    Therefore, effective resource allocation must jointly optimize computational resources (CPU/GPU cycles) and network resources such as bandwidth ($B$) and Round-Trip Time ($RTT$) to satisfy some SLA constraint~\cite{8016573}.

    Traditional resource allocation schemes often rely on reactive scaling, which is insufficient for the strict uRLLC requirements for xG networks.
    Recent research has shifted toward predictive approaches using real-time telemetry.
    Shah et al. investigated SLA compliance in edge environments where multiple Deep Neural Networks (DNNs) contend for resources on a single server.
    By characterizing response times as a function of request rate, processor allocation, and memory availability, they demonstrated that pre-interface telemetry (e.g., baseline latency, utilization features) is sufficient to anticipate SLA breaches before they occur~\cite{11073603}.
    Their Random Forest-based predictor achieved 90.2\% accuracy in forecasting latency violations, establishing a precedent for telemetry-aware control.

    While Shah et al. focused on inter-model interference (multiple models fighting for resources) on a CPU-bound server, this proposal extends the concepts to the intra-model scaling behavior of a single distributed semantic pipeline within the O-RAN architecture.
    This work proposes that by monitoring CSI and compute telemetry via the RIC, the system can proactively adjust the semantic compression ratio to maintain the total $l_{\mathrm{E2E}}$ within some SLA boundaries.
    This ensures that the system dynamically balances the compute-communication trade-off to meet stringent O-RAN timing requirements.
    
    Within this context, semantic communication serves not merely as a compression technique but as an SLA-enabling control mechanism, allowing distributed edge-cloud systems to dynamically balance computation cost, transport delay, and semantic fidelity to remain within prescribed performance envelopes.

\subsection{System Model and Design Objectives}
\label{ch02:system}

    \subsubsection{Architecture Overview}
\begin{figure}[h]
            \centering
            \includegraphics[width=\linewidth]{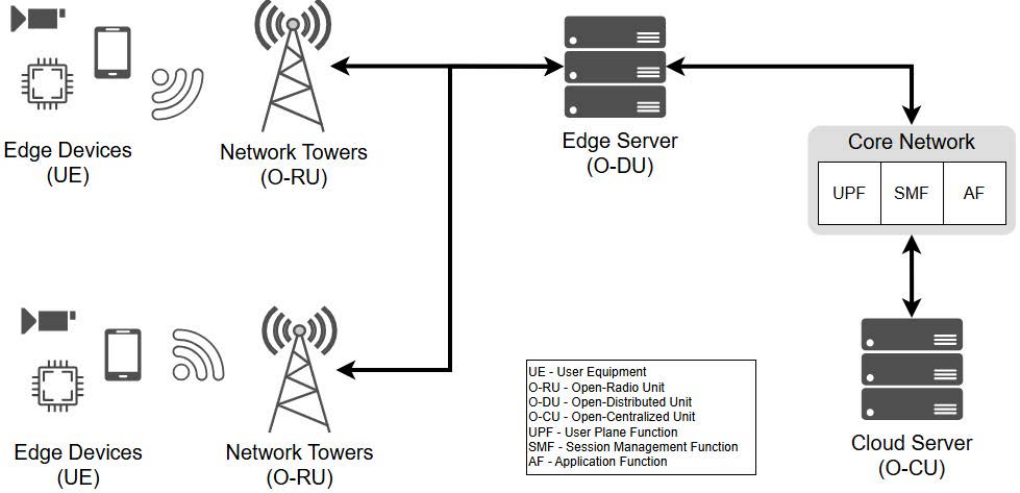}
            \caption{Three-tier edge-cloud semantic communication system model}
            \label{fig:ch02:system_model}
        \end{figure}

        This proposal considers a three-tier architecture comprising edge devices (e.g., cameras, sensors), an Edge Server (Multi-Access Edge Computing), and a cloud server.
        Fig.~\ref{fig:ch02:system_model} illustrates this end-to-end model, spanning from edge devices at the O-RU and the edge servers at the O-DU, to the cloud infrastructure located beyond the O-CU.

        The operational roles within this architecture form a sequential pipeline.
        Edge devices are responsible for generating high-resolution visual data (e.g., 4K/8K frames) and forwarding packets over the (O-)RAN to the edge. 
        Upon receipt, the edge server (MEC) utilizes a deep semantic encoder to transform the raw input into compact feature representations. 
        These latent features are then transmitted over the backhaul network, serving as a compressed semantic payload for the next stage. 
        Finally, the cloud server receives the masked latent features and executes diverse downstream task-driven applications, such as reconstruction, classification, segmentation, or advanced semantic analysis. 
        This separation of concerns preserves low latency at the edge while leveraging the scalability of cloud resources for computationally intensive inference.
        
    \subsubsection{Optimization Framework and Design Objectives}
        The primary variable subject to optimization in this pipeline is the transmission latency over the backhaul.
        While encoding and radio delays are largely fixed by hardware and RAN constraints, the backhaul transmission latency is directly proportional to the semantic payload $S$ and the available link bandwidth $B$.
        Given a round-trip propagation time $l_{\mathrm{RTT}}$, the transmission latency ($l_{\mathrm{tx}}$) is modeled as:
        
        \begin{equation}
            l_{\mathrm{tx}} \;\approx\; \frac{S}{B} \;+\; \frac{l_{\mathrm{RTT}}}{2}.
            \label{eq:ch02:tx_latency}
        \end{equation}

        Here the semantic payload $S$ acts as the control variable.
        By dynamically adjusting some masking ratio to select the most salient features within the latent representation, the system can linearly reduce $l_{\mathrm{tx}}$, thereby minimizing the total end-to-end latency $l_{\mathrm{E2E}}$.

        This optimization framework is grounded in three key design assumptions. 
        First, the core network is assumed to be effectively noise-free, meaning that edge-cloud performance is dominated by congestion, throughput variability, and latency rather than packet loss. 
        Second, the telemetry data, specifically throughput $B$ and latency $l_{\mathrm{RTT}}$ is accessible per batch via mechanisms such as SDN, NWDAF, or passive probes, or can be sampled from realistic distributions for evaluation. 
        Third, the model assumes task-driven decoding, where the cloud server possesses the capability to either reconstruct pixels or run task heads directly on the latent features.

        Building upon these assumptions, the proposed edge-cloud semantic communication system is designed with a set of tightly coupled objectives that jointly address latency, semantic fidelity, computational feasibility, and operational robustness. 
        First, the system aims to minimize the semantic payload size transmitted over the edge-cloud link, as transmission latency is directly proportional to payload size and available bandwidth. 
        By selectively transmitting only the most informative latent features, the system reduces backhaul and core-network delay while alleviating congestion and queuing effects under limited or fluctuating throughput conditions.
        
        Second, payload reduction must not compromise semantic integrity or downstream task performance. 
        The system therefore prioritizes preserving task-relevant information, ensuring that reconstructed content or task outputs (e.g., classification, segmentation, or description) maintain high fidelity despite aggressive compression. 
        This objective reflects the core principle of semantic communication: efficiency gains are meaningful only if semantic utility is preserved.
        
        Third, the system explicitly accounts for the computational constraints of edge servers. 
        While deep transformer-based encoders offer strong semantic representation capabilities, their execution cost can be prohibitive on resource-constrained edge hardware. 
        Accordingly, the architecture bounds edge-side computation by offloading heavy inference tasks to the cloud, leveraging cloud elasticity while keeping edge latency predictable and manageable.
        
        Finally, the system is designed to support deterministic, telemetry-driven adaptation to network dynamics without requiring online retraining or iterative feedback loops. 
        By relying on readily available network telemetry—such as bandwidth and round-trip latency—the semantic encoding process can be adjusted in real time using lightweight and interpretable control policies. 
        This enables stable and reproducible behavior under changing network conditions, a critical requirement for SLA compliance and practical deployment in operational edge-cloud environments.
        
\subsection{section Summary}
\label{ch02:summary}
    This section reviewed the evolution of semantic communication, transformer-based visual encoders, and modern xG network infrastructure. 
    It highlighted the shift from physical-layer optimization to end-to-end, SLA-aware system design in edge-cloud environments. 
    Building on these foundations, the section introduced a three-tier system model that enables network-aware semantic adaptation using real-time telemetry. 
    The next chapters leverage this model to analyze SLA feasibility, design semantic transcoding mechanisms, and develop network-aware adaptive control strategies.

\section{System Modeling Assumptions, Data Pipelines, and Methodology}
\label{sec:methodology}
\noindent
This section consolidates system-level modeling assumptions, dataset construction, telemetry pipelines, and experimental methodology for evaluating semantic communication across heterogeneous network conditions and compute profiles.


\label{ch03}

\subsection{Introduction}
\label{ch03:intro}
    
    section~\ref{ch02} introduced the end-to-end edge-cloud semantic communication architecture shown in Fig.~\ref{fig:ch02:system_model}. 
    This section focuses on the edge access segment of that architecture, specifically the edge device and the proximal edge server (MEC) tier, where compute heterogeneity and latency variability are most consequential. 
    While the cloud tier can host complex post-processing and large-scale inference workloads, the objective here is to determine whether a given configuration of \textit{(i)} edge device resources, \textit{(ii)} edge server resources, and \textit{(iii)} edge device-to-server network conditions can satisfy predefined Service Level Agreements (SLAs).
    
    In communication systems, SLAs define measurable performance targets. 
    Typically, latency, throughput, and availability—that ensure predictable and reliable service delivery between network entities. 
    They serve as an operational contract linking user expectations with system capability and provide the basis for QoS and QoE guarantees~\cite{ibm_sla_2024,aakbari_2023}. 
    In distributed edge environments, SLA compliance is especially challenging because end-to-end performance depends jointly on compute latency and communication latency. 
    Embedding SLA awareness into semantic communication pipelines, therefore, enables performance assurance (pre-deployment feasibility screening) and adaptive control (runtime policy selection).
    
    Prior studies of SLA-driven performance predictors have focused exclusively on CPU-only deployments and do not capture heterogeneous pipelines in which computation is divided between a resource-constrained edge node and a GPU-accelerated server~\cite{11073603}. 
    To address this limitation, this section develops a predictive framework that models end-to-end behavior under realistic bandwidth and latency conditions and predicts SLA class membership for candidate configurations, enabling proactive compliance assessment prior to deployment.
    
\subsection{Modeling Objective}
\label{ch03:model}
    The objective is to train a predictive model that estimates whether a specific semantic encoder-decoder configuration satisfies latency targets under specific compute and network constraints.
    The encoder is quantized and deployed, emulating an edge device, while the decoder will run at full capacity on a GPU-equipped server.
    Rather than exhaustively benchmarking all configurations online, the predictor learns the relationship between utilization signals, network conditions, and the resulting latency.

    \subsubsection{Latency Model and Prediction Target}
    \label{ch03:latency_model}
        For this experiment, the latency parameters that are of interest are everything before the core network.
        These parameters are $l_{\mathrm{edge}},\space l_{\mathrm{tx}},\space l_{\mathrm{cloud}}$ from Eq.~\ref{eq:ch02:e2e_latency}.
        The encoder-decoder pair will be deployed in a distributed system at the edge, hence the latency model here is:
        \begin{equation}
            l_{\mathrm{E2E}} \;=\; l_{\mathrm{edge}} \;+\; l_{\mathrm{tx}} \;+\; l_{\mathrm{cloud}}
            \label{eq:ch03:edge_latency}
        \end{equation}

\begin{figure}[h]
            \centering
            \includegraphics[width=\linewidth]{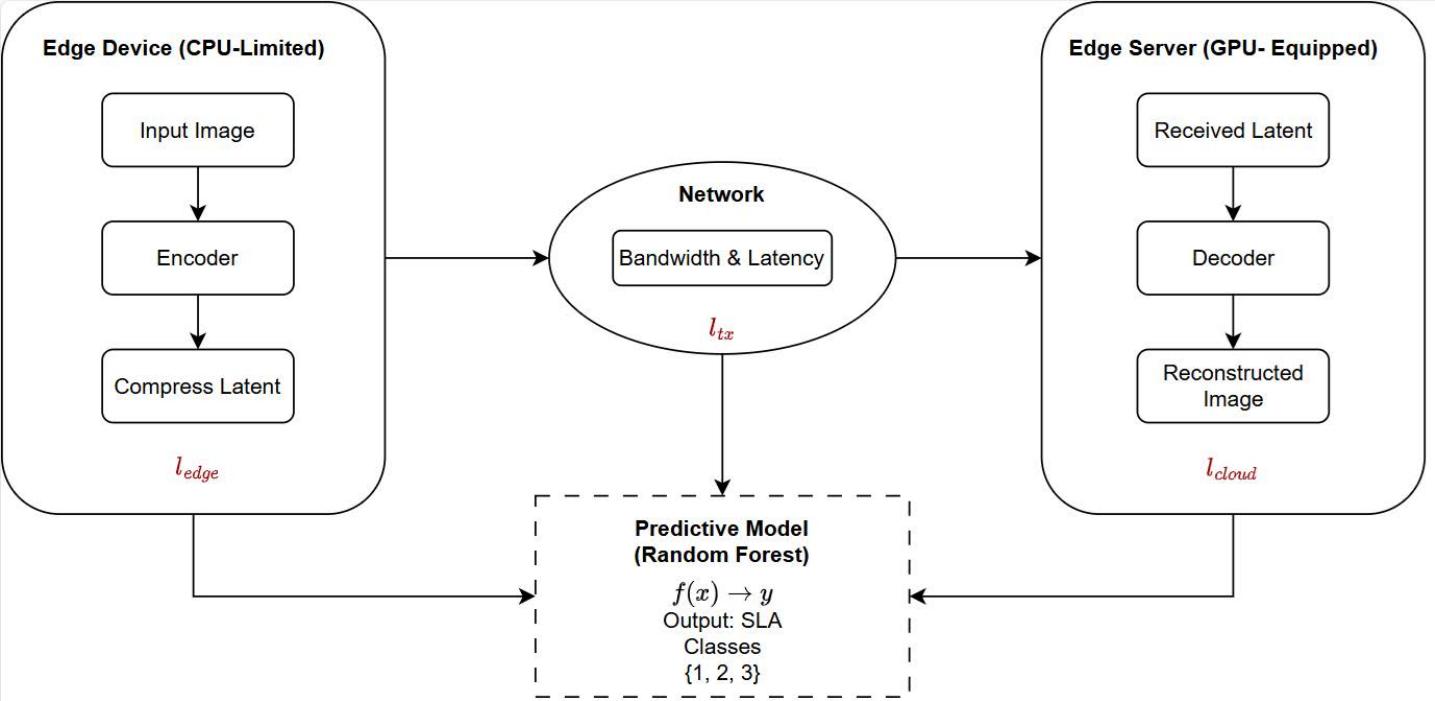}
            \caption{SLA compliance predictor scope: The latency model consists of Encoder (Device), Transmission (Network), and Decoder (Server)}
            \label{fig:ch03:edge_model}
        \end{figure}

        From Fig.~\ref{fig:ch03:edge_model}, the model takes in relevant features and learns a mapping  $f(x) \xrightarrow{}y$, where $x$ is a vector of system, network, and workload features, and $y \in \{1, 2, 3\}$ indicating the predicted SLA class.

    \subsubsection{Assumptions}
        The access-domain latency component (RAN scheduling/air-interface), denoted here as $l_{\mathrm{RAN}}$, is omitted to isolate compute and transport effects and because it varies across hardware implementations and deployment configurations. 
        Nonetheless, prior work reports typical access-domain latencies of 5-10~ms (up to 15-20~ms at the 95th percentile) for eMBB scenarios and approximately 1-4~ms for uRLLC configurations employing grant-free or minislot scheduling~\cite{10287312,pravez2018}. 
        
        To evaluate compliance under heterogeneous compute and network conditions, the measured latency $l_{\mathrm{E2E}}$ is mapped to three SLA tiers as shown in Table~\ref{tab:ch03:sla_levels}, which serve as the classification labels.
        
        \begin{table}[h]
            \centering
            \caption{SLA tiers categorized by latency constraints and operational requirements}
            \label{tab:ch03:sla_levels}
            \begin{tabular}{lll}
                \toprule
                \textbf{Level} & \textbf{Latency Threshold} & \textbf{Operational Mode} \\
                \midrule
                SLA~1 & \(l_{\mathrm{E2E}} \le 30~\mathrm{ms}\) & Real-time processing \\
                SLA~2 & \(30 < l_{\mathrm{E2E}} \le 100~\mathrm{ms}\) & Near real-time / interactive \\
                SLA~3 & \(l_{\mathrm{E2E}} > 100~\mathrm{ms}\) & Off-site or batch processing \\
                \bottomrule
            \end{tabular}
        \end{table}

\subsection{Data Collection and Feature Design}
\label{ch03:dc_fd}
    \subsubsection{Dataset Overview}
        Each data record in the constructed dataset corresponds to a single end-to-end execution of the distributed encoder-decoder pipeline for one image under a specific network condition configuration. 
        For every run, the system captures a comprehensive snapshot of both computational and communication behavior, enabling fine-grained analysis of end-to-end performance.
        
        Specifically, edge-side metrics include semantic encoder latency, CPU utilization, memory usage, and statistics on active CPU cores, reflecting the computational cost of semantic extraction under different model configurations. 
        On the server side, the dataset records decoder latency along with GPU utilization and memory consumption, capturing the resource footprint of cloud-side reconstruction or downstream processing.
        
        In addition, network metrics such as available bandwidth, observed throughput, and round-trip latency are logged. 
        These parameters are synthetically varied across experiments to emulate realistic edge-cloud operating regimes and to expose the model to a wide range of transport conditions. 
        Finally, outcome metrics are collected, including the resulting semantic payload size and, where applicable, reconstruction fidelity measures such as MS-SSIM.
        
        Together, these measurements capture the coupled effects of computation and communication on total end-to-end latency and provide labeled samples suitable for supervised learning tasks, including SLA tier classification and performance prediction.

    \subsubsection{Feature Categories}
        The feature set considered in this survey is organized into three complementary categories that collectively capture the interaction between computation, communication, and application-level performance. 
        First, system and workload features characterize the computational state of the edge and server platforms and include CPU utilization, the number of active CPU cores, memory footprint, GPU allocation and utilization, as well as model-specific attributes such as the selected architecture variant (e.g., small, base, or large) and quantization mode. 
        These features reflect the computational load imposed by semantic encoding and decoding and directly influence processing latency and resource availability.
        
        Second, network features describe the instantaneous state of the communication substrate and include the available bandwidth $B$, the measured round-trip time $l_{\mathrm{RTT}}$, and, where available, observed end-to-end throughput. 
        These features capture the dominant transport-layer dynamics that affect payload delivery time in edge-cloud systems and provide the primary inputs for network-aware adaptation.
        
        Finally, performance and fidelity features quantify the end-to-end impact of computation and communication decisions. 
        This category includes edge-side processing latency $l_{\mathrm{edge}}$, server-side latency $l_{\mathrm{cloud}}$, semantic payload size, and reconstruction or task-quality metrics such as MS-SSIM. 
        Together, these features enable contextual analysis of the trade-offs between computational effort, transport latency, and semantic fidelity, and serve as key indicators for evaluating SLA compliance.

\subsection{Predictive Modeling Framework}
\label{ch03:modeling_framework}

    \subsubsection{Model Selection}
        Following the predictive SLA compliance approach in~\cite{11073603}, a Random Forest (RF) classifier is employed. 
        RF models capture non-linear interactions prevalent in edge systems, including the coupled effects of CPU saturation, memory contention, and network variability. 
        Tree-based ensembles operate effectively on heterogeneous feature types without strict normalization requirements and provide intrinsic feature-importance scores, which are useful for identifying the parameters most associated with SLA violations.
    
    \subsubsection{Training Procedure}
        The predictor is trained on feature--label pairs generated from repeated pipeline runs. 
        The dataset is partitioned into training (80\%) and validation (20\%) subsets. 
        Cross-validation is used to tune hyperparameters such as the number of trees $N_{\mathrm{trees}}$, maximum depth $D_{\mathrm{max}}$, and minimum split size $s_{\mathrm{min}}$. 
    
    Each tree learns decision boundaries that minimize classification impurity (Gini index). 
    The final prediction is determined by majority vote:
    \begin{equation}
        \hat{y} = \mathrm{mode}\bigl(T_1(\mathbf{x}), T_2(\mathbf{x}), \dots, T_{N_{\mathrm{trees}}}(\mathbf{x})\bigr).
    \end{equation}
    
    \subsubsection{Evaluation Metrics}
        Performance is assessed using accuracy and macro-averaged F1-score to account for potential class imbalance. 
        A confusion matrix is used to inspect per-class error patterns, and feature-importance analysis ranks input variables by their contribution to the model’s decisions, identifying the dominant drivers of SLA compliance (or violation) under heterogeneous conditions.
    
    \subsection{Experimental Setup and Analysis}
    \label{ch03:exp_setup_ana}
    
    This section describes the hardware/software environment and analyzes latency and utilization results under varying bandwidth regimes, quantization modes, and model sizes.
    
    \subsubsection{Experimental Setup}
        All runs were executed on a workstation configured to emulate the edge-device $\rightarrow$ edge-server inference pipeline. 
        The host system includes an Intel Core i7-13700K CPU and 32~GB RAM. 
        To emulate a resource-constrained edge device, the quantized encoder was restricted to 6 CPU cores and 8~GB RAM to emulate an iPhone 17. 
        The decoding stage was accelerated using an NVIDIA RTX 3050 GPU, representing a modest accelerator available at an edge server/MEC. 
        The software environment ran Ubuntu 24.04 under Windows Subsystem for Linux (WSL). 
        Swin Transformer variants were implemented in Python 3.11.10 and PyTorch 2.5.1, leveraging native support for quantization and transformer inference.
    
    \subsubsection{Experimental Results and Discussion}
        This subsection evaluates the feasibility of meeting the SLA tiers in Table~\ref{tab:ch03:sla_levels}. 
        Contrary to the initial expectation that quantization and smaller transformer variants might enable near real-time operation, the results show that the computational cost of the Swin-based encoder dominates the end-to-end latency in the evaluated edge configuration.
        
        \begin{figure}[h]
            \centering
            \begin{subfigure}[]{0.47\textwidth}
                \centering
                \includegraphics[width=\linewidth]{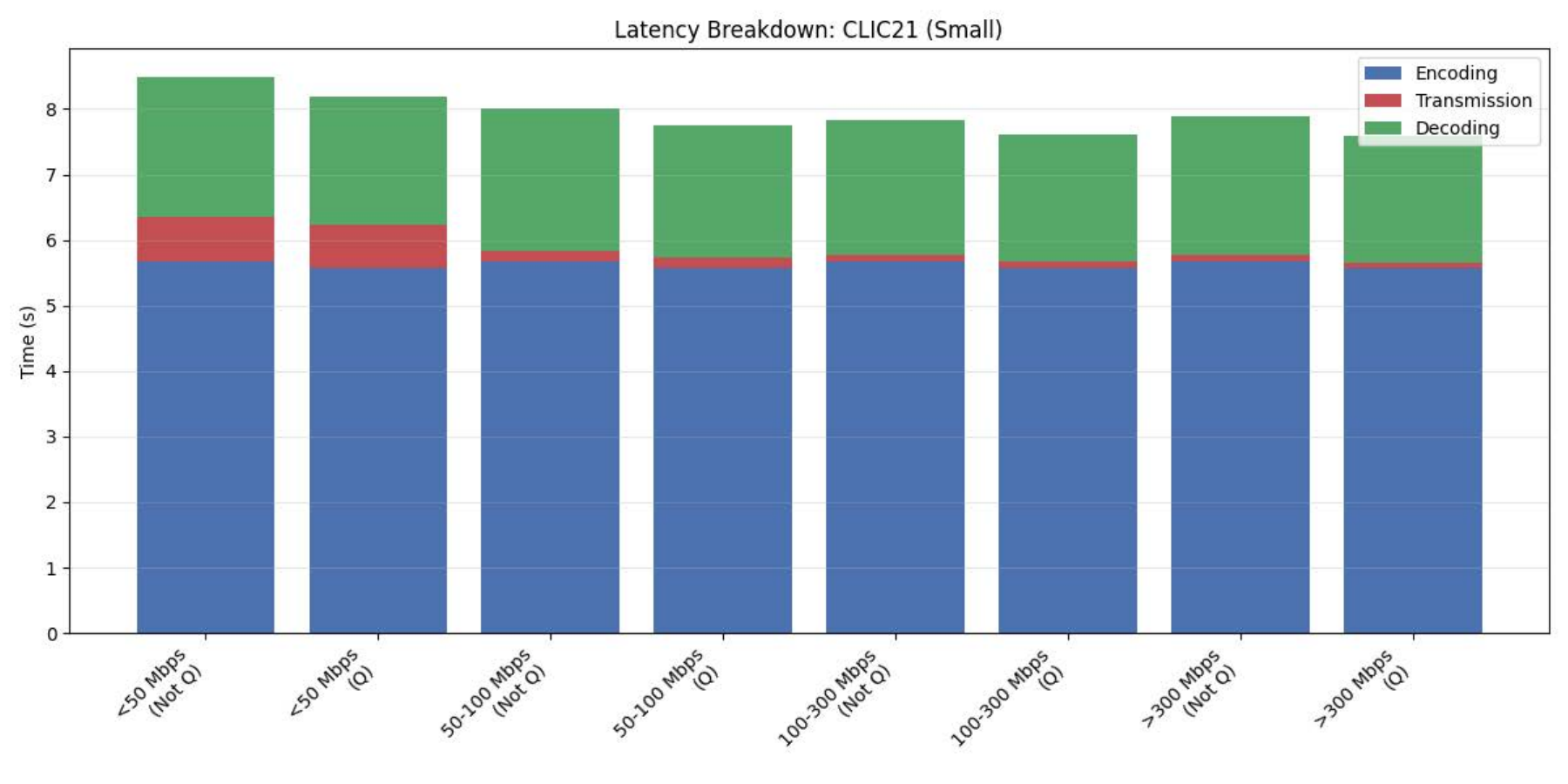}
                \caption{Small model}
            \end{subfigure}
            \par\bigskip 
            \begin{subfigure}[]{0.47\textwidth}
                \centering
                \includegraphics[width=\linewidth]{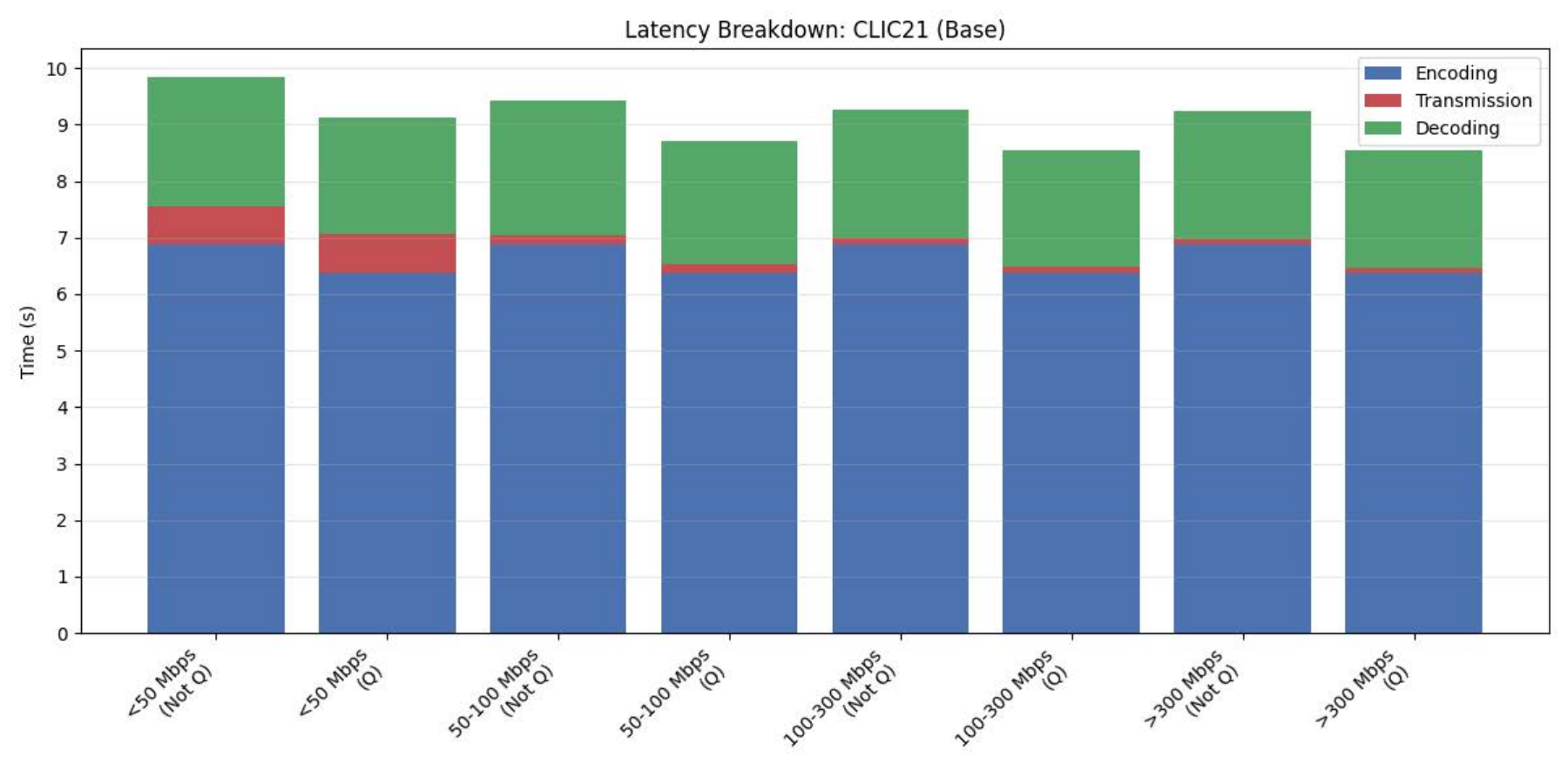}
                \caption{Base model}
            \end{subfigure}
            \par\bigskip
            \begin{subfigure}[]{0.47\textwidth}
                \centering
                \includegraphics[width=\linewidth]{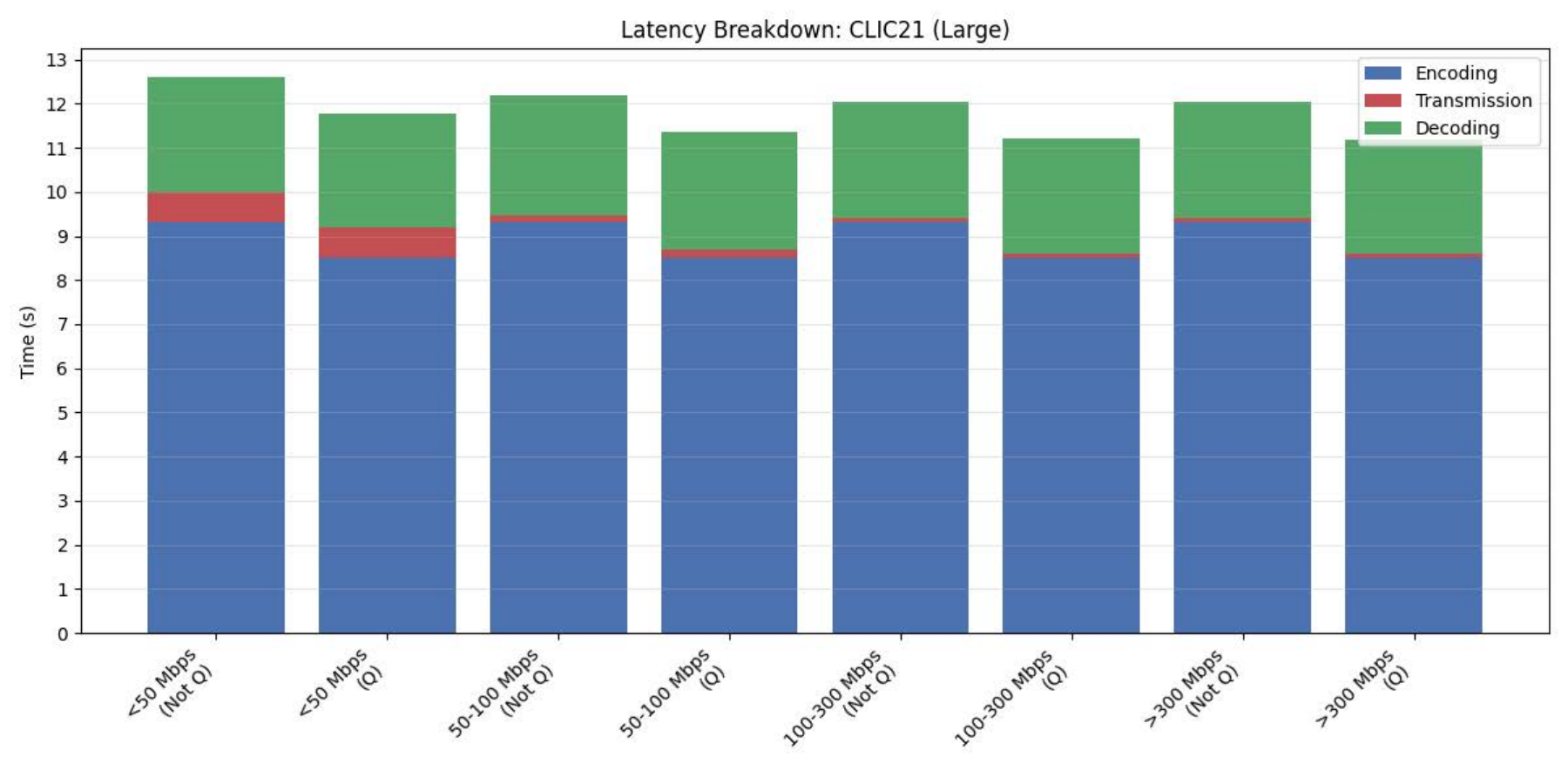}
                \caption{Large model}
            \end{subfigure}
            \caption{Latency breakdown by model size (CLIC21). The ``latency floor'' increases substantially with model complexity, shifting the system from transmission-bound to compute-bound behavior}
            \label{fig:ch03:clic21_latency}
        \end{figure}
        
        The working hypothesis was that applying quantization (e.g., dynamic linear quantization \texttt{dyn\_linear}) to the Swin encoder would reduce inference latency sufficiently to meet real-time (SLA~1, $\le$ 30ms) or interactive (SLA~2, $\le$ 100ms) operation under moderate-to-high bandwidth. 
        However, while quantization provides measurable speedups, it does not close the order-of-magnitude gap required for these SLA targets. 
        For example, for the CLIC21 dataset and the base model at $B \ge$ 300Mbps, total latency decreases from 9.34s to 8.64s with quantization, which remains far above the 100ms interactive threshold.
        
        As defined in Eq.~\ref{eq:ch03:edge_latency}, total latency is the sum of edge compute, transport, and server compute. 
        Fig.~\ref{fig:ch03:clic21_latency} shows that $l_{\mathrm{edge}}$ dominates across bandwidth tiers, indicating a compute-limited regime:
        \begin{itemize}
            \item \textbf{Low bandwidth ($<$50Mbps):} semantic compression reduces payload size, keeping transport latency relatively bounded (e.g., $l_{\mathrm{tx}} \approx$ 0.7s for CLIC21). 
            However, encoder latency remains very large (e.g., $l_{\mathrm{edge}} \approx$ 6-9s), accounting for $>$85\% of total latency.
            \item \textbf{High bandwidth ($>$300Mbps):} $l_{\mathrm{tx}}$ becomes negligible ($<$0.2s), but total latency remains high due to the fixed edge compute cost. 
            This creates a practical latency floor that cannot be improved by network upgrades alone.
        \end{itemize}
        
        \begin{figure}[h] 
            \centering 
            \includegraphics[width=\linewidth]{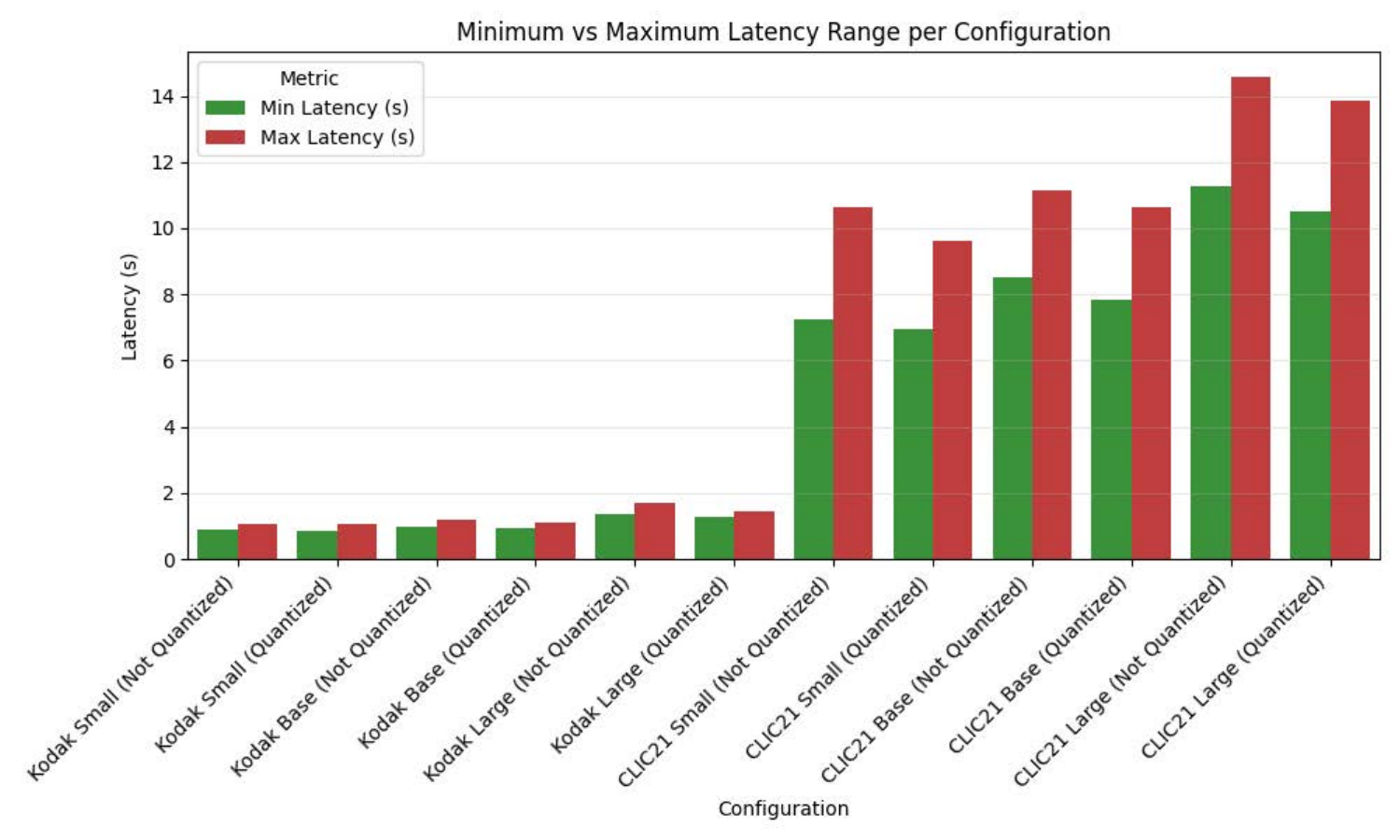}
            \caption{Minimum vs. maximum latency range per configuration. Latency scales non-linearly with model complexity; quantization yields modest gains without changing the overall magnitude. }
            \label{fig:ch03:min_max_latency} 
        \end{figure}
        
        Fig.~\ref{fig:ch03:min_max_latency} further characterizes the scaling behavior of the proposed semantic encoder under different configurations and datasets. 
        First, the impact of \emph{quantization} is evident: dynamic linear quantization consistently reduces encoding latency by approximately 0.5-1s across all evaluated scenarios. 
        While this reduction is non-negligible and confirms the effectiveness of quantization as a latency optimization technique, it remains insufficient to bridge the order-of-magnitude gap required to satisfy the 100ms interactive SLA threshold.
        
        Second, the results highlight a strong \emph{dataset sensitivity}. 
        Higher-resolution datasets, such as CLIC21 incur substantially larger end-to-end latencies compared to lower-resolution datasets such as Kodak. 
        Although the shifted-window attention mechanism in the Swin Transformer reduces the asymptotic complexity relative to global self-attention, the absolute computational cost still scales sharply with pixel count. 
        As a result, high-definition imagery remains prohibitively expensive to process on resource-constrained edge CPUs, even when network conditions are favorable.
        
        Overall, these observations demonstrate that the evaluated semantic encoder operates in a compute-dominated regime. 
        Under the current edge CPU constraints, neither quantization nor improved transport conditions are sufficient to achieve SLA~1 or SLA~2 compliance, reinforcing the need for offloading strategies or fundamentally more efficient semantic encoder architectures.

\subsection{section Summary}
\label{ch03:summary}
    This section introduced an SLA-driven performance prediction framework for the edge device $\rightarrow$ edge server segment of an edge-cloud semantic communication system. 
    A latency model decomposing end-to-end delay into edge compute, transport, and server compute components was defined, and SLA tiers were formulated as a multi-class prediction target. 
    Using empirical measurements across multiple model sizes, quantization modes, and bandwidth regimes, the analysis showed that edge-side Swin-based encoding remains the dominant bottleneck and prevents meeting real-time and interactive SLA thresholds under the evaluated resource constraints. 
    These findings motivate the need for compute-aware adaptation and network-aware semantic mechanisms developed in subsequent chapters.

    

\section{Edge--Cloud Interface: Measurements and Experimental Evaluation}
\label{sec:evaluation}
\noindent
This section reviews measurement methodology and evaluation metrics for semantic edge--cloud systems, including transport utilization, effective throughput, reconstruction quality, and latency accounting.


\label{ch04}

\subsection{Introduction}
\label{ch04:intro}
    The system model established in section~\ref{ch02} identified the edge-to-cloud backhaul as a critical bottleneck in distributed semantic communication pipelines.
    While the rapid evolution of xG networks promises high-throughput radio access for smart devices and Internet of Things (IoT) endpoints~\cite{shafi2017_5G-tutorial}, the core network and backhaul links faces immense pressure from the escalating volume of data traffic generated by high-fidelity vision applications (e.g., AR/VR, autonomous surveillance)~\cite{1}.
    These systems are required to meet the ever-increasing user demands for higher data raetes, reduced latency and ensuring seamless data exchanges~\cite{4}.
    As defined in the latency decomposition model in Section~\ref{ch02:latency_sla}, the end-to-end performance ($l_{\mathrm{E2E}}$) is heavily dependent on the transmission latency ($l_{\mathrm{tx}}$) across these links.

    Edge computing addresses part of this challenge by processing data closer to the user, thereby minimizing the distance data must travel.
    However, as discussed in section~\ref{ch03}, resource-constrained edge devices often lack the capacity to perform a full-scale inference for complex tasks.
    This necessitates a split-computing approach where the edge server performs initial semantic extraction, and the cloud handles the heavy-lifting task execution, all while keeping the edge devices transmitting data to the edge server.
    In this specific architecture, the efficiency of the Edge-to-Cloud data exchange becomes the defining factor for system performance.

    This section introduces Semantic Transcoding, a mechanism designed to optimize the data exchange between the edge and the cloud.
    Unlike traditional compression, which is agnostic to content, or the wireless-focused semantic communication systems like DeepJSCC discussed in Section~\ref{ch02:visual_semcom}, which optimize for channel noise, Semantic Transcoding optimizes for throughput and congestion.
    It leverages a Swin Transformer-based encoder deployed at the edge to filter out the latent features, transmitting only the "semantic meaning" required by the cloud.

    A significant gap exists in current semantic communication literature regarding the edge-cloud paradigm.
    The majority of existing works, including the standard implementations of the SwinJSCC~\cite{yang2023_swinjscc}, focus on the device-to-edge wireless link, aiming to combat signal fading and noise.
    There is a notable scarcity of research addressing the edge-to-cloud segment, where the challenge is not bit-errors, but rather bandwidth fluctuations and queuing delays, as highlighted in Section~\ref{ch01:research_gap}.

    This section addresses this gap by extending the semantic communication paradigm to the wired backhaul.
    It proposes a transcoding framework that allows the edge server to act not merely as a pass-through relay, but as an intelligent semantic gateway.
    By interpreting the data locally and pruning redundant features before transmission~\cite{7, 12}, the system can actively manage the transmission latency $l_{\mathrm{tx}}$ component of the latency budget.

    The work in this section functions as the control mechanism complementing the predictive analysis of section~\ref{ch03}.
    While section~\ref{ch03} provided the methodology to predict whether a specific configuration would violate an SLA rule, this section provides the transcoding framework required to prevent or mitigate those violations.
    By integrating the theoretical foundations in section~\ref{ch02} with the operational mechanisms proposed here, we establish a closed-loop system capable of maintaining QoS and reliability in dynamic edge-cloud environments~\cite{3, 6}.
    
    The rest of the section is outlined as follows: 
    Section~\ref{ch04:sys_model} introduces the proposed semantic-based optimized latency framework.
    Section~\ref{ch04:exp_setup} outlines the tools utilized, the measured metrics, and the data collection process.
    Section~\ref{ch04:exp_analysis} lists experimental results, semantic models comparisons, and data interpretation.
    Section~\ref{ch04:summary} summarizes the findings in this section.

\subsection{Semantic Edge-Cloud Computing Model}
\label{ch04:sys_model}
\begin{figure}[H]
        \centering
        \includegraphics[width=1.0\columnwidth]{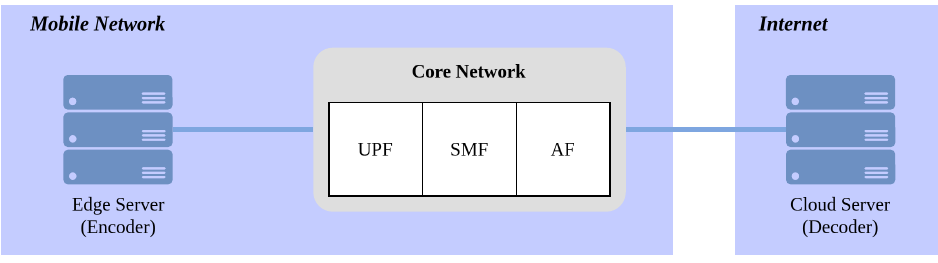}
        \caption{Overview of the semantic edge-cloud transcoding model}
        \label{fig:ch04:system_model}
    \end{figure}

    The system model (Fig.~\ref{fig:ch04:system_model}) considers transformer-based semantic encoding at the edge and semantic decoding at the cloud, inspired by the architectural principles of WITT and SwinJSCC~\cite{yang2022_witt,yang2023_swinjscc}. 
    The encoder extracts a latent semantic representation and applies a rate constraint (e.g., a target CBR), reducing payload size prior to transport. 
    The cloud-side decoder reconstructs semantic content from the received representation for downstream use.
    
\begin{figure}[h]
        \centering
        \includegraphics[width=1.0\columnwidth]{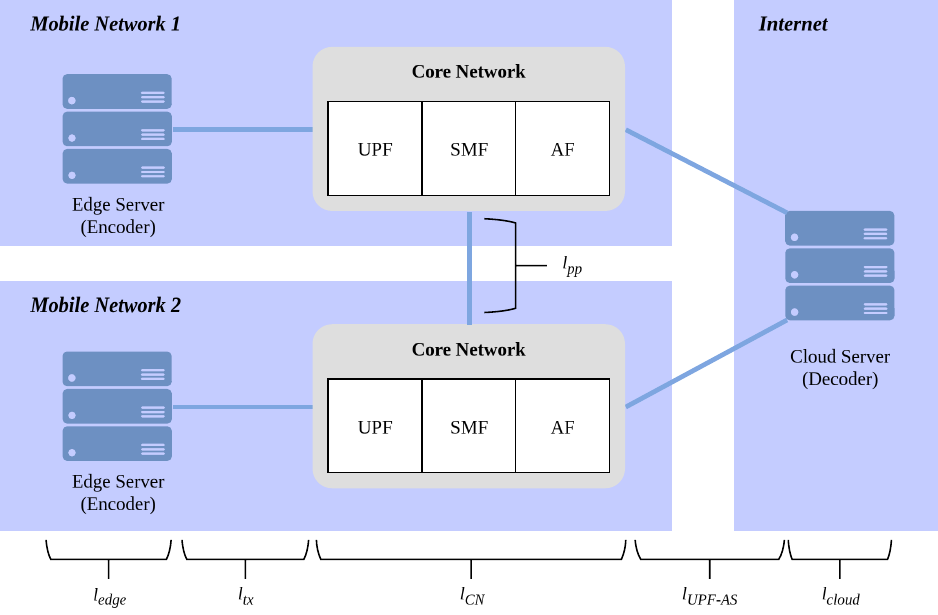}
        \caption{Latency components affecting end-to-end edge-to-cloud data transmission.}
        \label{fig:ch04:latency_components}
    \end{figure}

    The end-to-end communication path includes access, transport, and core-network components (Fig.~\ref{fig:ch04:latency_components}). 
    In the edge-cloud scenario, transport/core performance is frequently dominated by latency variability (queuing, congestion, and routing) rather than by channel noise on the wired path. 
    Accordingly, reducing semantic payload size is a direct mechanism for reducing the transport latency component and improving effective end-to-end service performance.
    
    Table~\ref{tab:ch04:latency_assumptions} provides a concise overview of the critical network latency assumptions used to evaluate communication delays within our experimental setup.
    Each entry represents a distinct aspect of latency within the network architecture, supported by references from the relevant literature.
    These assumed values, derived from the referenced papers, will form the basis for this experiment's benchmarking and evaluation.
    
    \begin{table}[!t]
        \centering
        \caption{Representative network latency values used for benchmarking.}
        \label{tab:ch04:latency_assumptions}
        \renewcommand{\arraystretch}{1.2}
        \begin{tabular}{lllc}
            \toprule    
            \textbf{Description} & \textbf{Parameter} & \textbf{Value/Range} & \textbf{Reference} \\
            \midrule   
             Core Network & $l_{\text{CN}}$ & $\sim$ 24 ms & \cite{shafi2017_5G-tutorial} \\
             Cloud-based (E2E) & $l_{\text{E2E}}$ & [53.8, 150] ms & \cite{CollPerales2023_E2E-V2X} \\
             UPF to AS & $l_{\text{UPF-AS}}$ & $\sim$ 20 ms & \cite{candela2020_covid19-impact} \\
             Edge / Cloud processing & $l_{\text{cloud}},\, l_{\text{edge}}$ & $\sim$ 13 ms (reported) & \cite{yang2023_swinjscc} \\
             Local Peering Point & $l_{\text{pp}}$ & 0.431 ms & \cite{CollPerales2023_E2E-V2X} \\
            \bottomrule
        \end{tabular}
    \end{table}

\subsection{Experimental Setup}
\label{ch04:exp_setup}
    This section describes the experimental environment used to benchmark semantic transcoding models and to profile their computational footprints.
    It includes details on the hardware setup, evaluation metrics, implementation details and the data collection process.
    
    \subsubsection{Hardware/Software Configuration and Metrics}
        The experiments were conducted on a high-performance computing system running Ubuntu 22.04.4 LTS x86. The hardware configuration included an Intel Xeon E5-2667 V3 operating at 3.196 GHz, 24 GB of RAM and a 150GB SSD. The implementation utilized libraries like PyTorch 1.9.0 with Python 3.8.19. Additional libraries included psutil for system monitoring, threading for concurrent execution, pandas for data manipulation, and seaborn for data visualization.
        
        System performance metrics focused on computational and transmission costs, specifically monitoring CPU and memory utilization over time and the transmission latency of a single latent representation at different compression ratios, to evaluate the resource efficiency of the models during encoding and decoding processes.

    \subsubsection{Data Collection}
        To obtain representative traces without excessive run time, we selected three random images from each of the Kodak and CLIC21 datasets and processed them using both WITT and SwinJSCC at a fixed coding rate (CBR 1/8). 
        For each run, CPU and memory traces were sampled periodically while the models executed the encode/decode pipeline.
        On the other hand, to analyze the transmission latency of a single latent representation, the input image will be compressed at four different compression rates.
        The observed latency will be compared among two different models and a baseline being transmitting the input image as-is.
    
    \subsubsection{Resources Utilization Measurement}
        CPU time statistics are read from \texttt{/proc/stat}, which provides cumulative time spent in multiple CPU states. 
        Table~\ref{tab:ch04:cpu_states} summarizes the relevant fields.
        
        \begin{table}[H]
            \centering
            \caption{CPU state fields in /proc/stat}
            \label{tab:ch04:cpu_states}
            \renewcommand{\arraystretch}{1.2}
            \begin{tabular}{>{\centering\arraybackslash}m{0.17\columnwidth} p{0.73\columnwidth}}
                \toprule    
                \textbf{State} & \multicolumn{1}{c}{\textbf{Description}} \\
                \midrule   
                 \textbf{user} & Time in user-mode processes (excluding nice) \\
                 \textbf{nice} & Time in lower-priority user-mode processes \\
                 \textbf{system} & Time in kernel-mode execution \\
                 \textbf{idle} & Time not executing any process \\
                 \textbf{iowait} & Time waiting for I/O operations \\
                 \textbf{irq/softirq} & Time handling interrupts \\
                \bottomrule
            \end{tabular}
        \end{table}
        
        The created logger reads the cumulative CPU times at two points in time separated by a specified interval. By calculating the difference between these two readings, it determines the time spent in each state during that interval.
            
        The percentage of CPU utilization is calculated as: $[\text{CPU Usage (\%)} = \frac{\text{Active Time}}{\text{Total Time}} \times 100,]$ where Active Time = user + nice + system + irq + softirq, and Total Time = Active Time + idle + iowait.
        
        Also, regarding memory utilization, it is calculated by reading the memory statistics provided by the \texttt{psutil} Python library, which uses data from system files like \texttt{/proc/meminfo}. The relevant memory statistics include the total system memory available, the memory currently used by all running processes, and the memory available for use without swapping.
            
        Memory utilization in gigabytes (GB) is calculated as: $[\text{Memory Usage (GB)} = \frac{\text{Used Memory (Bytes)}}{1024^3}].$

\subsection{Experimental Analysis}
\label{ch04:exp_analysis}
    This section compares CPU and memory utilization traces for WITT and SwinJSCC while transcoding images from Kodak and CLIC21 at CBR 1/8. 
    The results characterize the computational overhead of semantic transcoding as a function of model choice and dataset complexity.

    \subsubsection{CPU Utilization}
\begin{figure}[H]
            \centering
            \begin{subfigure}[b]{\columnwidth}
                \centering
                \includegraphics[width=\linewidth]{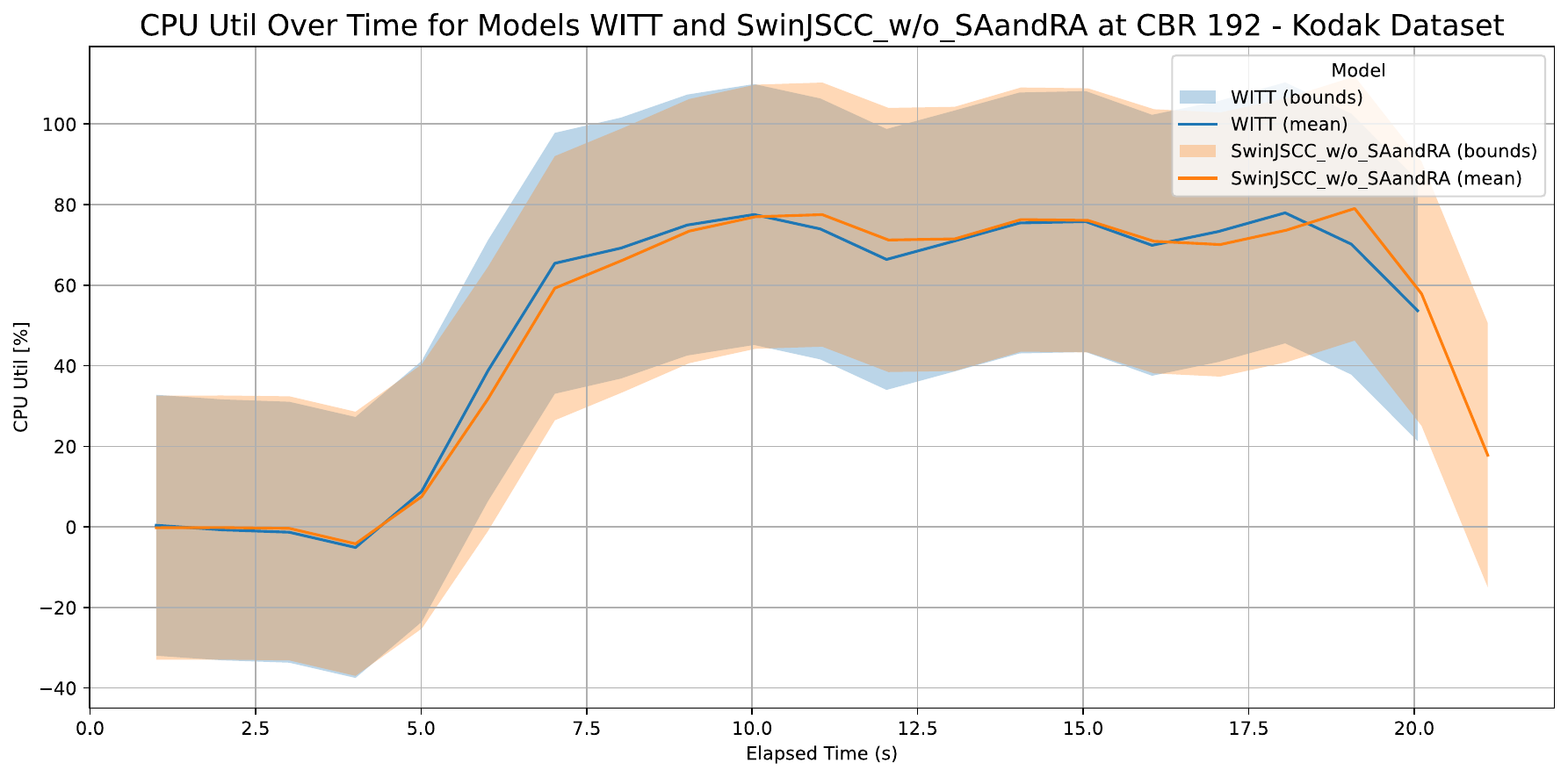}
                \caption{CPU utilization vs. time (Kodak)}
                \label{fig:ch04:swinVwitt_cpu_kodak}
            \end{subfigure}
            \hfill
            \begin{subfigure}[b]{\columnwidth}
                \centering
                \includegraphics[width=\linewidth]{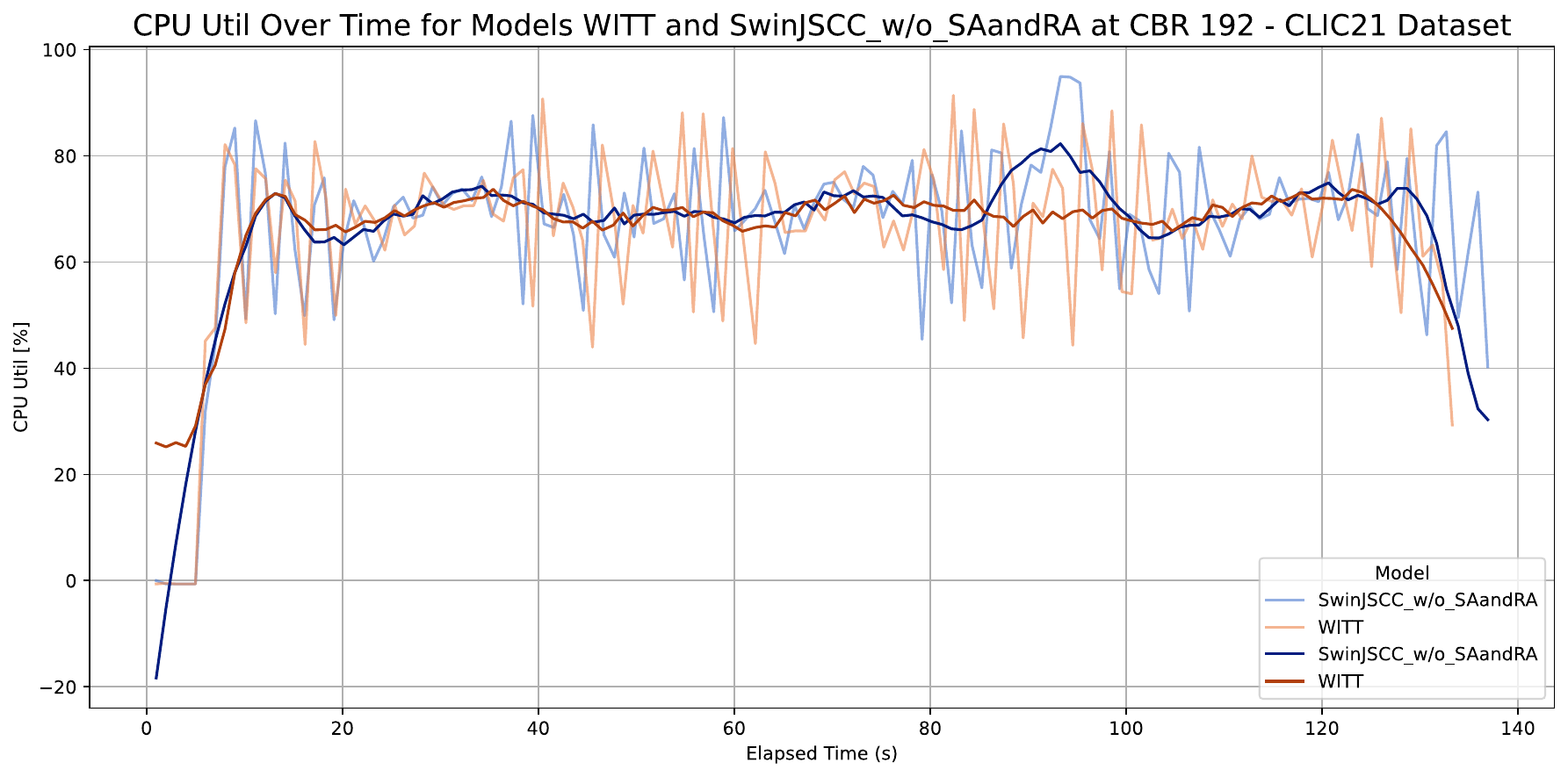}
                \caption{CPU utilization vs. time (CLIC21)}
                \label{fig:ch04:swinVwitt_cpu_clic21}
            \end{subfigure}
            \caption{CPU utilization traces for WITT and SwinJSCC while transcoding Kodak and CLIC21 images at CBR 1/8}
            \label{fig:ch04:cpu_util_comp}
        \end{figure}

        The CPU utilization plots for the WITT and SwinJSCC models reveal significant fluctuations in computational demands, with both models showing relatively similar performance. For the CLIC21 dataset, both models frequently hit peaks over 85\% CPU utilization (see Fig.~\ref{fig:ch04:swinVwitt_cpu_clic21}). These spikes result from complex operations like window partitioning, cyclic shifting, and adaptive modulation based on SNR values within the SwinTransformerBlock. The dynamic and conditional nature of these operations contributes to the variability in CPU usage. In contrast, while processing the Kodak dataset, both models exhibit slightly higher CPU utilization, about 5\% more than with the CLIC21 dataset (see Fig.~\ref{fig:ch04:swinVwitt_cpu_kodak}). The models in the Kodak dataset also tends to be relatively smoother compared to the CLIC21 dataset, this can be due to the Kodak dataset's simplicity. Despite this simplicity, both models still hit highs of over 80\% CPU utilization during processing, indicating the intensive nature of their operations.
        
        While Kodak processing is more stable, the CLIC21 dataset shows more fluctuations due to its complexity and the number of features to be processed. In both datasets, WITT processes images faster than SwinJSCC, with an interval of up to 3 seconds. The average processing time per image from the Kodak dataset is around 5-6 seconds, while it takes around 45 seconds for each image in the CLIC21 dataset. This substantial difference in processing time highlights the impact of image complexity on computational efficiency.
        
        It is also evident that CPU activity is intense during the encoding and decoding phases, as observed from the intervals where CPU utilization peaks. These observations underscore the importance of optimizing both models for better handling of complex datasets to achieve more consistent and efficient performance (see Fig.~\ref{fig:ch04:swinVwitt_cpu_clic21}).
        
    \subsubsection{Memory Utilization}
\begin{figure}[H]
            \centering
            \begin{subfigure}[b]{\columnwidth}
                \centering
                \includegraphics[width=\linewidth]{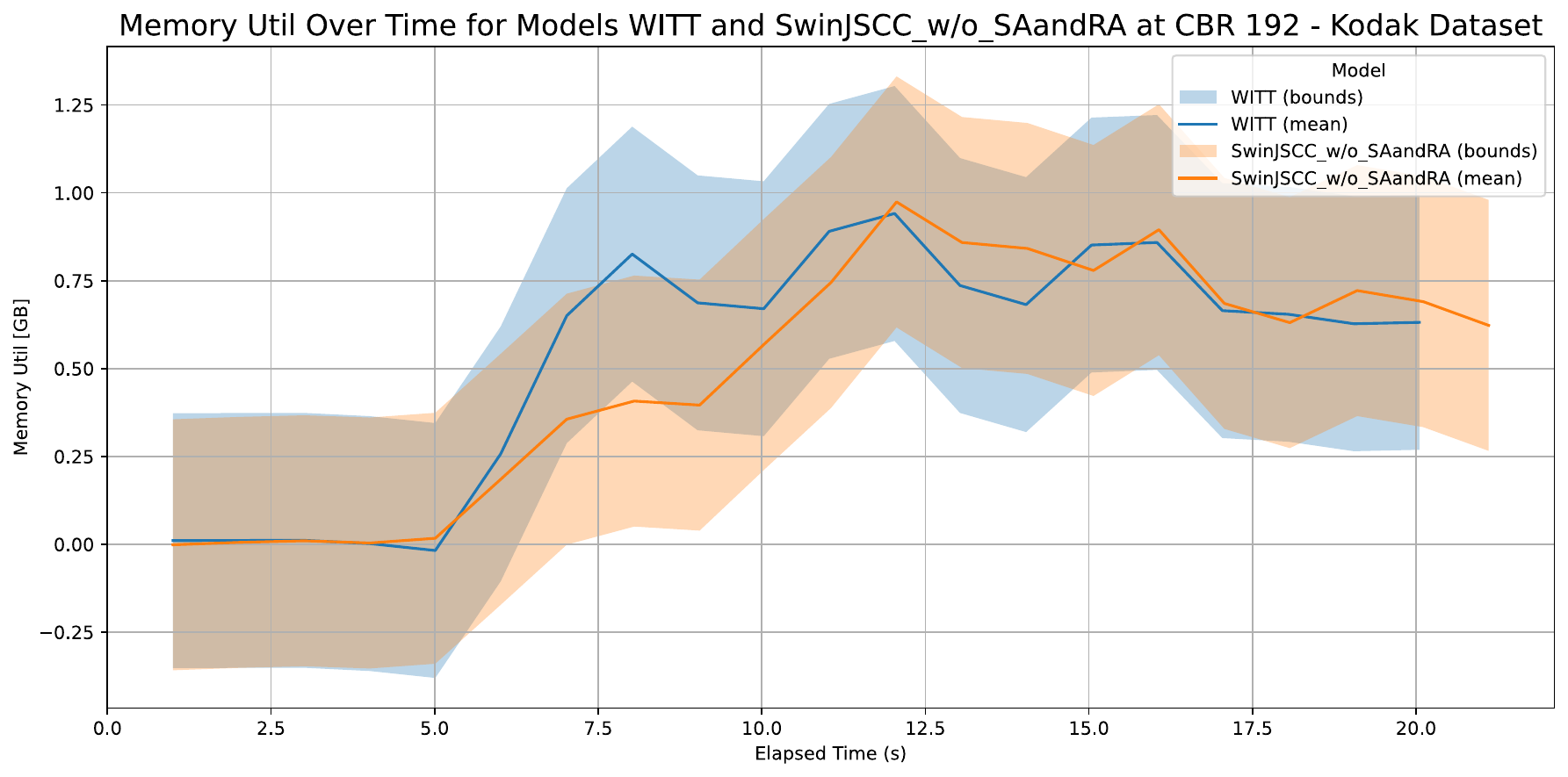}
                \caption{Memory utilization vs. time (Kodak)}
                \label{fig:ch04:swinVwitt_mem_kodak}
            \end{subfigure}
            \vfill
            \begin{subfigure}[b]{\columnwidth}
                \centering
                \includegraphics[width=\linewidth]{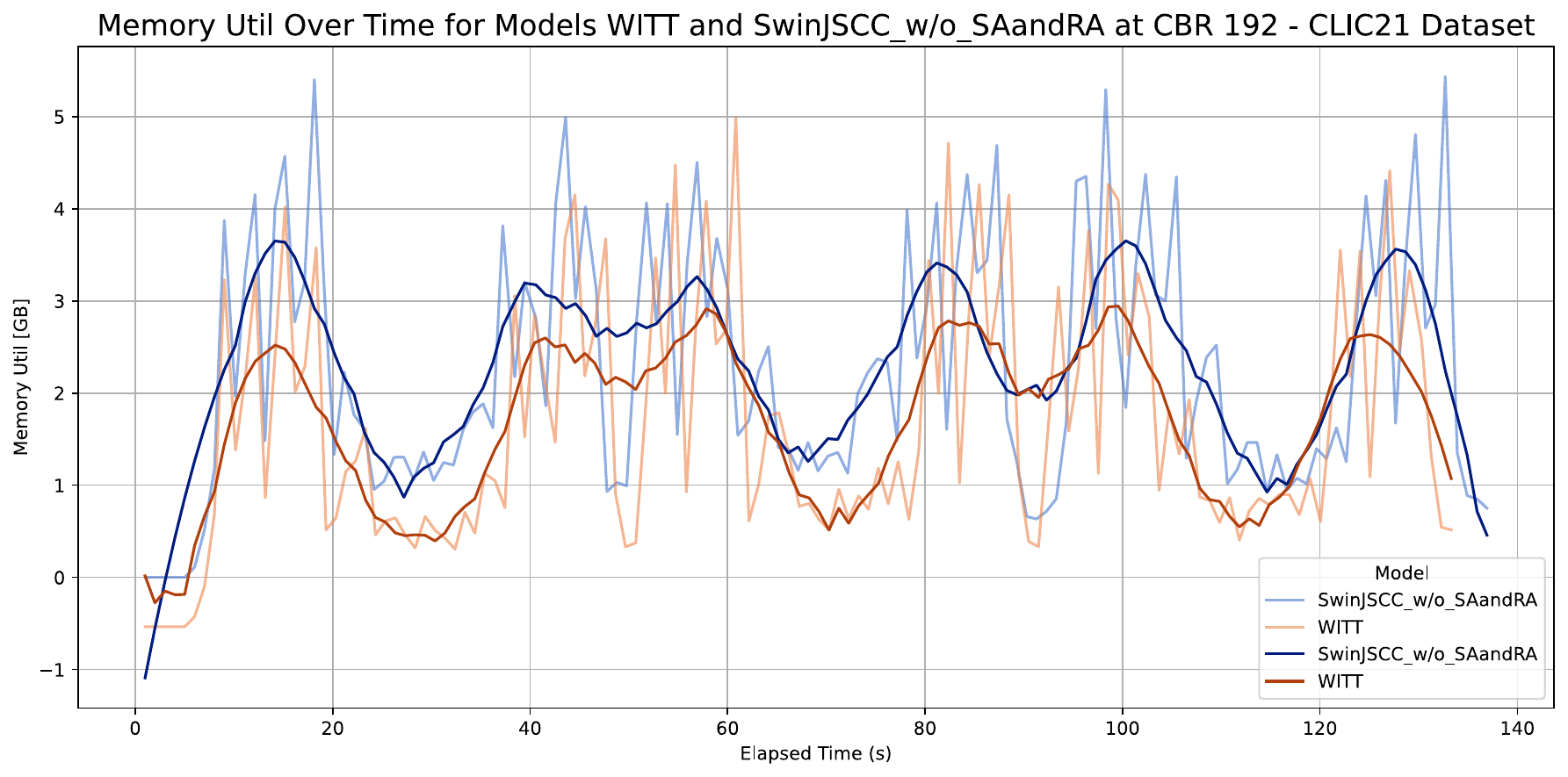}
                \caption{Memory utilization vs. time (CLIC21)}
                \label{fig:ch04:swinVwitt_mem_clic21}
            \end{subfigure}
            \caption{Memory utilization traces for WITT and SwinJSCC while transcoding Kodak and CLIC21 images at CBR 1/8}
            \label{fig:ch04:mem_util_comp}
        \end{figure}

        The memory utilization plots for the WITT and SwinJSCC models reveal distinct patterns when processing images from the Kodak and CLIC21 datasets (see Fig.\ref{fig:ch04:mem_util_comp}). For the Kodak dataset, the average memory utilization of the WITT model remains relatively stable (see Fig.\ref{fig:ch04:swinVwitt_mem_kodak}). This stability can be attributed to efficient memory handling within the SwinTransformerBlock and the model's structured processing.
        
        The complexity of the datasets plays a significant role in memory usage. Throughout the entire Kodak dataset processing, the maximum memory utilization for WITT did not exceed 1.5 GB, while for the CLIC21 dataset, it averaged around 4.5 GB, demonstrating the higher memory demands for more complex data (see Fig.~\ref{fig:ch04:swinVwitt_mem_clic21}). Despite SwinJSCC having similar memory utilization to WITT when processing the Kodak dataset, it consumes up to 40\% more memory on average compared to the WITT memory consumption (approximately 1 GB more) and exhibits more intense fluctuations.
        
        SwinJSCC uses approximately 40\% more memory than the WITT model, with a maximum of up to 5 GB of total memory usage, highlighting its higher memory demands during transcoding operations. However, the adaptability of the SwinJSCC attention mechanism, including the use of window-based attention and a shifted window scheme, allows it to adjust to varying data complexities dynamically. This adaptability enables the SwinJSCC model to handle complex patterns and dependencies effectively, making up for its higher memory consumption.
        
        These observations underscore the differences in memory utilization efficiency between the WITT and SwinJSCC models. While both models exhibit high computational demands, WITT manages memory usage more effectively, resulting in more stable consumption patterns. Nevertheless, the adaptability of the SwinJSCC model's attention mechanism provides a significant advantage in handling diverse and complex datasets. 
        
    \subsubsection{Transmission Latency Analysis}
\begin{figure}[H]
           \centering
           \begin{subfigure}[b]{0.48\textwidth}
               \centering
               \includegraphics[width=\linewidth]{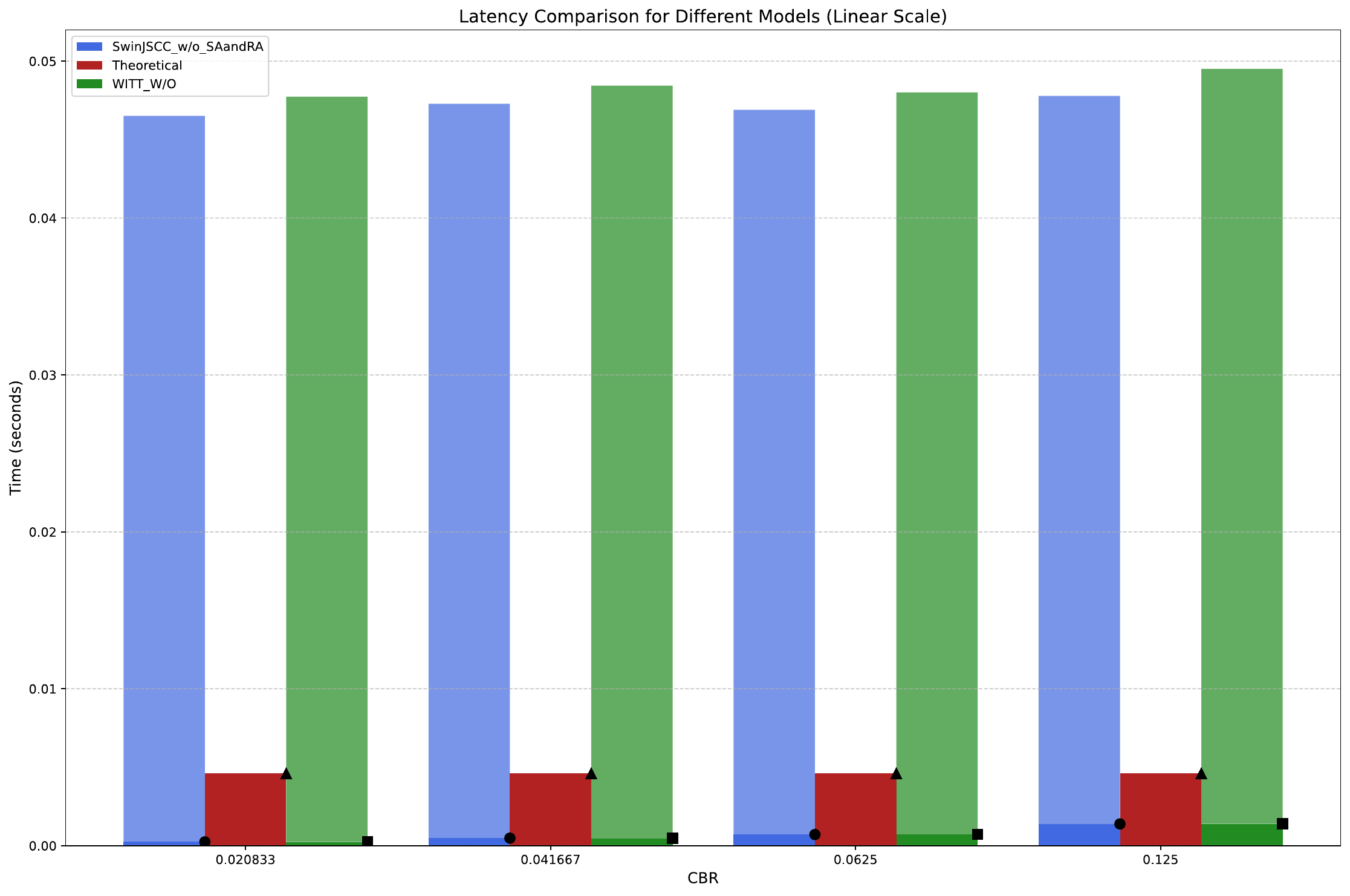}
               \caption{Linear Scale}
               \label{fig:ch04:lat_linear}
           \end{subfigure}
           \begin{subfigure}[b]{0.48\textwidth}
               \centering
               \includegraphics[width=\linewidth]{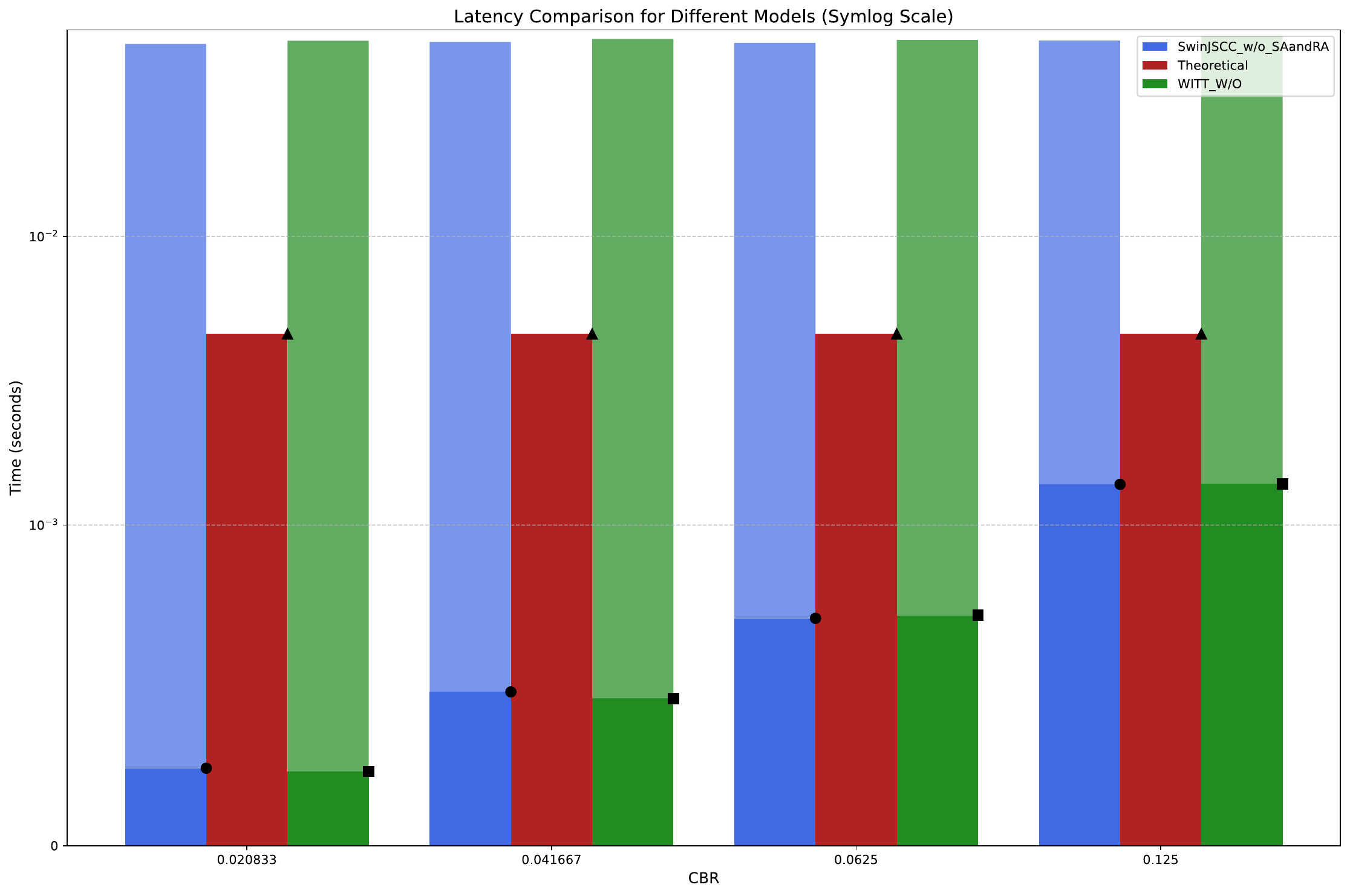}
               \caption{Symlog Scale}
               \label{fig:ch04:lat_symlog}
           \end{subfigure}
           \caption{Latency decomposition comparing Raw Transmission (Red) vs. Semantic Transcoding (Blue/Green). The darker lower sections represent the actual network transmission time, while the lighter upper sections represent computational overhead.}
           \label{fig:ch04:latency_breakdown}
        \end{figure}

        To evaluate the effectiveness of Semantic Transcoding in alleviating backhaul congestion, we benchmarked the total end-to-end latency against a theoretical baseline of transmitting raw, non-semantic data.
        Fig.~\ref{fig:ch04:latency_breakdown} presents the latency breakdown for SwinJSCC and WITT across varying Channel Bandwidth Ratios (CBR).
        
        To visualize the orders-of-magnitude difference between processing and transmission times, we present the data in both Symlog scale (Fig.~\ref{fig:ch04:lat_symlog}) and Linear scale (Fig.~\ref{fig:ch04:lat_linear}).
        The bars are decomposed into two distinct components:
        \begin{itemize}
           \item \textbf{Processing Latency (Lighter Shade):} The time consumed by the Encoder and Decoder to extract and reconstruct features ($l_{\mathrm{encoder}} + l_{\mathrm{decoder}}$).
           \item \textbf{Transmission Latency (Darker/Shaded Region):} The actual time taken to transmit the latent features over the network ($l_{\mathrm{tx}}$).
        \end{itemize}
        The \textbf{Theoretical Baseline (Red Bar)} represents the transmission latency of the raw input image without any semantic compression, calculated based on the available bandwidth and original file size.
        
        Focusing on the transmission latency represented as the darker, shaded regions of the Blue (SwinJSCC) and Green (WITT) bars, the results confirm the core hypothesis of this survey: semantic transcoding significantly reduces network load.
        As observed in Fig.~\ref{fig:ch04:lat_symlog}, the semantic transmission time is consistently lower than the raw transmission baseline (red bar), particularly at lower compression ratios.

        This reduction is the direct result of the feature pruning mechanisms described in the system model.
        By transmitting only the semantic meaning rather than the raw data, the system effectively relieves pressure on the core network's backhaul.
        As the CBR increases, the semantic payload size grows, causing the semantic transmission time to approach the raw transmission baseline.
        This behavior aligns with the theoretical model in Eq.~\ref{eq:ch02:tx_latency}, where $l_{\mathrm{tx}}$ is linearly proportional to the payload size $S$.

\subsection{section Summary}
\label{ch04:summary}
    This section introduced \emph{Semantic Transcoding} as a specialized mechanism to optimize data transmission across the edge-cloud backhaul.
    By shifting the focus from wireless noise resilience to wired transport efficiency, we established a framework where the edge server operates as an intelligent gateway, pruning redundant data to actively manage the transmission latency component ($l_{\mathrm{tx}}$).

    The experimental analysis validated the core hypothesis: transmitting latent semantic features yields a significant reduction in network latency compared to raw data transmission.
    Specifically, the transmission latency analysis (Section~\ref{ch04:exp_analysis}) demonstrated that semantic payloads remain consistently efficient even as the Channel Bandwidth Ratio (CBR) increases, adhering to the theoretical throughput models.
    This confirms that semantic transcoding is an effective strategy for maintaining low-latency data exchange in bandwidth-constrained or congested edge-cloud environments.

    Additionally, we profiled the computational footprint of the WITT and SwinJSCC models.
    The resource utilization traces revealed that while these transformer-based encoders offer superior transmission efficiency, they impose distinct CPU and memory demands that scale with dataset complexity.
    Understanding these transmission gains alongside their operational resource costs provides the necessary data to design the network-aware adaptive control strategies that will be developed in section~\ref{ch05}.

\section{Network-Aware Semantic Transcoding: SwinNetCC Case Study}
\label{sec:swinnetcc}
\noindent
This section surveys network-aware semantic transcoding mechanisms at the edge-cloud interface, using SwinNetCC as a representative design point. The focus is on adaptive channel selection, latency-aware policies, and end-to-end modeling of transport and reconstruction.


\label{ch05}

\subsection{Introduction}
\label{ch05:intro}
    The surge in global data demand, driven by the proliferation of smart devices, the Internet of Things (IoT), and bandwidth-intensive applications such as 4K video streaming and virtual/augmented reality, places an unprecedented pressure on 5G networks \cite{s23083876} \cite{y}.
    Next-generation systems are expected to handle up to 100 times more traffic than current networks \cite{9585402}, with global mobile data traffic projected to grow from 130 exabytes per month in early 2023 to over 460 exabytes by 2029 \cite{ericsson2023mobile}.
    Meeting these demands requires ultra-high data rates ($>$100 Gbps), ultra-reliable low-latency communication, and significant improvements in the core network \cite{vodafone2021}.
    Future networks will also rely heavily on AI-driven management systems and edge computing to maintain high performance under extreme traffic conditions \cite{su16167039}.
    
    The limitations of centralized cloud computing have become evident in latency-sensitive applications such autonomous systems, augmented reality, and real-time analytics \cite{8972389}.
    Edge computing offers a compelling alternative by processing data closer to where ti is generated, enabling faster responses and reducing the burden on the core network \cite{3} \cite{x}.
    Integrating semantic transcoding into edge servers further improves efficiency by transmitting meaning rather than raw data, thereby reducing communication overhead and enabling context-aware data exchange \cite{10667089}.
    However, semantic communication also introduces computational complexity, which may offset some latency gains \cite{qin2022semantic}.
    
    Another challenge arises from the highly dynamic nature of the core network. Variations in available bandwidth and latency degrade system performance if the encoder/decoder cannot adapt in real time.
    Joint Source-Channel Coding (JSCC) has been proposed to jointly optimize transmission across source and channel and boosting end-to-end content delivery \cite{5563107}, but conventional JSCC schemes are typically trained offline and struggle under unseen conditions \cite{10158528}.
    Thus, there is a need for adaptive semantic communication frameworks capable of dynamically adjusting to fluctuating network states.

    This work addresses these challenges by proposing a modular Swin-Transformer-based framework for adaptive semantic communication.
    The framework can be integrated with existing encoder/decoder models, enabling the system to selectively transmit the most important semantic features depending on the available network resources.
    This approach effectively reduces end-to-end latency while maintaining high-fidelity data reconstruction.

    The contributions of this paper can be summarized as follows:
    \begin{itemize}
        \item Design a modular semantic coding framework that adapts to varying core network conditions using Swin-Transformer encoders and decoders
        \item Introduce adaptive policies to selectively transmit semantic features, achieving a trade-off between latency and fidelity
        \item Evaluate the framework on high-resolution image transmission tasks and demonstrate improvements in metrics such as Multiscale Structural Similarity Index Metric (MS-SSIM) under varying bandwidth and latency constraints.
    \end{itemize}

\subsection{\textsc{SwinNetCC}: Swin Network-Aware Compressed Communication}
\label{ch05:sys-model}
\begin{figure}[!ht]
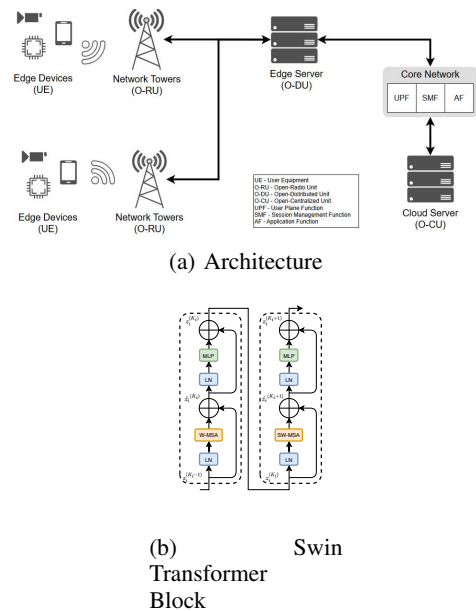

        \centering
        \begin{subfigure}{0.7\linewidth}
            \centering
            \includegraphics[width=\linewidth]{figs/system_model.pdf}
            \caption{Architecture}
            \label{fig:ch05:system_model}
        \end{subfigure}
        \begin{subfigure}{0.29\linewidth}
            \centering
            \includegraphics[width=\linewidth]{figs/swin_block.pdf}
            \caption{Swin Transformer Block}
            \label{fig:ch05:swin_block}
        \end{subfigure}
        \caption{(a) Overall architecture of \textsc{SwinNetCC} for edge-to-cloud communications; (b) internals of a Swin Transformer block}
        \label{fig:ch05:system_architecture}
    \end{figure}

    We consider a MEC node that must transmit high-resolution images to a cloud service over the core network as illustrated in Fig.~\ref{fig:ch05:system_architecture}. 
    Unlike wireless transmission, the edge-to-cloud path traverses a wired backhaul, where the dominant network impairments are available bandwidth, denoted by $B\,[\mathrm{Mbps}]$ and round-trip latency, denoted by $l_{\mathrm{RTT}}\,[\mathrm{ms}]$.
    \footnote{The pair $B,l_{\mathrm{RTT}}$ can be obtained from SDN counters, passive throughput monitors, or any telemetry fabric; our design does not assume NWDAF or PHY feedback.} 
    Packet loss is assumed to be managed by the transport layer and is therefore not explicitly modeled in our framework. 
    However, unlike classical separation coding (e.g., JPEG+TCP) which incurs retransmissions and increases end-to-end delay, JSCC/semantic coding embeds redundancy in feature space and degrades gracefully under congestion, preserving lower latency and acceptable reconstruction quality even without retransmission.

    \subsubsection{Swin-Encoder With Network-Aware Masking}
        At the sender edge node, each input image is first transformed into a latent representation tensor $y^\prime$, by a hierarchical Swin Transformer encoder. Starting from non-overlapping patch embeddings, multiple Swin blocks and patch-merging layers produce a feature map of shape $(b, N, C)$, where $(b)$ is the batch size, $(N)$ represents the number of tokens and $(C)$ is the number of channels.

        Before transmitting the latent representation $y^\prime$ across the network, we feed it to our \textit{NetAware-Modulator} module that adaptively compresses along the channel dimension based on the current network telemetry $(B, l_{\mathrm{RTT}})$. The mapping
        \begin{equation}
            k = h(B, l_{\mathrm{RTT}})
        \end{equation}
        determines how many channels are retained.
        In this work, latency is ignored in the mapping, where it is used as part of the latency computation, and $h(\cdot)$ depends only on $B$, defined as
        \begin{equation}
            h(B) = \texttt{levels}\!\Bigg[ 
                \min \Bigg(
                    \Big\lfloor 
                    \left( \eta(B) \right)^{\gamma} \cdot L 
                    \Big\rfloor, \; L-1
                \Bigg)
            \Bigg],
            \label{eq:ch05:k_estimator}
        \end{equation}
        
        \noindent
        where
        \begin{equation}
            \eta(B) = \frac{\log_{10}(B) - \log_{10}(B_{\min})}
                           {\log_{10}(B_{\max}) - \log_{10}(B_{\min})}.
        \end{equation}
        where $B \in [B_{\min},B_{\max}]$ is clamped to the telemetry window, $\gamma > 1$ is a tunable steepness parameter, $L = |\texttt{levels}|$ is the number of bins, and $\texttt{levels} = \{32, 64, 92, 128, 160, 192\}$ is a discrete set of channel budgets.
        The selected $k$ then directs the modulator to rank the channels by summed absolute activation, select the top-$k$, and zero out the remainder.

        \begin{figure}[h]
            \centering
            \includegraphics[width=0.7\linewidth]{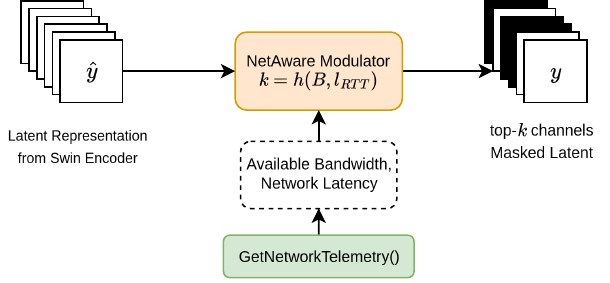}
            \caption{Showcasing NetAware Modulator retrieving network telemetry, and mask the latent representation}
            \label{fig:ch05:netaware_mod}
        \end{figure}

        The operational flow of this selection process is depicted in Fig.~\ref{fig:ch05:netaware_mod}.
        Once $k$ is determined by the telemetry-driven estimator, the NetAware-Modulator ranks the channels based on their semantic importance, calculated via the summed absolute activation across the spatial dimensions.
        A binary mask $\mathbf{m} \in \{0,1\}^C$ is generated to retain the top-$k$ channels while suppressing the rest.
        The final transmitted tensor $\hat{y}$ is obtained via element-wise multiplication, $\hat{y} = y^\prime \odot \mathbf{m}$, ensuring that bandwidth is allocated only to the most semantically salient features.
        
        A learnable neural mapping $h(B,l)$ was initially considered, but its design complexity and inference overhead made it impractical for real-time deployment. The deterministic log-based policy instead provides stability, interpretability, and lightweight adaptation. Future work will consider learned adaptive policies that incorporate latency $l_{\mathrm{RTT}}$ jointly with $B$.
        
    \subsubsection{Core-Network Channel Model}
        The wired 5G core path between edge and cloud is characterized solely by throughput $B$ and round‐trip latency $l_{\mathrm{RTT}}$, which can be obtained from SDN/NWDAF telemetry \cite{3gpp-nwdaf} or passive probes. In our implementation, throughput is sampled from a log‐uniform distribution (10 Mbps–1 Gbps) and latency from a truncated log‐normal distribution (5–300 ms). This choice reflects empirical observations: core-network throughput spans several orders of magnitude motivating a log-uniform model, while latency distributions are positively skewed with long tails motivating a log-normal model. Swapping in actual NWDAF or SNMP calls requires no changes to the upstream SwinNetCC pipeline, only the telemetry backend must be redirected.
    
    \subsubsection{Swin Decoder and Reconstruction}
        At the cloud receiver, the masked latent tensor $y$ (shape $(b \times N \times C)$) is fed into a hierarchical Swin Transformer decoder that inverts the encoder and reconstructs the original image. Each stage swaps window attention for patch \emph{un}merging, doubling spatial resolution while halving channels. After performing the patch un-merging stages, the feature map matches the original resolution. A final $1 \times 1$ projection and sigmoid yield the RGB reconstruction $\hat{x}$. Because masked channels are padded with zeros, tensor shapes stay fixed as $k$ varies, requiring no re-initialization.

\subsection{Mathematical Symbols \& Channel Semantics}
\label{ch05:math_sym}
    Let $\mathbf{x} \in \mathbb{R}^{H \times W \times 3}$ denote an input image.
    The Swin encoder $f_{\text{enc}}(\cdot;\phi)$ produces $\mathbf{y}' \in \mathbb{R^{N \times C}}$, where $N$ is the number of tokens and $C$ the number of channels.
    For each token $n \in \{1, \dots, N\}$, the vector $\mathbf{y}'_n = (y'_{n,1},\dots,y'_{n,C})$ contains channel coefficients corresponding to distinct semantic attributes.
    Channel importance is measured by
    \[
        \mathrm{importance}_c = \sum_{n=1}^N |y'_{n,c}|,
    \]
    and the top-$K$ channels are retained according to $k=h(B,l_{\mathrm{RTT}})$.
    The mask $\mathbf{m}\in \{0,1\}^C$ selects these channels, giving $\mathbf{y}=\mathbf{y}'\odot \mathbf{m}$.
    The Swin decoder $f_{\text{dec}}(\cdot;\theta)$ reconstructs $\hat{\mathbf{x}}$ from $(\mathbf{y},\mathbf{m})$, ensuring robust recovery even with partial features.

    The overall procedure is summarized in Algorithm~\ref{alg:network-aware-swin}.

    \begin{algorithm}[h]
        \caption{Network-Aware Adaptive Semantic Communication (MEC-to-Cloud over 5G Core)}
        \label{alg:network-aware-swin}
        \begin{algorithmic}[1]
            \Require Input data $\mathbf{x} \in \mathbb{R}^{H \times W \times 3}$
            \Require Swin Encoder $f_{\text{enc}}(\cdot;\phi)$, Decoder $f_{\text{dec}}(\cdot;\theta)$
            \Require Network telemetry API providing bandwidth $B$ and latency $L$
            \Ensure Reconstructed data $\hat{\mathbf{x}}$
    
            \State \textbf{// At MEC Transmitter}
                \Statex $\mathbf{y}' \gets f_{\text{enc}}(\mathbf{x};\phi)$ \Comment{Encode semantic features}
                \Statex $(B, L) \gets$ \text{GetNetworkTelemetry()} 
                \Statex $k \gets h(B, l)$ \Comment{$k\in\{32\dots192\}$}
                \Statex $\mathbf{m} \gets \text{TopK}(\text{importance}(\mathbf{y}'), k)$ 
                \Statex $\mathbf{y} \gets \mathbf{y}' \odot \mathbf{m}$ \Comment{Apply mask to latent features}
            \State \text{Transmit} $(\mathbf{y}, \mathbf{m})$ to Cloud Receiver
            \State \textbf{// At Cloud Receiver}
                \Statex $(\mathbf{y}, \mathbf{m}) \gets$ \text{ReceiveData()}
                \Statex $\hat{\mathbf{x}} \gets f_{\text{dec}}(\mathbf{y}, \mathbf{m};\theta)$ \Comment{Decode and reconstruct}
            \State \text{Deliver} $\hat{\mathbf{x}}$ to downstream application
        \end{algorithmic}
    \end{algorithm}

\subsection{Experimental Setup}
\label{ch05:exp_setup}
    SwinNetCC is evaluated across the 3GPP TS 23.501 core-network envelope, varying throughput $B \in [10,1000]$ Mbps and round-trip latency $l_{\mathrm{RTT}} \in [5,300]$ ms. 
    The one-way transmission delay for each snapshot is calculated using the model defined in section~\ref{ch02}:
    \begin{equation}
        l_{\mathrm{tx}} = S/B + l_{\mathrm{RTT}}/2,
        \tag{\ref{eq:ch02:tx_latency}}
    \end{equation}
    where $S$ is the payload size. 
    No runtime probing or NWDAF analytics are used; instead, the full envelope is swept to isolate codec behavior. 
    NWDAF statistics (e.g., mean, variance, percentiles) are accessible via standardized APIs \cite{3gpp-nwdaf}, and could be directly integrated in future live deployments.
    
    The model is trained on DIV2K (2K-resolution DSLR images) and tested on three benchmarks: Kodak (24 uncompressed film photographs), CLIC21 (web images up to 8 MP), and FLICKR2K (community images including 4K resolution). 
    These provide a range of natural and high-frequency details.
    
    For each test image, the system transmits (i) raw RGB (“Source”), (ii) full latent Swin output $y'$ (“Encoded”), and (iii) masked latent $y$ (“Masked”). 
    For every link snapshot $(B,l_{\mathrm{RTT}})$, the bandwidth utilization, transmission latency $l_{\mathrm{tx}}$, effective throughput, and reconstruction quality (MS-SSIM) are logged. 
    Figs.~\ref{fig:ch05:link_util}–\ref{fig:ch05:tx_lat} report these metrics.
    
    A rate and noise-agnostic Swin-JSCC variant is trained on DIV2K for 1,500 epochs using an Ubuntu 22.04 workstation (Ryzen 9 5900X, 32 GB RAM, RTX 3080 Ti).

\subsection{Experimental Analysis}
\label{ch05:exp_analysis}
        SwinNetCC is evaluated with the NetAware-Modulator on random frames drawn from Kodak and FLICKR2K. 
        During testing, the sample image receives a single snapshot of network telemetry, bandwidth $B$ and round-trip time $RTT$, sampled from the ranges recommended by 3GPP for core backbones (10 Mbps to 1000 Mbps, 5 ms to 300 ms). 

        \begin{figure}[H]
            \centering
            \begin{minipage}[H]{\linewidth}
                \centering
                \captionof{table}{Comparison of payload sizes at different states (Mb)}
                \label{tab:ch05:size_comp}
                \begin{tabular}{|c|c|c|c|}
                    \hline
                    & Source Image & Encoded & Masked Latent \\ \hline
                    Kodak    & 37.75  & 15.73  & 1.57–9.44 \\ \hline
                    CLIC21   & 251.66 & 104.86 & 10.49–62.91 \\ \hline
                    FLICKR2K & 235.93 & 98.30  & 9.83–58.98 \\ \hline
                \end{tabular}
            \end{minipage}
            \vspace{1em}
            
            \begin{minipage}[H]{\linewidth}
                \centering
                \includegraphics[width=0.6\linewidth]{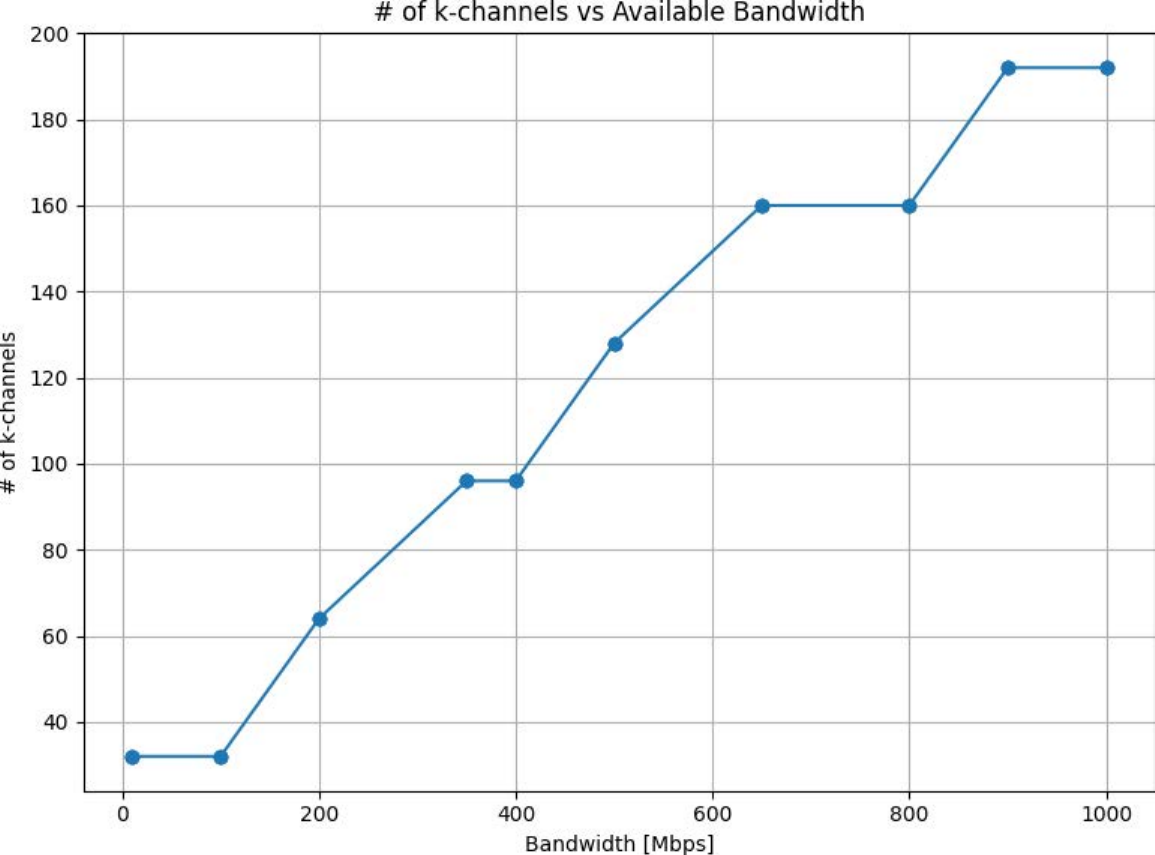}
                \caption{Chosen number of $k$ channels based on link capacity $B$}
                \label{fig:ch05:k_vs_bw}
            \end{minipage}
        \end{figure}
        
        Table~\ref{tab:ch05:size_comp} outlines the size of the sample image in Mbits, that was used during the testing at different stages: original size, after encoding by the Swin transformer and finally size of the payload after masking stage. 
        From Table~\ref{tab:ch05:size_comp} it is important to notice the size range under the \textit{Masked Latent} column. 
        This is due to the top-$k$ masking mechanism carried out in this work, and in the upcoming sections the masking effects will be discussed. 
        Fig.~\ref{fig:ch05:k_vs_bw} illustrates the relationship between the chosen number of $k$ channels with respect to the available bandwidth retrieved from the network telemetry.

        \subsubsection{Bandwidth Utilization}
\begin{figure}[h]
                \centering
                \includegraphics[width=0.8\linewidth]{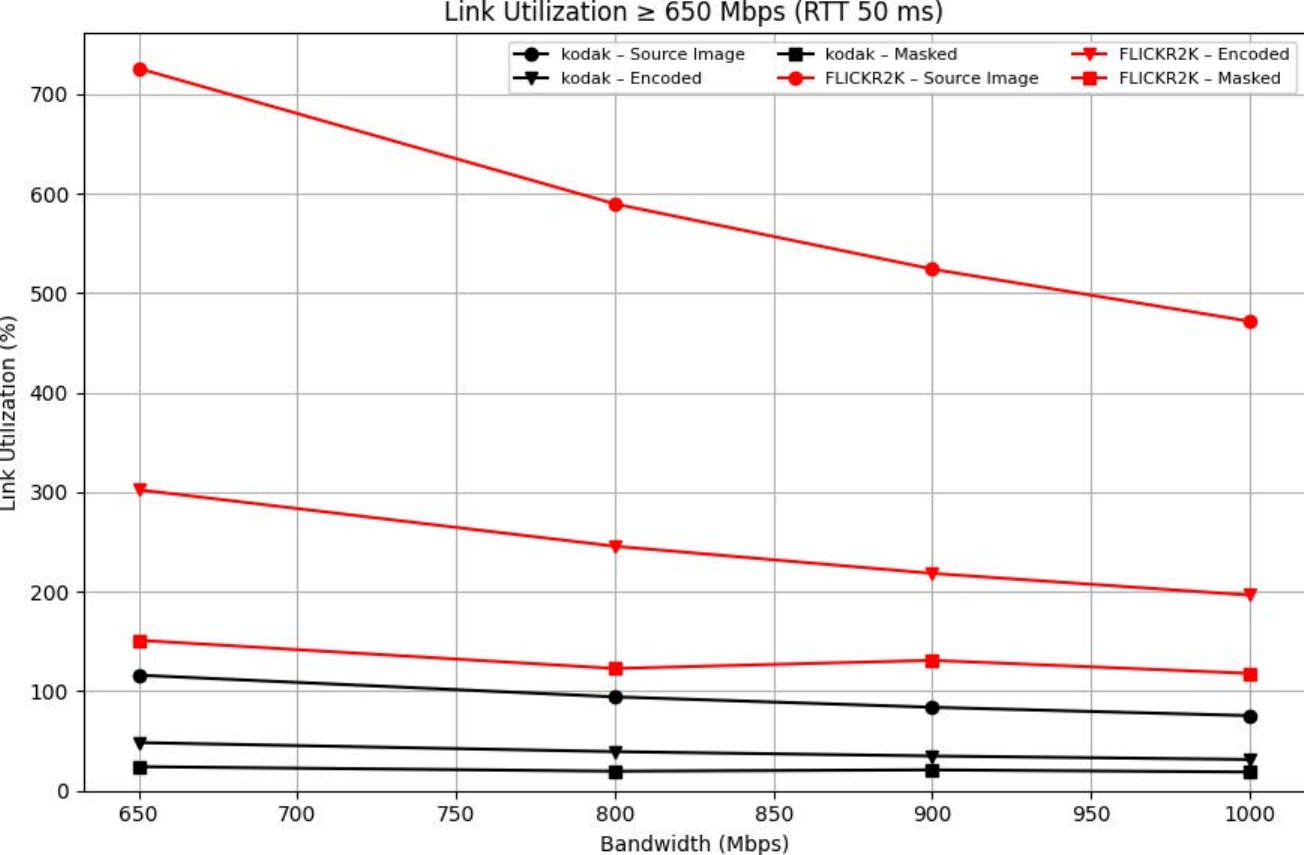}
                \caption{Bandwidth utilization as link capacity varies at $l_{\mathrm{RTT}} =$ 50.0 ms across Kodak and FLICKR2K datasets}
                \label{fig:ch05:link_util}
            \end{figure}

            Fig.~\ref{fig:ch05:link_util} plots, for a fixed $l_{\mathrm{RTT}} =$ 50 ms, the percentage of the link's bandwidth-delay product consumed by each payload as the available link capacity is varied from 10 Mbps to 1000 Mbps. 
            Payload magnitude used in the calculation appears in Table~\ref{tab:ch05:size_comp}. 
            
            The curves for the \textit{Source Image} and \textit{Encoded Source} fall within the expected $1/B$ scheme, as the payload sizes are constant so the utilization decreases exponentially as link capacity increases. 
            In contrast the \textit{Masked Latent} trace remains essentially flat ($\approx$ 15-20 \%) because the \textit{NetAware-Modulator} expands the latent tensor logarithmically to the bandwidth until the ceiling $k_{max} =$ 192 channels is reached. 
            If the policy’s bandwidth range were extended beyond 1000 Mbps, the masked curve would adopt the same inverse-capacity trend once the channel limit is saturated.

            Hence, within the 3GPP core-network envelope under study, channel masking delivers a bounded and resolution-independent bandwidth footprint, whereas transmitting the full latent or the raw image risks saturating the pipe at the lower end of the capacity sweep.
            
        \subsubsection{Network Throughput}
\begin{figure}[!h]
                \centering
                \includegraphics[width=0.8\linewidth]{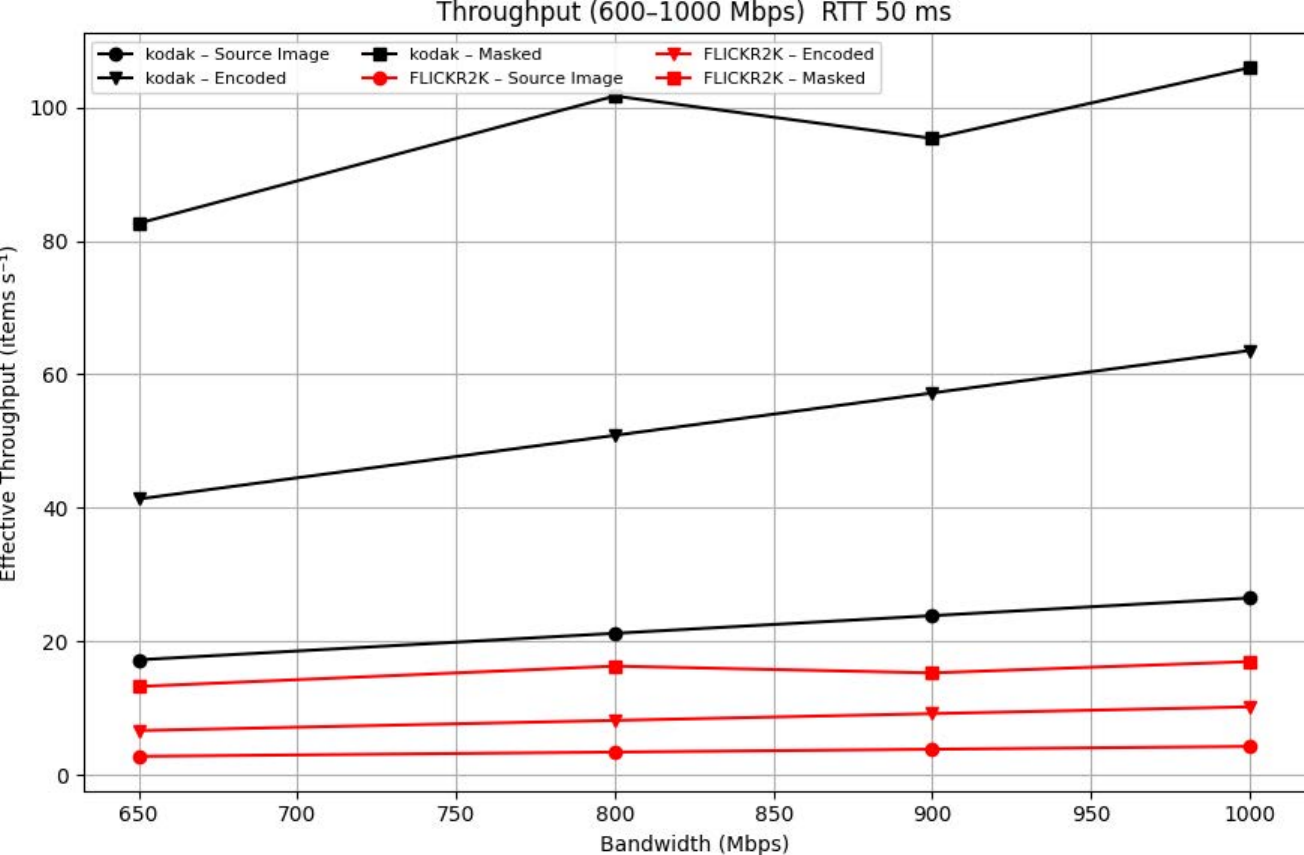}
                \caption{Effective throughput versus link bandwidth across Kodak and FLICKR2K datasets}
                \label{fig:ch05:net_thr}
            \end{figure}

            Fig.~\ref{fig:ch05:net_thr} plots the number of frames that can be delivered per second, as the link capacity rises from 10 Mbps to 1000 Mbps. Throughput is calculated as [$B/\text{payload size}$], where $B$ is the link capacity in Mbps and the payload size $S$ is in Mbits.

            For fixed-size \textit{Source Image} and \textit{Encoded Source} payloads, the curves grow linearly, so doubling the link rate simply doubles the frame rate. 
            The slopes differ only due to the constant size ratio found in Table~\ref{tab:ch05:size_comp}.

            The \textit{Masked Latent} behaves differently. 
            At low capacities the latent is aggressively pruned, so its small footprint yields a relatively higher rate than both \textit{Source Image} and \textit{Encoded Source} payloads and by an order of magnitude at 100 Mbps. 
            As the link's capacity improves the \textit{NetAware-Modulator} widens the latent (raising $k$), so the numerator and denominator scale together and the curve flattens. 
            Once the ceiling $k_{\max}=$ 192 channels is reached (near 900 Mbps to 1000 Mbps in these tests) the masked curve resumes the linear trend.
            The result is a codec that maximizes frame rate when the network is tight yet still exploits extra capacity by trading surplus bandwidth for quality rather than idle head-room.
     
        \subsubsection{Image Reconstruction Similarity}
\begin{figure}[H]
                \centering
              \includegraphics[width=0.77\linewidth]{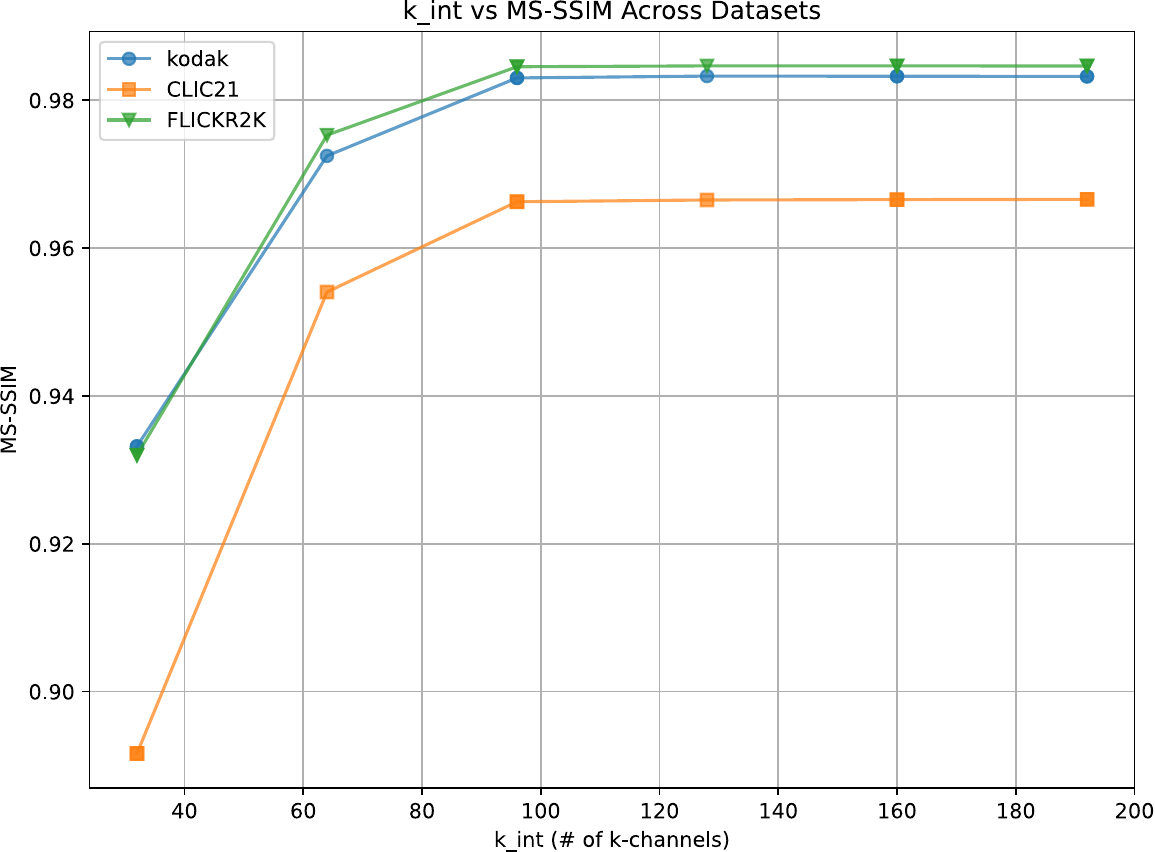}
                \caption{MS-SSIM scores of Kodak, CLIC21 and FLIC characteristically same sweep}
                \label{fig:ch05:combined_msssim}
            \end{figure}

            Fig.~\ref{fig:ch05:combined_msssim} plots MS-SSIM against the active channel budget $k$ at the x-axis. 
            All three datasets rise rapidly at small budgets, and by $k=$ 96 the curves effectively leveled off. 
            Beyond this point, additional channels yields only marginal improvements, signaling the transformer needs roughly half of the defined number of $k$ channels to achieve its best quality on a busy link. 
            It is worth noting that different autoencoder architectures may behave differently, in other words, the decoder might need more channels to accurately reconstruct the \textit{Masked Latent} and this performance is strictly associated to the Swin Transformer.

            Although the three curves converge in slope after $k =$ 96, their ceilings differ in a characteristic way. 
            \textit{CLIC21} records the lowest MS-SSIM score ($\approx$ 0.967). 
            Its lightly compressed web photos that is characterized by a mild ringing and edge softness results in a lower structural similarity. 
            By contrast, \textit{FLICKR2K} posts the highest MS-SSIM ($\approx$ 0.985). 
            Its sharper, highly textured scenes preserve local structure but incur small color/exposure shifts that inflate mean-square error. The \textit{Kodak} film scans sit between the two extremes, finishing at $\approx$ 0.983.

            The inversion lower MS-SSIM \textit{CLIC21} but higher for \textit{FLICKR2K}, highlights the complementary sensitivities of the metric rather than any weakness of the codec.
            Most importantly, once $k \ge$ 96 every dataset remains greater than the 0.96 MS-SSIM range across the entire bandwidth sweep, confirming that the adaptive channel mask delivers consistently high perceptual fidelity with only half of the latent channels.
      
        \subsubsection{Transmission Latency}
\begin{figure}[h]
                \centering
                \includegraphics[width=0.8\linewidth]{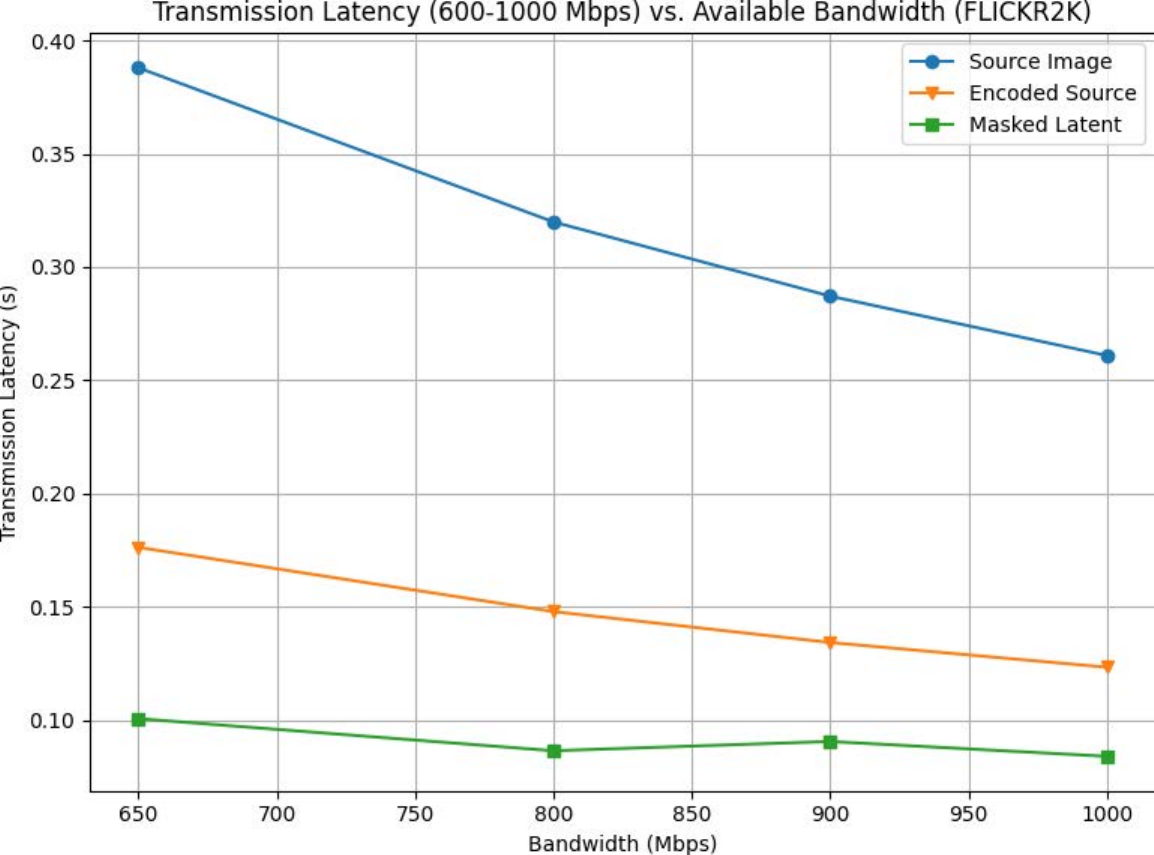}
                \caption{Zoomed transmission latency of masked latent representations vs original image from FLICKR2K dataset}
                \label{fig:ch05:tx_lat}
            \end{figure}

            Fig.~\ref{fig:ch05:tx_lat} zooms in on the \textit{FLICKR2K} results around the 650 Mbps to 1000 Mbps region contrasting the \textit{Source Image} and the \textit{Masked Latent} payloads.

            The raw frame size is 236 Mb resulting in a transmission latency that exceeds 250 ms even at 1000 Mbps link, while the \textit{Masked Latent} whose sizes ranges from 9.8 Mb @ $k=$ 32 to 59 Mb at $k=$ 192, falls under 60 ms under the same conditions. 
            The channel reduction remains substantial across the sweep: 96 \% at 10 Mbps, 88 \% at 400 Mbps, and 75 \% at 1000 Mbps. 
            Latency therefore tracks the channel budget: each time the \textit{NetAware-Modulator} increments \(k\) the curve steps upward, yet even at the ceiling $k_{\max}=$ 192 the masked payload needs only one-quarter of the transmission time of the original image.

            These results confirm that adaptive channel masking converts additional bandwidth primarily into lower delay rather than idle capacity, keeping end-to-end latency below 100 ms for every link rate above 650 Mbps and below one second down to 10 Mbps well within the interactive threshold for mobile-edge applications.

    \subsubsection{Runtime Complexity, Energy and Computational Processing Power}
        Table~\ref{tab:ch05:runtime_complexity} reports the encoder and decoder parameter counts and FLOPs, which serve as proxies for runtime complexity. 
        Both modules are symmetric in size, requiring roughly 9M parameters each. 
        The encoder performs 17.50 GFLOPs per frame, and the decoder 17.02 GFLOPs, for a combined cost of $\approx$34.5 GFLOPs per image.

        \begin{table}[h]
            \centering
            \caption{Encoder/Decoder complexity of SwinNetCC}
            \label{tab:ch05:runtime_complexity}
            \begin{tabular}{|c|c|c|}
                \hline
                Module & Parameters (M) & FLOPs (G) \\ \hline
                Encoder & 9.14 & 17.50 \\ \hline
                Decoder & 9.14 & 17.02 \\ \hline
                Total   & 18.28 & 34.52 \\ \hline
            \end{tabular}
        \end{table}
        
        The total latency of SwinNetCC is the sum of encoder time, transmission delay, and decoder time. 
        On modern GPUs, encoding and decoding together contribute a nearly constant overhead of $\approx$ 0.26-0.77 ms per frame. 
        At low bandwidths, transmission dominates (e.g., several seconds at 10~Mbps), making the compute cost negligible. 
        At higher capacities, however, transmission latency shrinks (e.g., $<100$~ms at 1~Gbps), so the fixed codec overhead becomes a significant fraction of the end-to-end delay. 
        Thus, total latency ranges from seconds in constrained links to well below 1s at gigabit rates, with codec runtime defining the floor.

        Using a typical GPU efficiency of 0.1--0.3~nJ/FLOP, the combined 34.5~GFLOPs per frame corresponds to $\approx$3.5-10.4~J per image. 
        At a streaming rate of 30~fps, this translates to an average power draw of $\approx$105-312~W, consistent with high-end GPU operation. 
        While feasible for data-center MEC servers, this footprint highlights the need for compression and quantization to enable energy-efficient deployment on constrained edge devices.
        
        SwinNetCC requires $\approx$34.5 GFLOPs per frame. 
        For real-time streaming at 30 fps, the sustained processing requirement is $\approx$1.0 TFLOP/s, achievable on commodity GPUs but challenging for resource-constrained MEC devices or mobile SoCs. 
        This highlights the importance of future work on pruning, quantization, and hardware-aware model adaptation to reduce the computational burden.

\subsection{section Summary}
\label{ch05:summary}
    This section proposed \textsc{SwinNetCC}, a network-aware adaptive semantic transcoding framework designed to optimize edge-to-cloud image transmission over fluctuating wired backhaul links.
    Addressing the limitations of static compression schemes, the proposed system integrates a Swin Transformer-based autoencoder with a novel Network-Aware Modulator.
    This modulator dynamically adjusts the semantic payload size by selecting the top-$k$ most relevant feature channels based on real-time network telemetry, effectively trading distinct semantic fidelity for transmission latency.

    Experimental validation across the 3GPP core-network envelope demonstrated that the framework significantly outperforms raw data transmission in bandwidth-constrained scenarios. 
    The results indicated that the adaptive masking policy maintains a bounded bandwidth utilization and achieves an effective throughput order of magnitude higher than standard encoding at low link capacities. 
    Finally, the analysis of reconstruction quality confirmed that retaining only the most salient semantic channels (approximately half of the total budget) is sufficient to maintain high perceptual fidelity (MS-SSIM $> 0.96$) while keeping end-to-end transmission latency below 100 ms in gigabit environments.




\section{Security, Trust, and Operationalization Considerations}
\label{sec:security}
\noindent
Although semantic communication is often motivated by efficiency, practical deployments must incorporate security, trust, and resilience as first-class constraints. Semantic systems expose additional attack surfaces beyond bit-level integrity, including semantic manipulation, latent-space poisoning, and inference-time adversarial behavior. Consequently, survey-level system design should consider defenses that span physical attributes, access control, and runtime enforcement.

\subsection{Physical and Link-Layer Trust Anchors}
Physical-layer authentication and security mechanisms provide complementary trust anchors by leveraging channel- and hardware-dependent features. Representative approaches include time-varying CFO-based authentication and multilayer physical-attribute authentication, which illustrate how non-payload attributes can support identity verification under mobility and channel dynamics \cite{x1,x19}. In parallel, stochastic-geometry-based security analysis has been used to quantify physical-layer security performance in IoT and hybrid access settings \cite{x7,x30}.

\subsection{Zero-Trust and Software-Defined Perimeter for Edge-Cloud Systems}
Zero-trust conceptualizations and SDP architectures are directly relevant to semantic edge-cloud systems, because they emphasize continuous verification, micro-segmentation, and policy-driven access control across dynamic service graphs \cite{x2,x10}. SDP has been explored as a secure solution for modern networks, NFV environments, SDN control planes, and IaaS settings \cite{x2,x21,x17,x14}, including IoT messaging ecosystems \cite{x20}. Extensions that combine SDP with moving target defense further strengthen resilience for decentralized and learning-enabled environments \cite{x5,x6}.

\subsection{Distributed Trust and Blockchain-Assisted Coordination}
Blockchain frameworks can provide auditable coordination and secure task sharing across MEC nodes, particularly when semantic workloads are distributed and require provenance and policy enforcement \cite{x31,x29}. Related approaches explore blockchain-enabled healthcare systems and policy and charging control for cellular roaming, illustrating how programmable trust can be integrated with edge/cloud workflows \cite{x25,x26}.

\subsection{Orchestration, Availability, and Resource Sharing}
Operationalization also depends on resilient orchestration and elastic resource management. Predictive auto-scaling for cellular core components \cite{x11}, secure migration under partial observability \cite{x28}, RL-based placement in O-RAN environments \cite{x27}, and consensus-based resource sharing using Raft \cite{x4} provide system-level mechanisms that can be layered with semantic communication to meet availability and performance objectives.

\section{Synthesis and Research Directions}
\label{sec:future}
\noindent
This section consolidates high-level synthesis and actionable research directions, focusing on standardization, cross-layer co-design, benchmarking, and hardware-software co-optimization for AI-native and 6G-ready semantic networking.


\label{ch06}

\subsection{section Overview}
\label{ch06:overview}
    This section consolidates the survey outcomes and positions the work within the broader trajectory of edge-cloud semantic communication for next-generation networks. 
    It first summarizes the principal findings and contributions across Chapters~\ref{ch03}-\ref{ch05}, emphasizing how telemetry-driven semantic transcoding enables a practical latency-fidelity trade-off at the edge-cloud interface. 
    It then outlines concrete future research directions required to transition from controlled experiments to deployable, SLA-governed semantic communication services.

\subsection{Summary of the survey}
\label{ch06:proposal_summary}
    This survey explores the integration of semantic communication into edge-cloud environments, shifting the focus from physical-layer noise resilience to transport-layer latency optimization and Service Level Agreement (SLA) compliance. 
    Leveraging the O-RAN architecture and Swin Transformers, the research establishes a three-tier system model that decomposes end-to-end latency into computational and transmission components, identifying the edge-to-cloud backhaul as a critical bottleneck. 
    Through an SLA-driven performance analysis, the work demonstrates that while edge-side encoding constitutes a significant computational burden, semantic transcoding effectively mitigates network congestion by pruning redundant features prior to transmission. 
    The study culminates in the development of \textsc{SwinNetCC}, a network-aware adaptive framework that utilizes real-time telemetry to dynamically modulate the semantic payload size via a channel masking mechanism. 
    This approach ensures predictable end-to-end performance, enabling the system to balance high-fidelity reconstruction with the stringent latency requirements of next-generation networks.

    \subsubsection{section 3: SLA-Driven Performance Prediction and Deployment Feasibility}
        section~\ref{ch03} developed a predictive framework for assessing Service Level Agreement (SLA) compliance, modeling end-to-end latency as a composite of edge processing, network transmission, and server-side reconstruction.
        The experimental analysis revealed that the system operates in a heavily compute-bound regime; despite the application of dynamic quantization, the Swin Transformer-based encoder on the edge device remained the dominant bottleneck, consistently preventing the system from meeting sub-100~ms latency targets regardless of network conditions.

        Consequently, the study concluded that for latency-sensitive applications, the edge device must function primarily as a transmission node, offloading intensive semantic feature extraction to the proximal edge server (MEC). 
        These findings underscore that while semantic communication reduces transmission load, its effective deployment requires a rigorous co-design of neural architecture and workload placement, motivating the adaptive trade-off mechanisms explored in the subsequent chapters.
    
    \subsubsection{section 4: Semantic Transcoding and Compute Footprint Characterization}
        section~\ref{ch04} introduced a semantic transcoding method designed to reduce latency in core network systems by optimizing bandwidth usage and minimizing unnecessary data transmissions. 
        By utilizing a transformer-based model for edge-to-cloud data transmission, the proposed approach demonstrated a significant reduction in transmission volume and lower overall latency compared to raw data transfer. 
        This reduction is critical for supporting next-generation, latency-sensitive applications such as AR, VR, and IoT, where the backhaul often becomes a bottleneck. 
        Through experimental analysis within the Edge-Cloud computing paradigm, this research validated the efficacy of semantic transcoding, demonstrating its capability to handle complex datasets more efficiently than conventional transmission methods.
    
    \subsubsection{section 5: Network-Aware Adaptive Semantic Transcoding at the Edge-Cloud Interface}
        This section presented \textsc{SwinNetCC}, a novel adaptive semantic communication framework designed to address the dynamic nature of edge-to-cloud backhaul links. 
        By integrating a Swin Transformer-based autoencoder with a deterministic Network-Aware Modulator, the system establishes a mechanism to move beyond static compression ratios. 
        The core strategy of dynamically masking feature channels based on real-time telemetry allows the system to prioritize semantic relevance over raw pixel data, ensuring that transmission payloads are scaled proportionally to instantaneous network capacity to prevent congestion while maximizing throughput.
        
        Experimental analysis across varying bandwidth conditions confirmed the efficacy of this adaptive approach, demonstrating substantial reductions in transmission latency compared to raw data transfer without effectively compromising reconstruction quality. 
        The results specifically highlighted that retaining only the most salient feature channels is sufficient to yield high structural similarity scores (MS-SSIM $> 0.96$) while keeping end-to-end delays within interactive thresholds. 
        These findings validate that network-driven semantic pruning is a viable and robust methodology for guaranteeing Quality of Service in next-generation edge-cloud architectures.

\subsection{Future Research Directions}
\label{ch06:future_work}
    Building on the findings of this survey, several avenues for future research emerge to transition semantic communication from experimental frameworks to robust, deployable edge-cloud services.

    \subsubsection{Joint Bandwidth--Latency Control and Tail-Latency Awareness}
        The current deterministic mapping selects \(k\) primarily from bandwidth. A natural extension is a joint policy \(k=h(B,l_{\mathrm{RTT}})\) that explicitly accounts for queuing tails and jitter, optimizing not only mean latency but also high-percentile delay (e.g., P95/P99) required by strict SLAs. This can be formulated as constrained optimization:
        \[
        \min_{k}\; \mathbb{E}[d(\mathbf{x},\hat{\mathbf{x}})]\quad \text{s.t.}\quad \Pr(l_{\mathrm{E2E}}(k)>l_{\mathrm{thr}})\le \epsilon,
        \]
        where \(\epsilon\) is an allowable violation probability.
        Incorporating tail-latency awareness is particularly critical for the transition from interactive applications (SLA 2) to real-time mission-critical controls (SLA 1) identified in section~\ref{ch03}, where average performance metrics often mask transient violations that lead to system failure.

    \subsubsection{Beyond Transformers: Exploring Efficient State Space Models}
        The experimental analysis in section~\ref{ch03} and section~\ref{ch04} demonstrated that while Swin Transformers offer superior semantic representation, they impose a heavy computational burden that creates a "latency floor" on resource-constrained edge devices. 
        To address this compute-bound regime, future work should investigate emerging efficient architectures such as State Space Models (SSMs), specifically Mamba and MambaJSCC.
        Unlike Transformers, which suffer from quadratic complexity with respect to sequence length (or linear complexity with significant overhead in Swin's window shifting), SSMs offer linear time scaling and faster inference throughput.
        Benchmarking Mamba-based semantic encoders against the Swin-based results in this survey could reveal a more viable path for deploying semantic extraction directly on battery-operated IoT endpoints without necessitating immediate offloading.

    \subsubsection{Granular Model Splitting and Collaborative Inference}
        This survey primarily treated the encoder as a monolithic unit deployed either at the device or the edge server.
        However, the results in section~\ref{ch03} indicated that many edge devices (e.g., smartphones, sensors) lack the capacity for full-scale semantic encoding.
        Future research should explore granular "Split Computing" strategies where the semantic encoder is decomposed into sequential layers.
        By dynamically determining a split point, early feature extraction layers can run on the device to perform lightweight dimensionality reduction, while deeper semantic abstraction layers are offloaded to the O-DU or MEC.
        This approach aims to minimize the computational burden on the device while still reducing the payload size before it enters the backhaul, offering a middle ground between raw transmission and full edge-side semantic encoding.

    \subsubsection{Telemetry-Driven Learning for O-RAN Intelligent Control}
        The \textsc{SwinNetCC} framework presented in section~\ref{ch05} utilized a deterministic log-based modulator for transparency and stability.
        However, deterministic rules may fail to capture complex, non-linear network dynamics found in multi-tenant 5G core networks.
        A significant advancement would be to replace the fixed heuristic with a learnable policy, implemented as an xApp within the O-RAN Near-Real-Time RAN Intelligent Controller (RIC).
        Using Deep Reinforcement Learning (DRL), a "Semantic Scheduler" agent could learn optimal masking policies by interacting with the network environment, receiving rewards based on SLA compliance and reconstruction fidelity.
        This would allow the system to adapt not just to bandwidth changes, but to predictive congestion patterns and inter-slice interference, aligning with the O-RAN vision of intelligence-native networking.

    \subsubsection{Multimodal and Task-Oriented Agentic Systems}
        While this survey focused on visual reconstruction, the ultimate goal of semantic communication is to facilitate machine intelligence.
        Future extensions should broaden the framework to support multimodal data (e.g., combining vision, LiDAR, and text) and diverse downstream tasks beyond pixel-level recovery.
        By training the cloud-side decoder to output task-specific heads—such as semantic segmentation maps, object classification vectors, or scene descriptions—the system can further optimize the latent representation for utility rather than appearance[cite: 10].
        This is foundational for "Agentic Operations," where autonomous agents (e.g., drones, robots) must communicate semantic intent to coordinate actions with minimal latency, prioritizing decision-making data over high-fidelity visual streaming.

\subsection{Conclusion}
\label{ch06:conclusion}
    In conclusion, this survey bridges the gap between the theoretical promise of semantic communication and the operational realities of edge-cloud deployment. 
    By shifting the optimization objective from physical-layer noise resilience to transport-layer latency compliance, the research demonstrates that adaptive semantic transcoding is a prerequisite for reliable performance in bandwidth-constrained backhaul environments. 
    The proposed \textsc{SwinNetCC} framework validates that utilizing real-time network telemetry to modulate semantic fidelity allows distributed systems to actively navigate the trade-off between computational cost and transmission delay. 
    Ultimately, this work establishes a foundational architecture for SLA-governed intelligence, paving the way for next-generation networks where resource allocation is driven by the semantic utility of information rather than raw data volume.

\subsection{Open Problems}
\begin{itemize}
  \item \textbf{Benchmarking and reproducibility:} Community benchmarks should couple semantic tasks with realistic network traces, transport behavior, and standardized telemetry collection so that reported gains are comparable across studies.
  \item \textbf{Cross-layer semantic control:} Joint control of semantic encoder policies and transport-layer behavior remains underexplored, especially under packet loss, congestion control dynamics, and multi-path routing.
  \item \textbf{Security-by-design semantics:} Semantic objectives should be explicitly constrained by trust and policy models (e.g., zero-trust), with defenses covering latent-space manipulation, inference leakage, and adversarial inputs.
  \item \textbf{Hardware-aware deployment:} Practical systems require co-optimization across model architecture, quantization, memory bandwidth, and accelerator mapping, particularly for edge devices with strict energy constraints \cite{x9,x12}.
  \item \textbf{6G integration:} Semantic networking should be aligned with AI-native control, zero-touch operation, and service-based architectures envisioned for beyond-5G/6G deployments \cite{x8,x6}.
\end{itemize}

\section{Conclusion}
This survey presents a comprehensive and system-oriented review of network-aware semantic communication for edge-cloud collaborative intelligence. By consolidating architectural designs, semantic representation and abstraction methods, network-adaptive encoding strategies, and learning-driven optimization mechanisms, the survey clarifies how semantic communication departs from traditional bit-centric paradigms and enables task-aware information exchange under heterogeneous network and resource conditions. Particular emphasis was placed on the interaction between semantic processing and practical system considerations, including latency and bandwidth variability, computation offloading, orchestration across edge and cloud resources, and reproducible performance evaluation.

Beyond efficiency and adaptability, the survey contextualized semantic communication within operational requirements that are critical for real-world deployment, such as security, trust, resilience, and scalability. By drawing connections to zero-trust networking principles, physical-layer security techniques, and emerging control and management frameworks, the survey highlights the necessity of integrating semantic communication with end-to-end system protection and governance mechanisms. Collectively, the surveyed works and synthesized perspectives provide a structured reference for researchers and practitioners seeking to design, evaluate, and deploy semantic networking systems, while also identifying open challenges that must be addressed to realize AI-native, 6G-ready edge-cloud intelligence at scale.

\bibliographystyle{IEEEtran}
\bibliography{refs_updated}

@INPROCEEDINGS{11073603,
    author={Shah, Sarah and Amannejad, Yasaman and Krishnamurthy, Diwakar},
    booktitle={NOMS 2025-2025 IEEE Network Operations and Management Symposium}, 
    title={SLA-Driven Performance Prediction for Co-Hosted DNNs}, 
    year={2025},
    volume={},
    number={},
    pages={1-7},
    doi={10.1109/NOMS57970.2025.11073603}
}

@inproceedings{yang2022_witt,
    title={{WITT}: A wireless image transmission transformer for semantic communications},
    author={Yang, Ke and Wang, Sixian and Dai, Jincheng and Tan, Kailin and Niu, Kai and Zhang, Ping},
    booktitle={ICASSP 2023-2023 IEEE International Conference on Acoustics, Speech and Signal Processing (ICASSP)},
    pages={1--5},
    year={2023},
    organization={IEEE}
}

@inproceedings{CollPerales2023_E2E-V2X,
    author    = {Baldomero Coll-Perales and M. Carmen Lucas-Esta{\~n} and Takayuki Shimizu and Javier Gozalvez and Takamasa Higuchi and Sergei Avedisov and Onur Altintas and Miguel Sepulcre},
    title     = {End-to-End V2X Latency Modeling and Analysis in 5G Networks},
    booktitle = {Proceedings of the International Conference on Communications},
    year      = {2023},
    pages     = {1-15},
    publisher = {IEEE},
    note      = {Available online; accessed April 15, 2024}
}

@inproceedings{liu2021swin,
    author = {Liu, Ze and Lin, Yutong and Cao, Yue and Hu, Han and Wei, Yixuan and Zhang, Zheng and Lin, Stephen and Guo, Baining},
    title = {Swin Transformer: Hierarchical Vision Transformer Using Shifted Windows},
    booktitle = {ICCV 2021},
    year = {2021},
    month = {October},
    url = {https://www.microsoft.com/en-us/research/publication/swin-transformer-hierarchical-vision-transformer-using-shifted-windows/},
}

@article{1,
    author={Sharma, Shree Krishna and Woungang, Isaac and Anpalagan, Alagan and Chatzinotas, Symeon},
    journal={IEEE Access}, 
    title={Toward Tactile Internet in Beyond 5G Era: Recent Advances, Current Issues, and Future Directions}, 
    year={2020},
    volume={8},
    number={},
    pages={56948-56991},
    
    doi={10.1109/ACCESS.2020.2980369}}

@article{2,
    author = {Innovation Spotlight},
    title = {How Advanced 5G and Beyond Networks Can Unlock the IoT Revolution},
    journal = {IEEE Innovate},
    year = {2023},
    url = {https://innovate.ieee.org/how-advanced-5g-and-beyond-networks-can-unlock-the-iot-revolution/}
}

@article{3,
    author={Tang, Shunpu and Chen, Lunyuan and He, Ke and Xia, Junjuan and Fan, Lisheng and Nallanathan, Arumugam},
    journal={IEEE Transactions on Network Science and Engineering}, 
    title={Computational Intelligence and Deep Learning for Next-Generation Edge-Enabled Industrial IoT}, 
    year={2023},
    volume={10},
    number={5},
    pages={2881-2893},
    doi={10.1109/TNSE.2022.3180632}}

@article{7,
    author={Guo, Jie and Chen, Hao and Song, Bin and Chi, Yuhao and Yuen, Chau and Yu, Fei Richard and Li, Geoffrey Ye and Niyato, Dusit},
    journal={IEEE Communications Magazine}, 
    title={Distributed Task-Oriented Communication Networks with Multimodal Semantic Relay and Edge Intelligence}, 
    year={2024},
    volume={},
    number={},
    pages={1-7},
    doi={10.1109/MCOM.001.2300155}
}

@ARTICLE{12,
  author={Xu, Wei and Yang, Zhaohui and Ng, Derrick Wing Kwan and Levorato, Marco and Eldar, Yonina C. and Debbah, Mérouane},
  journal={IEEE Journal of Selected Topics in Signal Processing}, 
  title={Edge Learning for B5G Networks With Distributed Signal Processing: Semantic Communication, Edge Computing, and Wireless Sensing}, 
  year={2023},
  volume={17},
  number={1},
  pages={9-39},
  doi={10.1109/JSTSP.2023.3239189}}

@article{yang2023_swinjscc,
    author={Ke Yang and Sixian Wang and Jincheng Dai and Xiaoqi Qin and Kai Niu and Ping Zhang},
    title={SwinJSCC: Taming Swin Transformer for Deep Joint Source-Channel Coding}, 
    journal = {arXiv},
    year = {2023}
}

@article{qin2022semantic,     
    author={Zhijin Qin and Xiaoming Tao and Jianhua Lu and Wen Tong and Geoffrey Ye Li},
    title={Semantic Communications: Principles and Challenges}, 
    journal = {arXiv},
    year = {2022}
}

@ARTICLE{shafi2017_5G-tutorial,
    author={Shafi, Mansoor and Molisch, Andreas F. and Smith, Peter J. and Haustein, Thomas and Zhu, Peiying and De Silva, Prasan and Tufvesson, Fredrik and Benjebbour, Anass and Wunder, Gerhard},
    journal={IEEE Journal on Selected Areas in Communications}, 
    title={5G: A Tutorial Overview of Standards, Trials, Challenges, Deployment, and Practice}, 
    year={2017},
    volume={35},
    number={6},
    pages={1201-1221},
    doi={10.1109/JSAC.2017.2692307}
}

@article{candela2020_covid19-impact,
    title = {Impact of the COVID-19 pandemic on the Internet latency: A large-scale study},
    journal = {Computer Networks},
    volume = {182},
    pages = {107495},
    year = {2020},
    issn = {1389-1286},
    doi = {https://doi.org/10.1016/j.comnet.2020.107495},
    author = {Massimo Candela and Valerio Luconi and Alessio Vecchio},
}

@article{LIU2023,
    author = {Yating Liu and Xiaojie Wang and Zhaolong Ning and MengChu Zhou and Lei Guo and Behrouz Jedari},
    title = {A survey on semantic communications: technologies, solutions, applications and challenges},
    journal = {Digital Communications and Networks},
    year = {2023},
    issn = {2352-8648},
    doi = {https://doi.org/10.1016/j.dcan.2023.05.010},
}

@article{yang2022semantic,
    author={Yang, Wanting and Liew, Zi Qin and Lim, Wei Yang Bryan and Xiong, Zehui and Niyato, Dusit and Chi, Xuefen and Cao, Xianbin and Letaief, Khaled B.},
    journal={IEEE Wireless Communications}, 
    title={Semantic Communication Meets Edge Intelligence}, 
    year={2022},
    volume={29},
    number={5},
    pages={28-35},
    doi={10.1109/MWC.004.2200050}
}

@misc{4,
    author = {DataReportal},
    title = {Internet use in 2024 — DataReportal – Global Digital Insights},
    year = {2024},
    url = {https://datareportal.com/reports/digital-2024-global-overview-report}
}

@article{6,
    author={Chengsi Liang and Hongyang Du and Yao Sun and Dusit Niyato and Jiawen Kang and Dezong Zhao and Muhammad Ali Imran},
    title={Generative AI-driven Semantic Communication Networks: Architecture, Technologies and Applications}, 
    journal = {arXiv},
    year = {2024},
    url = {https://arxiv.org/abs/2401.00124}
}

@Article{s23083876,
    AUTHOR = {Pons, Mario and Valenzuela, Estuardo and Rodríguez, Brandon and Nolazco-Flores, Juan Arturo and Del-Valle-Soto, Carolina},
    TITLE = {Utilization of 5G Technologies in IoT Applications: Current Limitations by Interference and Network Optimization Difficulties—A Review},
    JOURNAL = {Sensors},
    VOLUME = {23},
    YEAR = {2023},
    NUMBER = {8},
    ARTICLE-NUMBER = {3876},
    URL = {https://www.mdpi.com/1424-8220/23/8/3876},
    PubMedID = {37112216},
    ISSN = {1424-8220},
    DOI = {10.3390/s23083876}
}

@online{ericsson2023mobile,
    title = {Mobile Traffic Forecast},
    author = {{Ericsson Mobility Report}},
    year = {2023},
    url = {https://www.ericsson.com},
    note = {Accessed: Sep. 13, 2024}
}

@ARTICLE{9585402,
    author={Berardinelli, Gilberto and Baracca, Paolo and Adeogun, Ramoni O. and Khosravirad, Saeed R. and Schaich, Frank and Upadhya, Karthik and Li, Dong and Tao, Tao and Viswanathan, Harish and Mogensen, Preben},
    journal={IEEE Open Journal of the Communications Society}, 
    title={Extreme Communication in 6G: Vision and Challenges for ‘in-X’ Subnetworks}, 
    year={2021},
    volume={2},
    number={},
    pages={2516-2535},
    doi={10.1109/OJCOMS.2021.3121530}
}

@online{vodafone2021,
    title = {Vodafone and Nokia showcase 100 Gigabit Next Generation Broadband Network},
    author = {{Vodafone Group}},
    year = {2021},
    url={https://www.vodafone.com},
    urldate = {Accessed: Sep. 13, 2024}
}

@Article{su16167039,
    AUTHOR = {Sharma, Sanjeev and Popli, Renu and Singh, Sajjan and Chhabra, Gunjan and Saini, Gurpreet Singh and Singh, Maninder and Sandhu, Archana and Sharma, Ashutosh and Kumar, Rajeev},
    TITLE = {The Role of 6G Technologies in Advancing Smart City Applications: Opportunities and Challenges},
    JOURNAL = {Sustainability},
    VOLUME = {16},
    YEAR = {2024},
    NUMBER = {16},
    ARTICLE-NUMBER = {7039},
    URL = {https://www.mdpi.com/2071-1050/16/16/7039},
    ISSN = {2071-1050},
    DOI = {10.3390/su16167039}
}

@INPROCEEDINGS{10667089,
    author={Nisyif, Murtadha and Refaey, Ahmed and Aboagye, Sylvester},
    booktitle={2024 IEEE Canadian Conference on Electrical and Computer Engineering (CCECE)}, 
    title={Boosting Edge-to-Cloud Data Transmission Efficiency with Semantic Transcoding}, 
    year={2024},
    volume={},
    number={},
    pages={730-734},
    doi={10.1109/CCECE59415.2024.10667089}
}

@ARTICLE{8972389,
    author={Shafique, Kinza and Khawaja, Bilal A. and Sabir, Farah and Qazi, Sameer and Mustaqim, Muhammad},
    journal={IEEE Access}, 
    title={Internet of Things (IoT) for Next-Generation Smart Systems: A Review of Current Challenges, Future Trends and Prospects for Emerging 5G-IoT Scenarios}, 
    year={2020},
    volume={8},
    number={},
    pages={23022-23040},
    doi={10.1109/ACCESS.2020.2970118}
}

@ARTICLE{5563107,
    author={Fresia, Maria and Peréz-Cruz, Fernando and Poor, H. Vincent and Verdú, Sergio},
    journal={IEEE Signal Processing Magazine}, 
    title={Joint Source and Channel Coding}, 
    year={2010},
    volume={27},
    number={6},
    pages={104-113},
    doi={10.1109/MSP.2010.938080}
}

@ARTICLE{10158528,
    author={Dai, Jincheng and Wang, Sixian and Yang, Ke and Tan, Kailin and Qin, Xiaoqi and Si, Zhongwei and Niu, Kai and Zhang, Ping},
    journal={IEEE Journal on Selected Areas in Communications}, 
    title={Toward Adaptive Semantic Communications: Efficient Data Transmission via Online Learned Nonlinear Transform Source-Channel Coding}, 
    year={2023},
    volume={41},
    number={8},
    pages={2609-2627},
    doi={10.1109/JSAC.2023.3288246}
}

@INPROCEEDINGS{9685667,
    author={Huang, Danlan and Tao, Xiaoming and Gao, Feifei and Lu, Jianhua},
    booktitle={2021 IEEE Global Communications Conference (GLOBECOM)}, 
    title={Deep Learning-Based Image Semantic Coding for Semantic Communications}, 
    year={2021},
    volume={},
    number={},
    pages={1-6},
    doi={10.1109/GLOBECOM46510.2021.9685667}
}

@misc{graves2017adaptivecomputationtimerecurrent,
    title={Adaptive Computation Time for Recurrent Neural Networks}, 
    author={Alex Graves},
    year={2017},
    eprint={1603.08983},
    archivePrefix={arXiv},
    primaryClass={cs.NE},
    url={https://arxiv.org/abs/1603.08983}, 
}

@misc{zhou2021semanticcommunicationadaptiveuniversal,
    title={Semantic Communication with Adaptive Universal Transformer}, 
    author={Qingyang Zhou and Rongpeng Li and Zhifeng Zhao and Chenghui Peng and Honggang Zhang},
    year={2021},
    eprint={2108.09119},
    archivePrefix={arXiv},
    primaryClass={cs.CL},
    url={https://arxiv.org/abs/2108.09119}, 
}

@techreport{3gpp-nwdaf,
  title        = {{Architecture enhancements for 5G System (5GS) to support network data analytics services}},
  institution  = {3GPP},
  number       = {TS 23.288 V18.2.0},
  year         = {2023},
  month        = {June},
  note         = {Release 18}
}

@misc{ibm_sla_2024,
  author = {{IBM Corporation}},
  title = {What is a Service Level Agreement (SLA)?},
  year = {2024},
  howpublished = {\url{https://www.ibm.com/think/topics/service-level-agreement}}
}

@misc{aakbari_2023,
      title={Service Level Agreements for Communication Networks: A Survey}, 
      author={Ayyoub Akbari-Moghanjoughi and José Roberto de Almeida Amazonas and Germán Santos-Boada and Josep Solé-Pareta},
      year={2023},
      eprint={2309.07272},
      archivePrefix={arXiv},
      primaryClass={cs.NI},
      url={https://arxiv.org/abs/2309.07272}, 
}

@misc{pravez2018,
      title={A Survey on Low Latency Towards 5G: RAN, Core Network and Caching Solutions}, 
      author={Imtiaz Parvez and Ali Rahmati and Ismail Guvenc and Arif I. Sarwat and Huaiyu Dai},
      year={2018},
      eprint={1708.02562},
      archivePrefix={arXiv},
      primaryClass={cs.NI},
      url={https://arxiv.org/abs/1708.02562}, 
}

@ARTICLE{10287312,
  author={Municio, Esteban and Garcia-Aviles, Gines and Garcia-Saavedra, Andres and Costa-Pérez, Xavier},
  journal={IEEE Communications Standards Magazine}, 
  title={O-RAN: Analysis of Latency-Critical Interfaces and Overview of Time Sensitive Networking Solutions}, 
  year={2023},
  volume={7},
  number={3},
  pages={82-89},
  doi={10.1109/MCOMSTD.0001.2200041}}

@misc{Ericsson2024,
  author       = {{Ericsson}},
  title        = {Ericsson Mobility Report: early movers pursue performance-based business models},
  year         = {2024},
  month        = nov,
  day          = {26},
  howpublished = {Press Release},
  url          = {https://www.ericsson.com/en/press-releases/2024/11/ericsson-mobility-report-early-movers-pursue-performance-based-business-models},
  note         = {Accessed: Nov. 28, 2025}
}

@article{xin_semantic_2024,
	title = {Semantic {Communication}: {A} {Survey} of {Its} {Theoretical} {Development}.},
	volume = {26},
	issn = {1099-4300},
	doi = {10.3390/e26020102},
	language = {eng},
	number = {2},
	journal = {Entropy (Basel, Switzerland)},
	author = {Xin, Gangtao and Fan, Pingyi and Letaief, Khaled B.},
	month = jan,
	year = {2024},
	pmid = {38392357},
	pmcid = {PMC10888479},
	note = {Place: Switzerland},
}

@ARTICLE{9398576,
  author={Xie, Huiqiang and Qin, Zhijin and Li, Geoffrey Ye and Juang, Biing-Hwang},
  journal={IEEE Transactions on Signal Processing}, 
  title={Deep Learning Enabled Semantic Communication Systems}, 
  year={2021},
  volume={69},
  number={},
  pages={2663-2675},
  doi={10.1109/TSP.2021.3071210}}

@misc{weng2023deeplearningenabledsemantic,
      title={Deep Learning Enabled Semantic Communications with Speech Recognition and Synthesis}, 
      author={Zhenzi Weng and Zhijin Qin and Xiaoming Tao and Chengkang Pan and Guangyi Liu and Geoffrey Ye Li},
      year={2023},
      eprint={2205.04603},
      archivePrefix={arXiv},
      primaryClass={eess.AS},
      url={https://arxiv.org/abs/2205.04603}, 
}

@article{Zhang_2025,
   title={Semantic Edge Computing and Semantic Communications in 6G networks: A unifying survey and research challenges},
   volume={270},
   ISSN={1389-1286},
   url={http://dx.doi.org/10.1016/j.comnet.2025.111531},
   DOI={10.1016/j.comnet.2025.111531},
   journal={Computer Networks},
   publisher={Elsevier BV},
   author={Zhang, Milin and Abdi, Mohammad and Dasari, Venkat R. and Restuccia, Francesco},
   year={2025},
   month=oct, pages={111531} }

@INPROCEEDINGS{x,
  author={Figetakis, Emanuel and Bello, Yahuza and Refaey, Ahmed and Shami, Abdallah},
  booktitle={ICC 2024 - IEEE International Conference on Communications}, 
  title={Decentralized Semantic Traffic Control in AVs Using RL and DQN for Dynamic Roadblocks}, 
  year={2024},
  volume={},
  number={},
  pages={5449-5454},
  doi={10.1109/ICC51166.2024.10622833}}

@ARTICLE{y,
  author={Bello, Yahuza and Hussein, Ahmed Refaey},
  journal={IEEE Transactions on Network and Service Management}, 
  title={Dynamic Policy Decision/Enforcement Security Zoning Through Stochastic Games and Meta Learning}, 
  year={2025},
  volume={22},
  number={1},
  pages={807-821},
  doi={10.1109/TNSM.2024.3481662}}

@article{Bourtsoulatze_2019,
   title={Deep Joint Source-Channel Coding for Wireless Image Transmission},
   volume={5},
   ISSN={2372-2045},
   url={http://dx.doi.org/10.1109/TCCN.2019.2919300},
   DOI={10.1109/tccn.2019.2919300},
   number={3},
   journal={IEEE Transactions on Cognitive Communications and Networking},
   publisher={Institute of Electrical and Electronics Engineers (IEEE)},
   author={Bourtsoulatze, Eirina and Burth Kurka, David and Gunduz, Deniz},
   year={2019},
   month=sep, pages={567–579} 
}

@INPROCEEDINGS{10901610,
  author={Tian, Wenkai and Dong, Biao and Cao, Bin},
  booktitle={GLOBECOM 2024 - 2024 IEEE Global Communications Conference}, 
  title={Joint ROI Guidance and Spatial Analysis for Task-Aware Distributed Deep Joint Source-Channel Coding}, 
  year={2024},
  volume={},
  number={},
  pages={1936-1941},
  doi={10.1109/GLOBECOM52923.2024.10901610}
}

@ARTICLE{10832517,
  author={Zheng, Guangyuan and Wen, Miaowen and Ning, Zhaolong and Ding, Zhiguo},
  journal={IEEE Transactions on Wireless Communications}, 
  title={Computation-Aware Offloading for DNN Inference Tasks in Semantic Communication Assisted MEC Systems}, 
  year={2025},
  volume={24},
  number={4},
  pages={2693-2706},
  doi={10.1109/TWC.2024.3523517}
}

@misc{dehghani2019,
      title={Universal Transformers}, 
      author={Mostafa Dehghani and Stephan Gouws and Oriol Vinyals and Jakob Uszkoreit and Łukasz Kaiser},
      year={2019},
      eprint={1807.03819},
      archivePrefix={arXiv},
      primaryClass={cs.CL},
      url={https://arxiv.org/abs/1807.03819}, 
}

@misc{dosovitskiy2021,
      title={An Image is Worth 16x16 Words: Transformers for Image Recognition at Scale}, 
      author={Alexey Dosovitskiy and Lucas Beyer and Alexander Kolesnikov and Dirk Weissenborn and Xiaohua Zhai and Thomas Unterthiner and Mostafa Dehghani and Matthias Minderer and Georg Heigold and Sylvain Gelly and Jakob Uszkoreit and Neil Houlsby},
      year={2021},
      eprint={2010.11929},
      archivePrefix={arXiv},
      primaryClass={cs.CV},
      url={https://arxiv.org/abs/2010.11929}, 
}

@ARTICLE{10024837,
  author={Polese, Michele and Bonati, Leonardo and D’Oro, Salvatore and Basagni, Stefano and Melodia, Tommaso},
  journal={IEEE Communications Surveys \& Tutorials}, 
  title={Understanding O-RAN: Architecture, Interfaces, Algorithms, Security, and Research Challenges}, 
  year={2023},
  volume={25},
  number={2},
  pages={1376-1411},
  doi={10.1109/COMST.2023.3239220}}

@article{gavrilovska2020,
    author = {Gavrilovska, Liljana and Rakovic, Valentin and Denkovski, Daniel},
    year = {2020},
    month = {08},
    pages = {},
    title = {From Cloud RAN to Open RAN},
    volume = {113},
    journal = {Wireless Personal Communications},
    doi = {10.1007/s11277-020-07231-3}
}

@ARTICLE{8479363,
  author={Larsen, Line M. P. and Checko, Aleksandra and Christiansen, Henrik L.},
  journal={IEEE Communications Surveys \& Tutorials}, 
  title={A Survey of the Functional Splits Proposed for 5G Mobile Crosshaul Networks}, 
  year={2019},
  volume={21},
  number={1},
  pages={146-172},
  doi={10.1109/COMST.2018.2868805}}

@inproceedings{10.1145/2428955.2429005,
    author = {Sun, Le and Singh, Jaipal and Hussain, Omar Khadeer},
    title = {Service level agreement (SLA) assurance for cloud services: a survey from a transactional risk perspective},
    year = {2012},
    isbn = {9781450313070},
    publisher = {Association for Computing Machinery},
    address = {New York, NY, USA},
    url = {https://doi.org/10.1145/2428955.2429005},
    doi = {10.1145/2428955.2429005},
    booktitle = {Proceedings of the 10th International Conference on Advances in Mobile Computing \& Multimedia},
    pages = {263–266},
    numpages = {4},
    location = {Bali, Indonesia},
    series = {MoMM '12}
}

@ARTICLE{8016573,
  author={Mao, Yuyi and You, Changsheng and Zhang, Jun and Huang, Kaibin and Letaief, Khaled B.},
  journal={IEEE Communications Surveys \& Tutorials}, 
  title={A Survey on Mobile Edge Computing: The Communication Perspective}, 
  year={2017},
  volume={19},
  number={4},
  pages={2322-2358},
  doi={10.1109/COMST.2017.2745201}}

@ARTICLE{10896925,
  author={Tortonesi, Mauro},
  journal={Computer}, 
  title={The Compute Continuum: Trends and Challenges}, 
  year={2025},
  volume={58},
  number={3},
  pages={105-108},
  doi={10.1109/MC.2024.3520255}}

@ARTICLE{x1,
  author={Hou, Weikun and Wang, Xianbin and Chouinard, Jean-Yves and Refaey, Ahmed},
  journal={IEEE Transactions on Communications}, 
  title={Physical Layer Authentication for Mobile Systems with Time-Varying Carrier Frequency Offsets}, 
  year={2014},
  volume={62},
  number={5},
  pages={1658-1667},
  doi={10.1109/TCOMM.2014.032914.120921}}

@ARTICLE{x2,
  author={Moubayed, Abdallah and Refaey, Ahmed and Shami, Abdallah},
  journal={IEEE Network}, 
  title={Software-Defined Perimeter (SDP): State of the Art Secure Solution for Modern Networks}, 
  year={2019},
  volume={33},
  number={5},
  pages={226-233},
  doi={10.1109/MNET.2019.1800324}}

@misc{x4,
      title={Raft Distributed System for Multi-access Edge Computing Sharing Resources}, 
      author={Zain Khaliq and Ahmed Refaey Hussein},
      year={2024},
      eprint={2412.16774},
      archivePrefix={arXiv},
      primaryClass={cs.DC},
      url={https://arxiv.org/abs/2412.16774}, 
}

@INPROCEEDINGS{x5,
  author={Abdelhay, Zeyad and Refaey, Ahmed},
  booktitle={2024 International Wireless Communications and Mobile Computing (IWCMC)}, 
  title={xG Security: Zero-Trust and Moving Target Defense in Decentralized Learning Environment}, 
  year={2024},
  pages={1820-1825},
  doi={10.1109/IWCMC61514.2024.10592368}}

@ARTICLE{x6,
  author={Abdelhay, Zeyad and Bello, Yahuza and Refaey, Ahmed},
  journal={IEEE Wireless Communications}, 
  title={Toward Zero-Trust 6GC: A Software Defined Perimeter Approach with Dynamic Moving Target Defense Mechanism}, 
  year={2024},
  volume={31},
  number={2},
  pages={74-80},
  doi={10.1109/MWC.001.2300358}}

@ARTICLE{x7,
  author={Chamkhia, Hela and Erbad, Aiman and Mohamed, Amr and Hussein, Ahmed Refaey and Al-Ali, Abdulla Khalid and Guizani, Mohsen},
  journal={IEEE Internet of Things Journal}, 
  title={Stochastic Geometry-Based Physical-Layer Security Performance Analysis of a Hybrid NOMA-PDM-Based IoT System}, 
  year={2024},
  volume={11},
  number={2},
  pages={2027-2042},
  doi={10.1109/JIOT.2023.3292262}}

@ARTICLE{x8,
  author={Baccour, Emna and Allahham, Mhd Saria and Erbad, Aiman and Mohamed, Amr and Hussein, Ahmed Refaey and Hamdi, Mounir},
  journal={IEEE Communications Magazine}, 
  title={Zero Touch Realization of Pervasive Artificial Intelligence as a Service in 6G Networks}, 
  year={2023},
  volume={61},
  number={2},
  pages={110-116},
  doi={10.1109/MCOM.001.2200508}}

@ARTICLE{x9,
  author={DeSantis, Christopher and Hussein, Ahmed Refaey},
  journal={IEEE Canadian Journal of Electrical and Computer Engineering}, 
  title={AI SoC-Based Accelerator for Speech Classification}, 
  year={2022},
  volume={45},
  number={3},
  pages={222-231},
  doi={10.1109/ICJECE.2022.3199563}}

@ARTICLE{x10,
  author={Bello, Yahuza and Hussein, Ahmed Refaey and Ulema, Mehmet and Koilpillai, Juanita},
  journal={IEEE Transactions on Network and Service Management}, 
  title={On Sustained Zero Trust Conceptualization Security for Mobile Core Networks in 5G and Beyond}, 
  year={2022},
  volume={19},
  number={2},
  pages={1876-1889},
  doi={10.1109/TNSM.2022.3157248}}

@ARTICLE{x11,
  author={Bello, Yahuza and Abdellatif, Alaa Awad and Allahham, Mhd Saria and Hussein, Ahmed Refaey and Erbad, Aiman and Mohamed, Amr and Guizani, Mohsen},
  journal={IEEE Access}, 
  title={B5G: Predictive Container Auto-Scaling for Cellular Evolved Packet Core}, 
  year={2021},
  volume={9},
  pages={158204-158214},
  doi={10.1109/ACCESS.2021.3126048}}

@INPROCEEDINGS{x12,
  author={Asad, Saad and Refaey, Ahmed},
  booktitle={2021 IEEE International Conference on Imaging Systems and Techniques (IST)}, 
  title={On IoT Edge Devices: Manifold Unsupervised Learning for SoM Platforms}, 
  year={2021},
  pages={1-5},
  doi={10.1109/IST50367.2021.9651345}}

@ARTICLE{x14,
  author={Singh, Jaspreet and Refaey, Ahmed and Koilpillai, Juanita},
  journal={Canadian Journal of Electrical and Computer Engineering}, 
  title={Adoption of the Software-Defined Perimeter (SDP) Architecture for Infrastructure as a Service}, 
  year={2020},
  volume={43},
  number={4},
  pages={357-363},
  doi={10.1109/CJECE.2020.3005316}}

@ARTICLE{x17,
  author={Sallam, Ahmed and Refaey, Ahmed and Shami, Abdallah},
  journal={IEEE Access}, 
  title={On the Security of SDN: A Completed Secure and Scalable Framework Using the Software-Defined Perimeter}, 
  year={2019},
  volume={7},
  pages={146577-146587},
  doi={10.1109/ACCESS.2019.2939780}}

@ARTICLE{x19,  
  author={A. {Refaey} and W. {Hou} and L. {Loukhaoukha}},  
  journal={Journal of Computer and Communications},  
  title={Multilayer Authentication for Communication Systems Based on Physical-Layer Attributes},  
  year={2014},  
  volume={2},  
  pages={64-75},
  doi={10.4236/jcc.2014.28007}}

@ARTICLE{x20,
  author={Refaey, A. and Sallam, A. and Shami, A.},
  journal={Electronics Letters},
  title={On IoT applications: a proposed SDP framework for MQTT},
  year={2019},
  month=oct,
  volume={55},
  number={22},
  pages={1201-1203},
  doi={10.1049/el.2019.2334}
}

@ARTICLE{x21,
  author={Singh, Jaspreet and Refaey, Ahmed and Shami, Abdallah},
  journal={IEEE Network}, 
  title={Multilevel Security Framework for NFV Based on Software Defined Perimeter}, 
  year={2020},
  volume={34},
  number={5},
  pages={114-119},
  doi={10.1109/MNET.011.1900563}}

@ARTICLE{x25,
  author={Abdellatif, Alaa Awad and Al-Marridi, Abeer Z. and Mohamed, Amr and Erbad, Aiman and Chiasserini, Carla Fabiana and Refaey, Ahmed},
  journal={IEEE Network}, 
  title={ssHealth: Toward Secure, Blockchain-Enabled Healthcare Systems}, 
  year={2020},
  volume={34},
  number={4},
  pages={312-319},
  doi={10.1109/MNET.011.1900553}}

@ARTICLE{x26,
  author={Refaey, Ahmed and Hammad, Karim and Magierowski, Sebastian and Hossain, Ekram},
  journal={IEEE Network}, 
  title={A Blockchain Policy and Charging Control Framework for Roaming in Cellular Networks}, 
  year={2020},
  volume={34},
  number={3},
  pages={170-177},
  doi={10.1109/MNET.001.1900336}}

@INPROCEEDINGS{x27,
  author={Tamim, Ibrahim and Shami, Abdallah and Refaey, Ahmed},
  booktitle={ICC 2024 - IEEE International Conference on Communications}, 
  title={Security and High-Availability While Upholding Network Defense Patterns: The Advantages of A2C in O-RAN VNF Placement}, 
  year={2024},
  pages={1072-1077},
  doi={10.1109/ICC51166.2024.10622252}}

@INPROCEEDINGS{x28,
  author={Bello, Yahuza and Refaey, Ahmed and Shami, Abdallah},
  booktitle={GLOBECOM 2023 - 2023 IEEE Global Communications Conference}, 
  title={Secure Migration in NGN: An Optimal Stopping Problem Approach with Partial Observability}, 
  year={2023},
  pages={2680-2685},
  doi={10.1109/GLOBECOM54140.2023.10437408}}

@ARTICLE{x29,
  author={Vera-Rivera, Angelo and Hossain, Ekram and Hussein, Ahmed Refaey},
  journal={IEEE Open Journal of the Communications Society}, 
  title={Exploring the Intersection of Consortium Blockchain Technologies and Multi-Access Edge Computing: Chronicles of a Proof of Concept Demo}, 
  year={2022},
  volume={3},
  pages={2203-2236},
  doi={10.1109/OJCOMS.2022.3221667}}

@ARTICLE{x30,
  author={Chamkhia, Hela and Erbad, Aiman and Al-Ali, Abdulla Khalid and Mohamed, Amr and Refaey, Ahmed and Guizani, Mohsen},
  journal={IEEE Internet of Things Journal}, 
  title={3-D Stochastic Geometry-Based Modeling and Performance Analysis of Efficient Security Enhancement Scheme for IoT Systems}, 
  year={2022},
  volume={9},
  number={9},
  pages={6663-6677},
  doi={10.1109/JIOT.2021.3112883}}

@ARTICLE{x31,
  author={Rivera, Angelo Vera and Refaey, Ahmed and Hossain, Ekram},
  journal={IEEE Network}, 
  title={A Blockchain Framework for Secure Task Sharing in Multi-Access Edge Computing}, 
  year={2021},
  volume={35},
  number={3},
  pages={176-183},
  doi={10.1109/MNET.011.2000497}}

\end{document}